\documentclass[preprint,nofootinbib]{revtex4}
\usepackage{graphicx}
\def\be{\begin{equation}}
\def\ee{\end{equation}}
\def\bea{\begin{eqnarray}}
\def\eea{\end{eqnarray}}
\def\w{{\rm w}}

\def\gev{~{\rm GeV}}
\def\tev{~{\rm TeV}}

\def\calO{{\mathcal{O}}}

\def\lsim{\mathrel{\raise.3ex\hbox{$<$\kern-.75em\lower1ex\hbox{$\sim$}}}}
\def\gsim{\mathrel{\raise.3ex\hbox{$>$\kern-.75em\lower1ex\hbox{$\sim$}}}}
\def\ifmath#1{\relax\ifmmode #1\else $#1$\fi}
\begin{document}

\bibliographystyle{revtex}

\preprint{
 {\vbox{
 \hbox{MADPH--02--1317}
 \hbox{hep-ph/0301040}}}}

\vspace*{2cm}

\title{Phenomenology of the Little Higgs Model}

\author{Tao Han, Heather E. Logan, Bob McElrath, and Lian-Tao 
Wang\footnote{than@pheno.physics.wisc.edu\\ logan@pheno.physics.wisc.edu\\
mcelrath@pheno.physics.wisc.edu\\ liantaow@pheno.physics.wisc.edu}}
\affiliation{\vspace*{0.1in}
Department of Physics, University of Wisconsin, 1150 University
Avenue, Madison, WI 53706} 

\vspace*{1.0cm}

\begin{abstract}
We study the low energy phenomenology of the little Higgs model.
We first discuss the linearized effective theory of the 
``littlest Higgs model"
and study the low energy constraints on the model parameters.
We identify sources of 
the corrections to low energy observables,
discuss model-dependent arbitrariness, and outline some
possible directions of extensions of the model in order
to evade the precision electroweak constraints.
We then explore the characteristic signatures to test the model in the current
and future collider experiments. We find that the LHC
has great potential to discover the new $SU(2)$ gauge bosons
and the possible new $U(1)$ gauge boson to the multi-TeV mass scale.
Other states such as the colored vector-like quark $T$ and
doubly-charged Higgs boson $\Phi^{++}$ may also provide
interesting signals. At a linear collider, precision measurements
on the triple gauge boson couplings could be sensitive to the new physics
scale of a few TeV. We provide a comprehensive list of the linearized
interactions and vertices for the littlest Higgs model in the appendices.
\end{abstract}


\maketitle

\section{Introduction}

One of the major motivations for physics beyond the Standard Model (SM)
is to resolve the hierarchy and fine-tuning problems  between the electroweak
scale and the Planck scale. Supersymmetric theories introduce an extended 
space-time symmetry and quadratically divergent quantum corrections 
are canceled 
due to the symmetry between the bosonic and fermionic  partners. 
This naturally stabilizes the electroweak scale against the large corrections 
in the ultra-violet (UV) regime. 
Technicolor theories introduce new strong dynamics at scales 
not much above the electroweak scale, thus defer the hierarchy problem. 
Theories with TeV scale quantum gravity reinterpret the problem completely 
by lowering the fundamental Planck scale. 
Current and future collider experiments will 
provide hints to tell us which may be the ultimately correct path.
 
Recently, there has been a new formulation for the physics of electroweak
symmetry breaking, dubbed the
``little Higgs" models
\cite{Arkani-Hamed:2001nc,Arkani-Hamed:2002pa,Arkani-Hamed:2002qx,
littlest,Low:2002ws,Schmaltz:2002wx}.
The key ideas of the little Higgs theory may be summarized by
the following points:
\begin{itemize}
\item
The Higgs fields are Goldstone bosons 
\cite{Dimopoulos:1981xc,Kaplan:1983fs,Kaplan:1983sm,Banks:1984gj,
Georgi:1984af,Georgi:ef,Dugan:1984hq},  associated with some global symmetry 
breaking at a higher scale $\Lambda_S$; 
\item
The Higgs fields acquire a mass and become pseudo-Goldstone
bosons via symmetry breaking (possibly radiatively) at the
electroweak scale;
\item
The Higgs fields remain light, being protected by the approximate global 
symmetry and free from 1-loop quadratic sensitivity to the cutoff scale
$\Lambda_S$.
\end{itemize}
The scalar mass in a generic quantum field
theory will receive quadratically divergent radiative corrections 
all the way up to the cut-off scale. 
The little Higgs model 
solves  this problem by eliminating the lowest order contributions 
via the presence of a partially broken global symmetry. 
The non-linear transformation of the Higgs fields 
under this global symmetry prohibits the existence 
of a Higgs mass term of the form $m^2 |h|^2$. 
This can also be illustrated in a more intuitive way:
Besides the Standard Model gauge bosons, there are a set of heavy
gauge bosons with the same gauge quantum numbers. 
The gauge couplings to the Higgs bosons are patterned in such a way that 
the quadratic divergence induced by the SM gauge boson loops are 
canceled by the quadratic divergence  induced by the heavy gauge
bosons at one loop level. 
One also introduces a heavy fermionic state which couples to the Higgs field 
in a specific way, so that the 1-loop quadratic divergence 
induced by the top-quark 
Yukawa  coupling to the Higgs boson is canceled. Furthermore, extra Higgs 
fields exist as the Goldstone boson multiplets from the global 
symmetry breaking.

It is interesting to note that, unlike the supersymmetry relations
between the bosons and fermions, the cancellations of the quadratic 
divergence in the little Higgs model occur between particles with the 
same statistics: divergences due to gauge bosons are canceled by
new gauge bosons and similarly for the heavy quarks.
A scale less than several TeV and the 
specification of the couplings 
to the Higgs boson are necessary requirements for the model 
to avoid fine-tuning. These features
could lead to distinctive experimental signatures, which is the subject 
for the current work. The paper is organized as follows. In Sec.~\ref{frame}, 
we lay out a concrete model as proposed in \cite{littlest}. 
We linearize the theory and discuss the important features. 
In Sec.~\ref{pheno}, we explore the characteristic phenomenology 
of this model. Regarding the constraints from the precision
electroweak data, we explore the properties associated
with the custodial $SU(2)$ breaking and the sources
which lead to the large corrections to low-energy observables in the model. 
We identify the arbitrariness in particular related to the $U(1)$ sector.
We then outline the possible fine-tunings or directions of extensions 
of the model in order to evade the precision electroweak constraints.
We also study the characteristic signals at the future collider 
experiments at the LHC and a linear collider.
We summarize our results in Sec.~\ref{summer}.
We present the detailed derivation and the
Feynman rules of the littlest Higgs model in two appendices.

\section{The Framework of the Littlest Higgs Model}
\label{frame}
 
An explicit  model has been constructed based on the 
idea of the little Higgs models,
dubbed the ``littlest Higgs model"  \cite{littlest}.
It begins with $SU(5)$ global symmetry, with a locally gauged subgroup 
$G_1\otimes G_2=[SU(2)_1\otimes U(1)_1]\otimes [SU(2)_2\otimes U(1)_2]$.  
The phase transitions associated with  the symmetry breaking
in this model proceed in two stages:  
\begin{enumerate}
\item At scale $\Lambda_S$, the global symmetry $SU(5)$ is
spontaneously broken down to its subgroup $SO(5)$ via a vev of order $f$.
Naive Dimensional Analysis
\cite{Manohar:1983md,Luty:1997fk,Cohen:1997rt} establishes a simple
relation $\Lambda_S \sim 4 \pi f$.  
At the same time, the gauge symmetry
$[SU(2)\otimes U(1)]^2$ is also broken 
into its diagonal subgroup $SU(2)_L \otimes
U(1)_Y$, identified as the SM gauge group.  
The global symmetry breaking leaves 14 massless Goldstone
bosons which transform under the electroweak gauge group as a real singlet
${\mathbf{1}}_0$, a real triplet ${\mathbf{3}}_0$, a complex doublet
${\mathbf{2}}_{\pm \frac{1}{2}}$, and a complex triplet
${\mathbf{3}}_{\pm 1}$. The real singlet and the real triplet become
the longitudinal components of the gauge bosons associated with
the broken gauge groups, giving them masses of the order $f$, 
while the complex doublet
and the complex triplet remain massless at this stage. 
\item The presence of gauge and Yukawa couplings that
break the global $SO(5)$ symmetry will induce a
Coleman-Weinberg \cite{Coleman:jx} type potential for the remaining
pseudo-Goldstone bosons. In particular, it will give the complex
triplet a heavy mass of the order $f$ and give the neutral component of the
complex doublet a non-vanishing vacuum expectation value (vev) $v$ 
which in turn triggers the electroweak symmetry breaking. 
\end{enumerate}

Before we lay out the effective field theory below the scale of $\Lambda_S$
for the littlest Higgs model, we note that some matching procedure for
the operators  
that are sensitive to physics at  higher energies will eventually be needed. 
Namely, one would need to consider the UV origin of the theory above
$\Lambda_S$. 
We will not attempt to explore the UV completion of the theory in this paper
but rather refer the reader to some discussions in the literature
\cite{littlest,UV}. 
 
\subsection{Scalar and Gauge Boson Sector}

\subsubsection{Gauge Bosons and Pseudo-Goldstone Bosons}

At the scale $\Lambda_S$, a vev $f$ breaks the assumed global $SU(5)$ symmetry into
its subgroup $SO(5)$, resulting in 14  Goldstone bosons.
The effective field theory of those Goldstone bosons is parameterized by a
non-linear $\sigma$-model with a gauge symmetry $[SU(2)\otimes U(1)]^2$,
spontaneously broken down to the Standard Model gauge group. In
particular, the Lagrangian will still preserve the full $[SU(2)\otimes
U(1)]^2$ gauge symmetry.
The leading order dimension-two term in the non-linear $\sigma$-model 
can be written for the scalar sector as \cite{littlest}
	\begin{equation}
	\mathcal{L}_{\Sigma} = \frac{1}{2} \frac{f^2}{4} 
	{\rm Tr} | \mathcal{D}_{\mu} \Sigma |^2.
\label{Sigma}
	\end{equation}
The numerical coefficients have been chosen so that the scalar kinetic terms
are canonically normalized. The covariant derivative is defined as
         \begin{equation}
           \mathcal{D}_\mu \Sigma=  \partial_\mu\Sigma - i \sum_{j=1}^2\left( 
g_j( W_j\Sigma +  \Sigma W_j^T) + g'_j (B_j\Sigma + \Sigma B_j^T) \right).
	\end{equation}
To linearize the theory, one can expand $\Sigma$  in powers of ${1}/{f}$
around its vacuum expectation value $\Sigma_0$  
\begin{equation}
\Sigma = \Sigma_0 + \frac{2 i}{f} \left( \begin{array}{ccccc}
\phi^{\dagger} & \frac{h^{\dagger}}{\sqrt{2}} & {\mathbf{0}}_{2\times
2} \\
\frac{h^{*}}{\sqrt{2}} & 0 & \frac{h}{\sqrt{2}} \\
{\mathbf{0}}_{2\times 2} & \frac{h^{T}}{\sqrt{2}} & \phi
\end{array} \right) + {\cal O}(\frac{1}{f^2}),
\end{equation}
where $h$ is a doublet and $\phi$ is a triplet under the unbroken $SU(2)$.
The appearance of the $\Sigma_0$ breaks the
local gauge symmetry $[SU(2)\otimes U(1)]^2$ into its diagonal subgroup
$[SU(2)\otimes U(1)]_{SM}$, giving rise to mass of order $f$ for  half of 
the gauge bosons
\begin{eqnarray}
	m_{W^{\prime}} = \frac{f}{2} \sqrt{g_1^2 + g_2^2}
	= \frac{g}{2sc} f , \qquad
	m_{B^{\prime}} =  \frac{f}{2\sqrt{5}} 
	\sqrt{g_1^{\prime 2} + g_2^{\prime 2}}
	= \frac{g^{\prime}}{2\sqrt{5}s^{\prime}c^{\prime}}f ,
\end{eqnarray}
with the field rotation to the mass eigenstates given by
\begin{eqnarray}
	W = s W_1 + c W_2, &\qquad& W^{\prime} = -c W_1 + s W_2 \nonumber \\
	B = s^{\prime} B_1 + c^{\prime} B_2, &\qquad&
	B^{\prime} = -c^{\prime} B_1 + s^{\prime} B_2.
\end{eqnarray}
The mixing angles are given by
\begin{eqnarray}
	s = \frac{g_2}{\sqrt{g_1^2+g_2^2}}, \qquad
	s^{\prime} = \frac{g_2^{\prime}}{\sqrt{g_1^{\prime 2}+g_2^{\prime 2}}}.
\end{eqnarray}
The $W$ and $B$ remain massless and are identified as the SM
gauge bosons, with couplings
\begin{equation}
	g = g_1 s=g_2 c, \qquad g^{\prime} = g_1^{\prime} s^{\prime}=g_2^{\prime} c^{\prime}.
\label{grelations}
\end{equation}

The couplings of $W$, $W^{\prime}$ to two scalars are given by:
\begin{eqnarray}
	\mathcal{L}_{\Sigma}(W\cdot W) &=& 
	\frac{g^2}{4} \left[  W^a_{\mu} W^{b \mu} 
	-  \frac{(c^2-s^2)}{sc} W^a_{\mu} W^{\prime b \mu}
	\right] 
	{\rm Tr} \left[ h^{\dagger}h \delta^{ab} 
	+ 2\phi^{\dagger}\phi \delta^{ab} 
	+ 2 \sigma^a \phi^{\dagger} \sigma^{bT} \phi \right]
	\nonumber \\
	&-& \frac{g^2}{4} \left[
	 W^{\prime a}_{\mu} W^{\prime a \mu}
	{\rm Tr} \left[ h^{\dagger}h + 2\phi^{\dagger}\phi \right]
	- \frac{(c^4+s^4)}{2s^2c^2}
		W^{\prime a}_{\mu} W^{\prime b \mu} 
	{\rm Tr} \left[2 \sigma^a \phi^{\dagger} \sigma^{bT} \phi \right]
	\right].
	\label{wwhh}
\end{eqnarray}
In the SM, the four-point couplings of the form $WW h^{\dagger}h$
lead to a quadratically divergent contribution to the Higgs mass.
In the littlest Higgs model, however, the $W^{\prime}W^{\prime} h^\dagger h$ 
coupling has an unusual form as seen in Eq.~(\ref{wwhh}),
which serves to exactly cancel the quadratic divergence in the Higgs 
mass arising from the seagull diagram involving a $W$ boson loop.
Similarly, the couplings of $B$, $B^{\prime}$ to two scalars are:
\begin{eqnarray}
	\mathcal{L}_{\Sigma}(B\cdot B) &=&
	g^{\prime 2}\left[  B_{\mu} B^{\mu}
	-  \frac{(c^{\prime 2}-s^{\prime 2})}{s^{\prime}c^{\prime}}
	B_{\mu} B^{\prime \mu}
	\right] {\rm Tr} \left[ \frac{1}{4} h^{\dagger} h + \phi^{\dagger}\phi
		\right] \nonumber \\
	&-& 
	 g^{\prime 2}\left[
	B_{\mu}^{\prime} B^{\prime \mu}
	{\rm Tr} \left[ \frac{1}{4} h^{\dagger} h \right]
	- \frac{(c^{\prime 2}-s^{\prime 2})^2}
		{4s^{\prime 2}c^{\prime 2}}
	B_{\mu}^{\prime} B^{\prime \mu} 
	{\rm Tr} \left[ \phi^{\dagger}\phi \right] \right].
\label{bbhh}
\end{eqnarray}
We see that the $B^{\prime}B^{\prime}$ coupling to $h^{\dagger}h$
serves to exactly cancel the quadratic divergence in the Higgs mass
arising from the seagull diagram involving a $B$ boson loop.
Note that terms of the form
$W \partial h h $ would not produce a quadratic divergence by power
counting. This absence of the quadratically divergent Higgs mass term
at one-loop order can also be understood by a set of global symmetries under
which the Higgs doublet transforms non-linearly and which is preserved
partially by the various interactions in the effective Lagrangian at
scale $\Lambda_S$ \cite{littlest}.

This cancellation may not follow one's intuition at first sight. It turns out
that the appearance of the different sign between the two $SU(2)'$s
(or $U(1)'$s) can be traced back to the unique pattern of gauge symmetry
breaking. For instance, the broken generators (associated with $W'$) are 
\begin{equation}
Q'^a=\frac{1}{\sqrt{g_1^4+g_2^4}} (g_1^2 Q_1^a - g_2^2 Q_2^a ),
\end{equation}
which do not satisfy the standard $SU(2)$ commutation relations. Technically,
this is the reason for the unusual negative sign of the gauge
couplings of the Higgs boson to $W'$ ($B'$).

\subsubsection{Higgs Bosons and the Electroweak Symmetry Breaking}

The electroweak symmetry breaking in this model is triggered by the
Higgs potential generated by one-loop radiative corrections. The
Higgs potential includes the parts generated by the gauge boson loops
as well as the fermion loops. They can be presented in the standard
form of the Coleman-Weinberg potential in terms of $M_V^2(\Sigma)$ and
$M_f^2(\Sigma)$. By expanding the nonlinear $\sigma$-model field
$\Sigma$ as usual, we obtain the Higgs potential
\begin{equation}
	V = \lambda_{\phi^2} f^2 {\rm Tr}(\phi^{\dagger} \phi)
	+ i \lambda_{h \phi h} f \left( h \phi^{\dagger} h^T
		- h^* \phi h^{\dagger} \right)
	- \mu^2 h h^{\dagger}
	+ \lambda_{h^4} ( h h^{\dagger} )^2,
\end{equation}
where the coefficients $\lambda_{\phi^2}$, $\lambda_{h \phi h}$, and
$\lambda_{h^4}$ are functions of the fundamental parameters in this model
(the gauge couplings, top-quark Yukawa coupling, and two new coefficients
$a,a'$ in the Coleman-Weinberg potential), 
as explicitly given in Eq.~(\ref{lambdas2}) in the appendix. 
There could exist important two-loop contributions to the Higgs
potential. A term like $\Lambda_S^2 h^2 /(16 \pi^2)^2  \sim f^2 h^2/(16
\pi^2)$ gives rise to a mass term for  $hh^\dagger$ which could be as
large as the one-loop Coleman-Weinberg potential contribution. 
We will not attempt to evaluate these two-loop contributions
explicitly in terms of model
parameters. Instead, the Higgs mass parameter $\mu^2$ should be
treated as a {\it new} free parameter of the order of $f^2/16 \pi^2$.

Minimizing the potential to obtain the doublet and triplet 
vevs $v$ and $v'$, it is easy to  arrive at a relation
(see the Appendix for details):
\begin{equation}
\frac{\lambda_{h \phi h}}{\lambda_{h^4}} = 
\frac{4\lambda_{h \phi h}}{\lambda_{\phi^2}} = 
\frac{8 v^{\prime}}{v}
\frac{f}{v}.
\end{equation}
Diagonalizing the Higgs mass matrix, we obtain Higgs masses to the
leading order
\begin{equation}
	M^2_{\Phi} \simeq \lambda_{\phi^2} f^2, \qquad
	m^2_{H} \simeq 2 \left( \lambda_{h^4} 
	- \lambda_{h \phi h}^2 / \lambda_{\phi^2} \right) v^2 = 2 \mu^2.
\end{equation}  
Note that we must have $\lambda_{\phi^2} > 0$ to avoid generating
a triplet vev of order $f$, in gross violation of experimental constraints.
Also, we must have $\lambda_{h^4} > \lambda_{h \phi h}^2 / \lambda_{\phi^2}$
in order to get the correct vacuum for the electro-weak symmetry breaking 
(EWSB)  with $m^2_H>0$.
The masses of the triplet states are degenerate at this order. 
We can further relate
the masses by
\begin{equation}
 M^2_\Phi=\frac{2m_H^2 f^2}{v^2}
\frac{1}{\left[1- (4v^{\prime} f / v^2)^2 \right]}.
\label{MPHI}
\end{equation}
We can thus express all four parameters in the Higgs potential, to
leading order, in terms of the physical parameters $f$, $m_H^2$, $v$, and
$v^{\prime}$. As a side product, we obtain a relation among the vevs
by demanding the triplet  mass squared to be positive definite
\begin{equation}
{v'^2\over v^2} < { v^2\over 16f^2}.
\label{vprime}
\end{equation}
It is informative to note that the couplings of the Higgs triplet
to  the massive gauge bosons are relatively suppressed by $v'/v$;
while the charged Higgs boson couplings to a photon are  of
the full electromagnetic strength.

Now let us estimate the naturalness bound on the scale $f$ for
keeping $m_H$ light. We have the generic expression for 
\begin{equation}
m^2_{H} = 2\mu^2 \simeq a_{{\text{1-loop}}} \frac{f^2}{16 \pi^2} +
a_{{\text{2-loop}}} \frac{f^2}{16 \pi^2} ,
\label{ft}
\end{equation}
with both $a_{{\text{1-loop}}}$ (we have absorbed a factor of $\log(16 \pi^2)$ 
into the definition of $a_{{\text{1-loop}}}$ ) and
$a_{{\text{2-loop}}}$ containing 
many contributions (terms) from different interactions. 
Assuming that there is no large cancellation and $m_H^2$ is no less than
$10\%$ of the magnitude of the largest term on the 
right-hand side of Eq.~(\ref{ft}), 
we obtain a rough estimate of the natural scale 
\begin{equation}
	f \leq \frac{4 \pi m_H}{\sqrt{0.1 a_{\rm max}}}
	\simeq \frac{8 \ {\rm TeV}}{\sqrt{a_{\rm max}}} 
	\left( \frac{m_H}{200 \ {\rm GeV}} \right),
\end{equation}
where $a_{\text{max}}$ denotes the largest coefficient of the terms
in Eq.~(\ref{ft})  which could be of the order of 10.

\subsubsection{Gauge Boson Mass Eigenstates}

The EWSB induces further mixing between the light and heavy gauge bosons.
The final mass eigenstates for the charged gauge bosons are 
$W_L$ (light) and $W_H$ (heavy), with masses to the order of $v^2/f^2$ given by
\begin{eqnarray}
	M_{W_L^{\pm}}^2 &=& m_w^2 \left[
	1 - \frac{v^2}{f^2} \left( \frac{1}{6} 
	+ \frac{1}{4} (c^2-s^2)^2
	\right) + 4 \frac{v^{\prime 2 }}{v^2}\right],
		\label{MWL}\\
	M_{W_H^{\pm}}^2 &=& m_w^2\left( \frac{f^2}{s^2c^2v^2}-1\right) ,
\label{MWH}
\end{eqnarray}
where the mass parameter $m_w\equiv gv/2$ approaches the SM $W$-boson
mass when $f\to \infty$.
Note that the $W_L$ mass gets a correction at order $v^2/f^2$, which
will modify the relation among the $W$ mass, $g$, and $G_F$.
The neutral gauge boson masses are similarly given by
\begin{eqnarray}
	M_{A_L}^2 &=& 0, \nonumber \\
	M_{Z_L}^2 &=& {m_z^2}
	\left[ 1 - \frac{v^2}{f^2} \left( \frac{1}{6} 
	+ \frac{1}{4} (c^2-s^2)^2
	+ \frac{5}{4} (c^{\prime 2}-s^{\prime 2})^2 \right)+ 8
	\frac{v^{\prime 2 }}{v^2}  \right],
	\label{MZL}\\
	M_{A_H}^2 &=& m_z^2 s_\w^2 \left(
	\frac{ f^2 }{5 s^{\prime 2} c^{\prime 2}v^2}
	- 1 + \frac{x_H c_\w^2}{4s^2c^2  s_\w^2} \right)
	\label{MAH} \\
	M_{Z_H}^2 &=& m_w^2 \left( \frac{f^2}{s^2c^2 v^2}
	- 1 -  \frac{x_H s_\w^2}{s^{\prime 2}c^{\prime 2}c_\w^2}\right) ,
	\label{MZH}
\end{eqnarray}
where $m_z\equiv {gv}/(2c^{}_\w)$ is the SM limit when $f\to \infty$.
Again, the $Z_L$ mass gets a correction at order $v^2/f^2$. 
$x_H^{}$ characterizes the heavy gauge boson mixing and
depends on the gauge couplings as given in the appendix.

The ratio of the $W_L$ and $Z_L$ boson masses (which are 
identified as those experimentally observed),  to order $v^2/f^2$, is:
\begin{equation}
	\frac{M_{W_L^{\pm}}^2}{M_{Z_L}^2}
	= c_\w^2 \left[ 1 + \frac{v^2}{f^2} \frac{5}{4} 
	(c^{\prime 2}-s^{\prime 2})^2 - 4 \frac{v^{\prime 2 }}{v^2}
	\right].
	\label{rho}
\end{equation}
The breaking of the custodial $SU(2)$ symmetry at order $v^2/f^2$ 
in this model is manifest.   The tree-level SM relation 
$M_{W}^2 / M_Z^2 = c^2_\w$ (or $\rho = 1$) is no longer valid.
This breaking of the custodial $SU(2)$ symmetry can be traced back
to the vacuum expectation value of $\Sigma$.
As shown in Eq.~(\ref{eq:Sigma}),
the $\mathcal{O}(1/f^2)$ term in the 
expansion has its $h^{\dagger}h$ vev in the position of the neutral 
component of the scalar triplet in the $\mathcal{O}(1/f)$ term in the 
expansion.  Thus the $h$ vev acts like
a triplet vev at order $v^2/f^2$.
The $U(1)$ gauge coupling of the triplet also
breaks the custodial $SU(2)$ at the order $v'^2/v^2$. 
It is also interesting to note that for the case of no mixing $s'=c'$ 
(or $g_1'=g_2'$)
and $v'=0$, the $W,Z$ mass ratio remains the SM form.
We will discuss the theoretical origin of the custodial $SU(2)$ symmetry  
breaking in more detail in Sec.~\ref{su2}.

\subsection{Fermions and Their Interactions} 

\subsubsection{Yukawa Interactions} 
\label{u1s}

The Standard Model fermions acquire their masses through the Higgs
mechanism via Yukawa interactions. Due to its large Yukawa coupling
to the Higgs field, the top quark
introduces a quadratic correction to the Higgs boson
mass of the order ${y_t^2} \Lambda_S^2/(16 \pi^2) \sim
f^2$ and spoils the naturalness of a light Higgs boson. 
In the littlest Higgs model
\cite{littlest}, this problem is resolved by introducing a new set of
heavy fermions with couplings to the Higgs field such that it cancels
the quadratic divergence due to the top quark. 
The new fermions come in as a vector-like
pair, $\tilde t$ and $\tilde t^{\prime c}$, with quantum numbers
$({\mathbf{3,1}})_{Y_i}$ and $({\mathbf{\bar{3},1}})_{-Y_i}$. 
Therefore, they are allowed to have a
bare mass term which is {\it chosen} to be of order $f$. 
The coupling of the Standard Model top quark
to the pseudo-Goldstone bosons and the heavy vector pair
in the littlest Higgs model is chosen to be 
\begin{equation}
{\mathcal{L}}_Y = {1\over 2}\lambda_1 f \epsilon_{ijk} \epsilon_{xy} \chi_i
\Sigma_{jx} \Sigma_{ky} u^{\prime c}_3 
+ \lambda_2 f \tilde{t} \tilde{t}^{\prime c}
+ {\rm h.c.},  
\label{yuk}
\end{equation}
where $\chi_i=(b_3, t_3, \tilde{t})$ and $\epsilon_{ijk}$ and $\epsilon_{xy}$
are antisymmetric tensors. It is now straightforward to work out the Higgs-heavy
quark interactions, as given in Appendix \ref{yukawa-appen}.
The most important consequence is the cancellation of the quadratically divergent
corrections to the Higgs mass due to $t_3,\tilde t, u^{\prime c}_3$ at the one-loop order
explicitly seen in Eq.~(\ref{top-cancel}),
\begin{eqnarray}
	 -i \lambda_1 ( \sqrt{2} h^0 t_3 + i f \tilde t 
		- \frac{i}{f}  h^0h^{0*} \tilde t ) u_3^{\prime c}
			+ {\rm h.c.} 
	\nonumber
\end{eqnarray}
This is due to the flavor (anti)symmetry introduced
in Eq.~(\ref{yuk}). The mass of the vector-like quark is arbitrary in principle. 
It is chosen by hand as $\lambda_2 f$ to preserve naturalness. 
It is a tuning in this sense. However, once we make this
choice, it is stable against radiative corrections. The new model-parameters
$\lambda_1,\ \lambda_2$ are supposed to be of the order of unity.

Expanding the $\Sigma$ field and diagonalizing the mass matrix, 
we obtain our physical states $t_L,\ t_R^c,\ T_L$ and $T_R^c$ with masses 
\begin{equation}
	{\mathcal{L}}_f = -m_t t_L t_R^c - M_T T_L T_R^c
\end{equation}
where, up to order $v^2/f^2$ relative to the leading term,
\begin{eqnarray}
	m_t &=& \frac{i\lambda_1\lambda_2}{\sqrt{\lambda_1^2 + \lambda_2^2}}
	v \left\{ 1 + 
	\frac{v^2}{f^2} \left[ -\frac{1}{3} + \frac{fv^{\prime}}{v^2}
	+ \frac{1}{2}\frac{\lambda_1^2}{\lambda_1^2+\lambda_2^2}
	\left(1 - \frac{\lambda_1^2}{\lambda_1^2+\lambda_2^2}
	\right) \right] \right\},
	\nonumber \\
	M_T &=& -f\sqrt{\lambda_1^2 + \lambda_2^2}
	\left[ 1 + \mathcal{O}(v^2/f^2) \right].
	\label{MTH}
\end{eqnarray}
Since the top-quark mass is already known in the SM, we have the approximate
relation
\begin{equation}
	{1\over \lambda_1^2} + {1\over \lambda_2^2}\approx \left(  {v}\over m_t \right)^2.
	\label{mtrelation1}
\end{equation}
which gives the absolute bounds on the couplings
\begin{equation}
	{\lambda_1} \ge  {m_t \over  v},\quad 
	{\lambda_2} \ge  {m_t \over v},\quad  {\rm or}\quad 
	{\lambda_1 \lambda_2} \ge {2}\left(  {m_t}\over v \right)^2.
	\label{mtrelation2}
\end{equation}

The  scalar interactions with the up-type quarks of the first two generations
can be chosen to
take the same form as in Eq.~(\ref{yuk}), except that there is no need for
the extra vector-like  quarks $\tilde t, \tilde t^{\prime c}$.
The interactions with the down-type quarks and leptons of the three
generations are generated by a similar Lagrangian, as given in
Appendix \ref{yukawa-appen}.

We choose the Standard Model fermions to be charged only under $SU(2)_1$
with generator $Q_1^a$. The $SU(2)$ gauge invariance of Eq.~(\ref{yuk})
is transparent: The first term is actually invariant under
an $SU(3)$ rotation under which $\chi_i$ transforms like a vector and
$\Sigma$ is transformed by a $3\times 3$ unitary rotation embedded in
the upper corner of the $5 \times 5$ matrix. 

\begin{table}[tbh]
\begin{tabular}{|c||c|c|c|c|c||c|c|}
\hline
 & $Q$ & $u^{\prime c}$ & $d^c$ & $L$ & $e^c$ 
	& $\tilde t$ & $\tilde t^{\prime c}$ \\
\hline
$Y_1$ & $-\frac{3}{10} - y_u$ & $y_u$ & $\frac{3}{5} + y_u$ 
	& $\frac{3}{10} - y_e$ & $y_e$ 
	& $\frac{1}{5} - y_u$ & $-\frac{1}{5} + y_u$ \\
\hline
$Y_2$ & $\frac{7}{15} + y_u$ & $-\frac{2}{3} - y_u$ & $-\frac{4}{15} - y_u$
	& $-\frac{4}{5} + y_e$ & $1-y_e$ 
	& $\frac{7}{15} + y_u$ & $-\frac{7}{15} - y_u$ \\
\hline
\end{tabular}
\caption{Fermion $U(1)$ hypercharge assignments required for fermion couplings
	to scalars to be invariant under $[SU(2)\otimes U(1)]_1 \otimes
[SU(2)\otimes U(1)]_2 $.
}
\label{tab:hypercharge}
\end{table}

The embedding of the two $U(1)$'s in this model can also be constructed 
by the gauge invariance of ${\mathcal{L}}_Y$. The basic requirement is to
reproduce the diagonal $U(1)_Y$ as the SM hypercharge
\begin{equation}
Y_1 + Y_2=Y.
\end{equation}
Some remarks are in order:
\begin{enumerate}
\item 
The gauge invariance of Eq.~(\ref{yuk}) under $U(1)_1\otimes U(1)_2$ 
dictates the hypercharge assignments of the fermions, which is the major
difference between our scheme and those in the literature \cite{ewdata,joA},
where the SM fermions are assumed to be 
charged only under one $U(1)$ gauge group.
Giving up the requirement of gauge invariance of 
 $U(1)_1 \otimes U(1)_2$ should be acceptable 
in principle, for example, by introducing extra fields breaking the $U(1)$'s 
at a scale $\Lambda_S$. However, this is an additional complication of
the littlest Higgs model which may need extra arguments for its naturalness. 
\item 
The gauge invariance of Eq.~(\ref{yuk}) alone cannot unambiguously fix all the
 $U(1)$ charge values.  We list them  in Table~\ref{tab:hypercharge}. 
Two parameters $y_u$ and $y_e$ are undetermined.
They can be fixed by requiring that the $U(1)$ charge
assignments be anomaly free, {\it i.e.}, $ Y_1=xY,\ Y_2=(1-x)Y$.
This leads to the particular values
\begin{equation}
y_u=-\frac{2}{5}, \hspace{1cm} y_e = \frac{3}{5}.
\end{equation}  
However, as an effective field theory below a
cutoff, it is unnecessary to be completely anomaly free,
although it is certainly a desirable property from a model
building point of view since we do not have to introduce specific type
of extra matter at the cutoff scale. In this sense, $y_u$ and $y_e$
can be thought of as partially parameterizing the model dependence of
the $U(1)$ sector of some extension of the littlest Higgs model.  In
our current study, we choose 
not to be limited by the requirement of anomaly cancellation and indicate
the anomaly free assignment as a special case.  
\item It is convenient and simple to assume that the first two 
generations of quarks also obtain
their masses through a coupling similar to the first term in
Eq.~(\ref{yuk}).
However, this
requires some tuning of the parameters to get the correct fermion mass
hierarchy. It is certainly no worse than the tuning of the Yukawa
couplings in the Standard Model. On the other hand, it might be
interesting to postulate that the mass terms of the first two generations 
actually come from
higher dimensional operators of the form $\bar{q} q {\mathcal{O}} /
\Lambda^n_S$, where ${\mathcal{O}}$ is some operator obtaining a vev of
mass dimension $(n+1)$. The fields in operator ${\mathcal{O}}$ can 
have different origins for different generations. 
Depending on its form and field content, we can
again have some relations of the $U(1)$ charge assignments different
from the third generation. In particular, if the operator  ${\mathcal{O}}$
is composed of the vevs of the pseudo-Goldstone bosons in $\Sigma$,
some relations can be derived. However, these relations are much less
constrained so that we can effectively treat the hypercharge of the
first two generations of fermions as free parameters with the only
constraint $Y_1 + Y_2 = Y$. If we further impose 
the anomaly cancellation condition, we can assign only
a discrete set of possible values.  
\end{enumerate}
With these possibilities in mind, the phenomenological
study of the couplings of the $U(1)$ currents may contain
large uncertainties due to model-dependence. We choose to
study the rather simple case in which all three generations obtain their
masses from the same type of gauge invariant operator as the first term in Eq.~(\ref{yuk}).

\subsubsection{Fermion Gauge Interactions} 

Assuming the fermions transform under $G_1\otimes G_2$ analogous to the
SM, the fermion gauge interactions can be constructed in a standard way, as
given in the appendix. The SM weak-boson couplings to fermions receive corrections
of the order $v^2/f^2$, while the electromagnetic coupling remains unchanged,
as required by the unbroken electromagnetic 
gauge interaction. There are new heavy gauge bosons
to mediate new gauge interactions.

For the gauge couplings involving the top quark, we must include the 
mixing between the chiral $t_3$ and the vector-like $\tilde t$.
Since these fermions have different $SU(2)\otimes U(1)$ quantum numbers,
their mixing will lead to flavor changing neutral currents
mediated by the $Z_L$ boson formally at the order of $v/f$.
The two right-handed fermions, $u_3^c$ and $\tilde t^c$, have the 
same quantum numbers under the Standard Model $SU(2)\otimes U(1)$ gauge
groups, so that their mixing does not cause any FCNC gauge couplings
involving the light gauge bosons.
A similar argument is applicable to the charged current, which gets modified as
\begin{equation}
	J^{+\mu} = \frac{1}{\sqrt{2}} \left[
	c_L \bar t_L \gamma^{\mu} b_L 
	+ s_L \bar T_L \gamma^{\mu} b_L \right],
\end{equation}
where $c_L$, $s_L$ are given in Eq.~(\ref{eq:Ttmixing}).
It is useful for future
phenomenological studies to write the mixing to order $v/f$ as
\begin{equation}
	s_L \simeq \frac{\lambda_1}{\lambda_2} \frac{m_t}{M_T}.
\end{equation}

We will also assume that the first two generations get their masses 
through normal Yukawa couplings which reproduce, to the
leading order in $v/f$, the usual CKM matrix.  However,
because of the mixing of the $SU(2)$ doublet state $t_3$ into the heavier
mass eigenstate $T_L$, the CKM matrix involving only the SM
quarks is no longer unitary and the leading deviation occurs at the order $v^2/f^2$,
as given by
\begin{eqnarray}
	V_{tb} &=& c_L V_{tb}^{\rm SM}
	= V_{tb}^{\rm SM} \left(1 - \frac{1}{2}
	\frac{\lambda_1^2}{\lambda_2^2} \frac{m_t^2}{M_T^2} \right), 
	\nonumber \\
	V_{Tb} &=& s_L V_{tb}^{\rm SM}
	= V_{tb}^{\rm SM} \frac{\lambda_1}{\lambda_2} \frac{m_t}{M_T}.
\end{eqnarray}
It is apparent from the Feynman rules in Appendix B how the
heavy fermion $T$ couples to other particles.  
$\overline{T_R} t_L H$ has a coupling of order one (not suppressed by
any powers of $v/f$), and the couplings to SM gauge bosons 
are formally suppressed by $v/f$. However the couplings to the longitudinally
polarized gauge bosons gain an enhanced factor $f/v$, resulting in an effective
coupling the same strength as that to $H$. We will discuss 
the phenomenological implications in Sec.~\ref{Tpheno}.

\subsection{On the $SU(2)$ Custodial Symmetry}
\label{su2}

An immediate question for an extended model is the possible tree-level violation of the
$SU(2)$ custodial symmetry and therefore potentially large deviations from the SM
prediction for the $\rho$ parameter. We briefly touched upon this issue when we
presented the $W,Z$ masses with Eq.~(\ref{rho}).
We now comment on the general features of little Higgs models
in this regard, but will discuss the numerical constraints on 
the littlest Higgs model and the possible ways to evade the constraints
in the next phenomenology section.

It is instructive to compare the littlest Higgs model with the Georgi-Kaplan 
composite Higgs model \cite{Georgi:1984af}. In that model,
which also has a global symmetry breaking pattern $SU(5)\rightarrow
SO(5)$, only one of the $SU(2)$ symmetries is gauged while the other is
used as the $SU(2)$ custodial global symmetry. Therefore, we might
expect that the littlest Higgs model, where both of the $SU(2)'$s are
gauged, will violate the custodial symmetry that protects the  
tree-level relation of the $W$ and $Z$ masses, and $\rho=1$. The
absence of the custodial $SU(2)$ symmetry is indeed true: 
Within the framework of  $SU(5)\rightarrow SO(5)$ and gauging
both $SU(2)$ subgroups, it is not possible to have another global 
custodial $SU(2)$ symmetry. 

There are three sources of custodial symmetry violation in this model. 
First, it is very interesting to note that although the masses of both the SM-like 
gauge bosons $W$ and $Z$ are shifted due to their mixing with the heavy
$SU(2)$ gauge bosons, see {\it e.g.}~the $c^2,s^2$ terms in
Eqs.~(\ref{MWL}) and (\ref{MZL}), 
the mass ratio still remains unchanged from that 
in the Standard Model. Therefore, gauging the second $SU(2)$ does
not give rise to tree-level corrections to the mass ratio.
However, there are indeed some new tree-level contributions to the 
effective $\rho$ parameter defined through the neutral current
couplings, coming from the exchange of the new heavy  
gauge bosons which in turn induce new four-fermion interactions. 

Second, the two $U(1)$'s in the littlest Higgs model violate the
custodial symmetry. One combination of them is actually the Standard
Model hypercharge so the violation is similar to that in the Standard
Model. However, the other combination does introduce new tree-level
custodial symmetry violation. It is interesting to compare it
with the Georgi-Kaplan model \cite{Georgi:1984af},
where there is an explicitly conserved $SU(2)$ custodial symmetry
broken only by the usual $U(1)_Y$. One important difference is
that the $U(1)_A$ introduced in Ref.~\cite{Georgi:1984af}
to drive the electroweak symmetry breaking is chosen to 
actually preserve the custodial $SU(2)$
symmetry. Therefore, in their model even radiative corrections from
the new $U(1)$ do not introduce new custodial symmetry violation.

Finally,  the Higgs triplet coupling to the $SU(2)$ gauge bosons does not have
the $SU(2)$ custodial symmetry. However, for the parameter space of
this model, the triplet only gets a much smaller vev ($v'$) than
the doublet and the correction
from the triplet in the form of ${v^{\prime 2}}/{v^2}$ is
smaller than that from the Higgs doublet vev.

We will discuss the implications of the exchange of heavy $SU(2)\otimes
U(1)$ gauge bosons and the existence of the triplet vev on the
electroweak precision measurements, as well as possible modifications
of the littlest Higgs model, in Sec.\ref{precise}.


\section{Phenomenology of the littlest Higgs model}
\label{pheno}

We have presented a linearized theory based on the littlest Higgs
model \cite{littlest} and discussed some of its main features.  
It would be ultimately desirable if some of its qualitative 
features can be experimentally verified.  For this purpose,
we explore the phenomenology in this section.
We first summarize the model parameters and their relevant ranges.
We then discuss the low energy constraints and possible directions of
extension of the model to evade the constraints. 
With the possible and even desirable extensions of the model in mind, 
we will focus our phenomenological studies 
on the generic features which will likely be present 
even in some extensions of the model. 
The existence of the heavy gauge bosons is generic if the
one-loop quadratic divergence is canceled using the little Higgs idea. 
The presence of the heavy fermions is also a necessary ingredient 
to control the contribution of the top loop. 
For a model in which the Higgs doublet is a
pseudo-Goldstone boson resulting from a global symmetry breaking, most
likely the Higgs sector would not be minimal and extra scalar states 
will be present. 

\subsection{Parameters in the Littlest Higgs Model}
\label{para}

\subsubsection{Couplings}

Due to the enlarged gauge group
$[SU(2)_1\otimes U(1)_1]\otimes [SU(2)_2\otimes U(1)_2]$, there are 
four gauge couplings $g_1,\ g_2,\ g'_1,$ and $g'_2$. Upon identifying
the diagonal  part as the SM gauge group, whose couplings are
experimentally determined, we obtain the relations:
\begin{equation}
{1\over g_1^2} +  {1\over g_2^2} =  {1\over g^2}\approx {1\over 0.43},\qquad
{1\over g_1'^{2}} +  {1\over g_2'^{2}} =  {1\over g'^2}\approx {1\over 0.12},
\end{equation}
or equivalently given by  Eq.~(\ref{grelations}).

Top quark and heavy vector-like quark couplings
$\lambda_1$ and $\lambda_2$ are related to give the SM
top Yukawa coupling and thereby the correct top-quark mass.
This is given by Eq.~(\ref{mtrelation1}) as
\begin{equation}
{1\over \lambda_1^2} + {1\over \lambda_2^2}\approx  2.
\end{equation}

The proportionality constants $a$ and $a'$ in the EWSB sector from
the Coleman-Weinberg potential, as discussed in appendix \ref{ewsb-appendix}, 
remain as free parameters. They enter
the Higgs potential which determines 
the Higgs boson masses and their couplings.
Their values again depend on the
matching condition  performed at the scale $\Lambda_S$ which in turn
depends on the details of the UV completion. Neither the limits
on $a$ and $a'$ nor the direct measurements of both of them can tell us
directly what the UV completion is. However, some numerical knowledge about 
them will give us very useful hints of the possible structure of 
the UV theory.  As we will discuss later, they can be traded for other physical
observable parameters.

\subsubsection{Heavy masses}

\begin{figure}[tb]
    \includegraphics[scale=0.8]{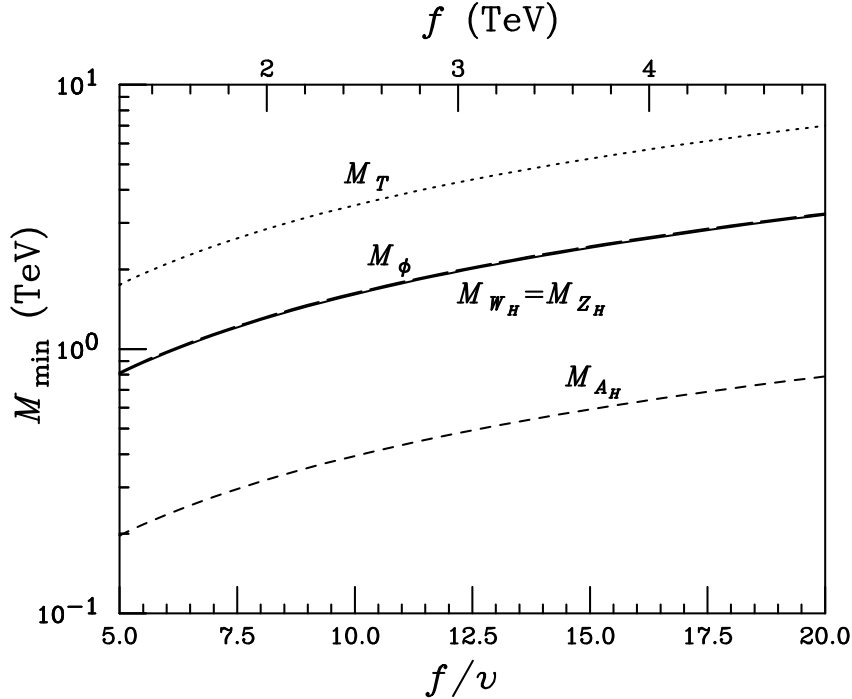}
    \caption {Theoretical lower bounds on the heavy state masses 
versus the scale  $f/v$ (bottom axis) or $f$ in TeV (top axis). 
For $M_\Phi$, we obtain the lower bound by assuming $m_H\ge 115$ GeV;
the long-dashed curve is indistinguishable from that of $M_{W_H}$.
}
    \label{MASSES}
\end{figure}

By construction, all of the new states (the heavy gauge bosons, 
the vector-like top quark, and the triplet Higgs boson) acquire
masses of the order $f$, modulo their couplings to the
mass-generation sector. These mass terms were discussed
in the previous section. To gain a qualitative understanding,
based on Eqs.~(\ref{MWH}), (\ref{MAH}) and (\ref{MZH}),
we approximate the mass relations for the heavy gauge bosons
as
\begin{eqnarray}
	M_{W_H}^2 \approx  M_{Z_H} ^2 \gsim 
	{m_w^2} {4f^2\over v^2},\qquad
	M_{A_H}^2 \gsim  m_w^2 \tan^2\theta_\w \frac{ 4 f^2 }{5 v^2} .
	\label{MAHa} 
\end{eqnarray}
As for the heavy quark mass in Eq.~(\ref{MTH}), we have
\begin{equation}
        M_T  \approx   {v\over m_t} {\lambda_1 \lambda_2}  f\ge {2m_t\over v} f.
\label{MTbound}
\end{equation}
The masses of the heavy triplet Higgs bosons are given in Eq.~(\ref{MPHI}).
At leading order all three physical states $\Phi^0$, $\Phi^+$, and $\Phi^{++}$
are degenerate in mass. The lower bound can be obtained as
\begin{equation}
	M_\Phi \gsim {{\sqrt 2}\ m_H\over v}  f.
\end{equation}

We plot the lower bounds on the masses  in Fig.~\ref{MASSES}
versus $f$ (top axis) or versus $f/v$ (bottom axis). 
We see that the $U(1)$ gauge boson $A_H$ can be as light as a few
hundred GeV due to the weaker hypercharge coupling. The $SU(2)$ gauge
bosons $W_H$ and $Z_H$ are mass-degenerate and are of the order of a
TeV. The vector-like new quark $T$ is typically heavier and is easily in
the range of multi-TeV.
The bound on $M_\Phi$ depends on the light Higgs mass $m_H$.
We obtain the lower bound  by assuming $m_H\ge 115$ GeV. 
The curve is indistinguishable from that of $M_{W_H}$.

To summarize this section, the new independent parameters in the
littlest Higgs  model are listed as:
\begin{enumerate}
\item gauge couplings $g_2,\ g_2'$ or equivalently $s,\ s'$. For convenience in
our phenomenological studies, we will take
\begin{equation}
{1\over 10} \le \cot\theta={c\over s}\le 2,\quad
{1\over 10} \le \tan\theta'={s'\over c'}\le 2;
\end{equation}
\item the symmetry breaking scales $f$ and $v'$: 
These are roughly related to the
(approximate) SM vev by ${v' / v} \lsim {v/ 4f}$;
\item new couplings in the Higgs potential $a,\ a'$ and the $\mu$
parameter: In principle, 
these can be traded for $m_H$ (and $v$, $v'$) after minimizing the potential;
\item new top Yukawa coupling $\lambda_2$: We trade it for $M_T$.
\end{enumerate}

\subsection{Low Energy Effects}

The little Higgs model contains new matter content and
interactions which will contribute to the electroweak precision
observables. 
Due to the excellent agreement between the Standard Model theory
and the precision measurements at energies below the electroweak 
scale, one would expect to put significant constraints on the little
Higgs models. Indeed, stringent constraints have been obtained 
in recent studies \cite{ewdata,joA}. 
Here we discuss the origin of some of the most stringent  constraints, 
identify the arbitrariness in particular related to the $U(1)$ sector,
and suggest possible ways to suppress those extra contributions
either by tuning the parameters of the model or by extending it.

\subsubsection{ Low-energy constraints on and possible 
directions of extension to the littlest Higgs model} 
\label{precise}

We begin with a schematic review of how extra corrections from
the little Higgs model may be computed. While there are many parameters in
the electroweak sector in the littlest Higgs model,
we will use the measured values of $M_Z$, $G_F$ and $\alpha$ as input to
the fit.
Consider an electroweak
precision observable $O_i$. In the Standard Model, it can be written 
as a function $O_i^{SM}=O_i^{SM} (M_Z, G_F, \alpha )$. 
We then express the inputs $M_Z$, $G_F$ and $\alpha$ in terms of 
the parameters in the little Higgs model such as $v^2/f^2$, $c$, etc.,
and thus obtain the expressions for the
measured values of Standard Model observables $O^{SM}_i$ as a function of the
little Higgs model parameters. We then proceed to compute the
same observables from the little Higgs model. 
In general, we will get a
different  function of the parameters $O^{LH}_{i}=O^{LH}_i(v^2/f^2, c,{\text{
etc.}})$ due to the fact that there are extra contributions to the
electroweak processes beyond the Standard Model.  The difference,
\begin{equation}
\delta O_i =  O^{LH}_i(v^2/f^2, c, {\text{etc.}}) -
O^{SM}_i(v^2/f^2, c, {\text{etc.}}),
\end{equation}
is then the correction of the electroweak observable $O_i$ received
from the little Higgs model. 

The electroweak parameters in
the little Higgs model are: i) dimensionful parameters $M^2_{W_L}$,
$M^2_{Z_L}$, and $G_F$; ii) dimensionless parameters which are the
corrections to the vector and axial vector neutral current 
couplings $\delta g_{ff}$. An inspection of all the
$\delta g_{ff}$'s indicates that they only contain corrections proportional to
$c^2$ at the order ${v^2}/{f^2}$.  However, depending on the $U(1)$ charge 
assignments of the SM fermions, $\delta g_{ff}$'s  may receive constant
contributions. We find it informative to list the corrections to the 
electroweak 
parameters in a schematic manner according to the contributions from
the $SU(2)_H,\ U(1)_H,$ and from the triplet vev, 
as given in Table~\ref{tab:ew_contrib}.
Obviously, all the corrections in this model come in at the order of
$v^2/f^2$ or $v^{\prime 2}/v^2$. The corrections of $\calO(v'^2/v^2)$
are smaller and can be easily estimated with the help of 
Eq.~(\ref{vprime}).  We will thus not study the impact of $v'$
on the electroweak precision physics further.  
We now discuss how the corrections show up in the 
electroweak precision observables and consider how the corrections
may be suppressed to evade the constraints from the electroweak precision 
measurements. 

\begin{table}[tb]
\begin{tabular}{|c|c|c|c|}
\hline
 & $SU(2)_H$ & $U(1)_H$ & $<\phi^0>$ \\
\hline
$M^2_{W_L^{\pm}}$ & $-\frac{5}{12} \frac{v^2}{f^2} + c^2s^2 \frac{v^2}{f^2}$  & $0$ & $4
\frac{v^{\prime 2}}{v^2}$ \\
\hline
$M^2_{Z_L}$ & $-\frac{5}{12} \frac{v^2}{f^2} + c^2s^2 \frac{v^2}{f^2} $ &
$-\frac{5}{4} 
\frac{v^2}{f^2} (c^{\prime2} -s^{\prime2})^2$ & $8 \frac{v^{\prime
2 }}{v^2}$ \\ 
\hline
$G_F$ &  $\frac{5}{12} \frac{v^2}{f^2}$ & $0$ & $ -4
\frac{v^{\prime 2}}{v^2}$ \\
\hline
\hline
$M^2_{Z_L} G_F$ &  $c^2 s^2 \frac{v^2}{f^2} $ & $-\frac{5}{4}
\frac{v^2}{f^2}(c^{\prime 2}-s^{\prime2})^2$ & $4
\frac{v^{\prime 2}}{v^2}$ \\
\hline
\hline
$\delta{g_{ff}}$ &  $\propto c^{2} \frac{v^2}{f^2} $ & $
\propto (-c^{\prime 2}Y_1+s^{\prime2}Y_2) \frac{v^2}{f^2} $ & $0$   \\
\hline
\end{tabular}
\caption{Extra contributions to the electroweak parameters in the
littlest Higgs model. The first two columns denote the contributions
from the exchange of the heavy gauge bosons. The coefficient $-5/4$ in the $U(1)_H$
contribution to the $Z_L$ mass is determined by the $U(1)$ charge
assignment of the Higgs boson. $\delta{g_{ff}}$ collectively denotes
the modification of the neutral current couplings of the SM
fermions, including both $V$ and $A$ couplings. $Y_{1,2}$ are 
the $U(1)$ charge assignments of the SM fermions.  
The third column denotes the contribution of the triplet vev.
}
\label{tab:ew_contrib}
\end{table}

{\bf The effects of the $SU(2)_H$ gauge bosons:}
We cast the corrections due to the exchange of the heavy
$SU(2)_H$ gauge bosons into two types, 
all proportional to $v^2/f^2$: i) constants independent of the other model parameters;
ii) corrections proportional to $c^2$.
An important observation here is that the
dimensionless constant $M^2_{Z_L} G_F$ does not depend on the
constant term proportional to $v^2/f^2$. Actually, in all of the three
dimensionful parameters $M_{W^{\pm}_L}^2$, $M^2_{Z_L}$, and $G_F$, the
constant term is the same (up to a sign). We also notice that the correction to
$\delta g_{ff}$ due to the exchange of the $SU(2)_H$ gauge bosons are
all proportional to $c^2 v^2/f^2$. This is due to the fact that
the fermions transform under $SU(2)_1$  but not under $SU(2)_2$. 
A more detailed study of the electroweak observables shows that  
{\it the constant part of $SU(2)$ corrections to the parameters does not
contribute to the electroweak precision observables.} All of the
$SU(2)_H$ corrections are therefore proportional to $c^2$. 
Contributions through the exchange of the $SU(2)_H$ gauge
bosons can be thus suppressed systematically by choosing a smaller
$c$. This corresponds to making a significant difference between 
the $SU(2)_1 \otimes SU(2)_2$ gauge couplings, $g_1\ll g_2$.

{\bf The effects of the $U(1)_H$ gauge boson:}
Obviously, the results depend upon
the $U(1)$ charge assignments of the Higgs doublet ($Y_H$)
and the SM fermions ($Y_f$). It was assumed in Ref.~\cite{ewdata,joA} 
that fermions are only charged under one of the $U(1)$'s, 
and the Higgs doublet $U(1)$ charge assignment is kept as in the original 
littlest Higgs model.
This gives rise to some of the most stringent constraints on
the scale $f$ from, for example, $\Gamma_Z \sim \Gamma_Z^{SM}(1+1.7
{v^2}/{f^2})$  and $M_{W}\sim M_W^{SM}(1+0.89{v^2}/{f^2})$ \cite{ewdata},
leading to the conclusion that $f$ is greater than about $8$ TeV even when
both $c$ and $c'$ are small. 
There indeed exist some partial cancellations, resulting in $f\gsim 4$ TeV, 
which occurs near $c'=s'$.
Scrutinizing the properties of the heavy $U(1)_H$ gauge boson $A_H$,
one finds:
\begin{enumerate}
\item Modification of the $Z_L$ mass due to the mixing between the heavy and
light gauge bosons given in Eq.~(\ref{zlzh}).  The correction to the
$Z_L$ boson mass is proportional to $(c^{\prime 2}- s^{\prime 2})$,
and thus may be suppressed for
minimal mixing $c^{\prime }= s^{\prime }$, or equivalently $g'_1=g'_2$
($x_Z^{B^{\prime}} = 0$).
\item Modification of the neutral current  due to exchange of the heavy 
$A_H$ gauge boson.  As given in Eq.~(\ref{zcoupling}), the $A_H$ coupling to
fermions is proportional to $c^{\prime 2}Y_1 - s^{\prime 2}Y_2$.
Therefore, one can minimize the corrections to neutral current processes
due to $A_H$ exchange by setting $c^{\prime 2}Y_1 - s^{\prime 2}Y_2 = 0$,
or $\tan^2\theta^{\prime}=Y_1/Y_2$.
\item Modification of the $Z_L$ boson couplings to fermions due
to mixing between the heavy and light gauge bosons.
As given in Eq.~(\ref{zcoupling}), the correction to the $Z_L$ 
coupling to fermions is proportional to 
$(c^{\prime 2}- s^{\prime 2})(c^{\prime 2}Y_1 - s^{\prime 2}Y_2)$.
Therefore, the corrections to the $Z_L$ couplings to fermions
are minimized for either $c^{\prime }= s^{\prime }$ 
or  $c^{\prime 2}Y_1 - s^{\prime 2}Y_2 = 0$.
\end{enumerate}
Furthermore, one could consider combining the two conditions above to yield 
more suppression. This implies that $Y_1 \approx Y_2$. This is the maximum
cancellation of the extra contributions which can be achieved without
changing the $Y_H$ assignment. However, this optimal cancellation goes 
beyond the simple choice of Table~\ref{tab:hypercharge}, and can be achieved
only by using an alternative form of the fermion Yukawa couplings as
discussed in Sec.~\ref{u1s}.

In principle, changing the $U(1)$ assignments of the Higgs doublet
will also change the predictions for the electroweak precision
observables. In particular, the pattern 
$(c^{\prime 2}-s^{\prime 2}) v^2/f^2$ in the
expressions for the $W_L$ and $Z_L$ masses is a result of the particular $U(1)$
assignments of the Higgs doublet in the original littlest Higgs
model \cite{littlest}. Altering the Higgs $U(1)$ charge assignments will
certainly change this expression and change the result of the electroweak
precision fit.  However, the $U(1)$ charge assignment of the
original littlest Higgs model is fixed by the requirement that the $U(1)$s
are embedded in the global symmetry group $SU(5)$.
Giving up the requirement that the $U(1)$s are embedded in the $SU(5)$ will
require the introduction of extra $U(1)$ factors beyond the original
littlest Higgs model.
This embedding also leads to the 
cancellation of the quadratically divergent contributions to the Higgs
mass due to $U(1)$ gauge boson loops.
Giving up the cancellation of this quadratic divergence by changing 
the $U(1)$ assignments of the Higgs doublet will make this model less
natural.

We see from the discussions above that the contributions to the
electroweak precision observables can be suppressed by a certain
tuning of the parameters of the little Higgs model. However, this
situation is not satisfactory since we would like to have a natural mechanism
without too much careful adjustment of the model parameters. This
calls for an extension of the littlest Higgs model which 
can naturally give us some of the properties above.
In particular, we would like a model which makes the $U(1)$
structure more concrete so that some of the features
discussed above can be realized. We may also try to get away from the
$U(1)_H$ problem by gauging just one $U(1)$ and identifying it with
$U(1)_Y$. This $U(1)_Y$ will certainly regenerate the quadratic
divergence whose cancellation is one of the prime motivations of the
little Higgs model. It will make this model appear less
natural. However, numerically, this quadratic divergence is 
milder than that generated by $SU(2)$ interactions and might be
tolerable in considering the naturalness.  

There is another way of extending this model which will greatly
improve the situation. 
One would like to have a model which explicitly preserves the
$SU(2)$ custodial symmetry, which will remove the constraint from
$M_W$. The model could also have some extended symmetries which give
us the cancellation of the extra contributions to the electroweak
observables.  This will require a significant enlargement
of the current model and thus spoil its ``minimal'' nature. However, in
the light of the discussion above, some enlargement of the original
model seems desirable in order to make it a more natural mechanism
for electroweak symmetry breaking.

With those possible extensions of the littlest Higgs model in mind, we
will focus on studying some generic features of the little Higgs model. 
We will take the constraints
$f\gsim 4$ TeV  \cite{ewdata} or $f\gsim 3.5$ TeV \cite{joA}
as a general guide, but will not be confined by them 
since some variations of the model to evade the bounds are 
quite conceivable.

\subsubsection{Triple gauge boson couplings}

The triple gauge boson couplings can be written in the most 
general form \cite{Zeppenfeld}:
\begin{eqnarray}
	\mathcal{L}_{WWV} &=& i g_{WWV} \left[ 
	g_1^V (W^+_{\mu\nu} W^{- \mu} - W^{+ \mu} W^-_{\mu\nu} ) V^{\nu}
	+ \kappa_V W^+_{\mu} W^-_{\nu} V^{\mu\nu}
	+ \frac{\lambda_V}{m_W^2} W_{\mu}^{+\nu} W_{\nu}^{-\rho} V_{\rho}^{\mu}
	\right],
\end{eqnarray}
where in the Standard Model the overall couplings are $g_{WW\gamma} = -e$ and
$g_{WWZ} = -e c_\w/s_\w$, and 
$g_1^Z = g_1^{\gamma} = \kappa_Z = \kappa_{\gamma} = 1$ and 
$\lambda_Z = \lambda_{\gamma} = 0$.
In the littlest Higgs model, $\lambda_Z = \lambda_{\gamma} = 0$ is maintained
for all the gauge boson couplings.  The $W_L^+W_L^-\gamma$ couplings are not
modified from their Standard Model form.
The $W_L^+W_L^-Z_L$ couplings $g_1^Z$ and $\kappa_Z$ receive 
direct corrections only at order $v^4/f^4$ from gauge boson mixing.
However, they receive corrections at order $v^2/f^2$ when 
written in terms of the SM inputs $M_{Z_L}$, $G_F$ and $\alpha$:
\begin{equation}
	g_1^Z = \kappa_Z =
	1 + \frac{1}{2(c^2_\w-s^2_\w)} \left[
	\frac{1}{4} \frac{v^2}{f^2} \left(
	-1 + (c^2-s^2)^2 + 5 (c^{\prime 2}-s^{\prime 2})^2 \right)
	- 4 \frac{v^{\prime 2}}{v^2} \right].
\end{equation}
We see that there are several terms contributing to the anomalous coupling.
If we assume that there is no accidental cancellation among them, we may 
get an order-magnitude bound on the scale parameters. Taking the current bound of roughly 
$\pm 5\%$ on the deviation from the SM $\Delta g_1^Z$, we obtain
\begin{equation}
f \gsim {v\over \sqrt{4\times 5\% }}\approx 2.3\ v, \qquad 
v'\lsim {\sqrt{5\% }\ v\over 2}\approx {10\%}\ v.
\label{bounds}
\end{equation}
Although the bounds estimated on $f$ and $v'$ 
are not  close to the expected natural sizes,
it is conceivable that a future $e^+e^-$ linear collider will significantly
improve the accuracy on the triple gauge boson coupling measurements,
which can be as accurate as $10^{-3}-10^{-4}$ \cite{timb}. This would reach 
useful sensitivity to the parameters in the little Higgs models,
improving the bounds in Eq.~(\ref{bounds}) by more than an order of magnitude,
to $f \sim (15-50)v \sim 3.5-12$ TeV.

\subsection{New Heavy Gauge Bosons at the LHC}

The heavy  $SU(2)$ gauge bosons are crucial ingredients for little Higgs models.
The generic decay partial width for a vector to a fermion pair can be written
as, ignoring the fermion masses,
\begin{equation}
\Gamma(V\to f\bar f')={C\over 12\pi}(g_V^2+g_A^2) M_V,
\label{vtoff}
\end{equation}
where $C$ is the fermion color factor and $g_V,\ g_A$ the vector and axial 
vector couplings.  
As seen from the Feynman rules, the new gauge boson couplings 
to the SM fermions depend upon a mixing angle parameterized by $\cot\theta=c/s$.
To the leading order, we have
\begin{equation}
{\rm for}\ Z_H:\ g_V=-g_A={g\over 2} \cot\theta\ T_3;\qquad
{\rm for}\ W_H:\ g_V=-g_A=-{g\over 2\sqrt 2}\cot\theta.
\end{equation}
The couplings are purely left-handed, 
and universal to all fermions. In particular,
if the fermion masses can be ignored, the partial width to each species of 
fermion 
pair (one flavor and one color) would be the same, that is proportional to 
$M_V \cot^2\theta$. 

\begin{figure}[tb]
    \includegraphics[scale=0.6]{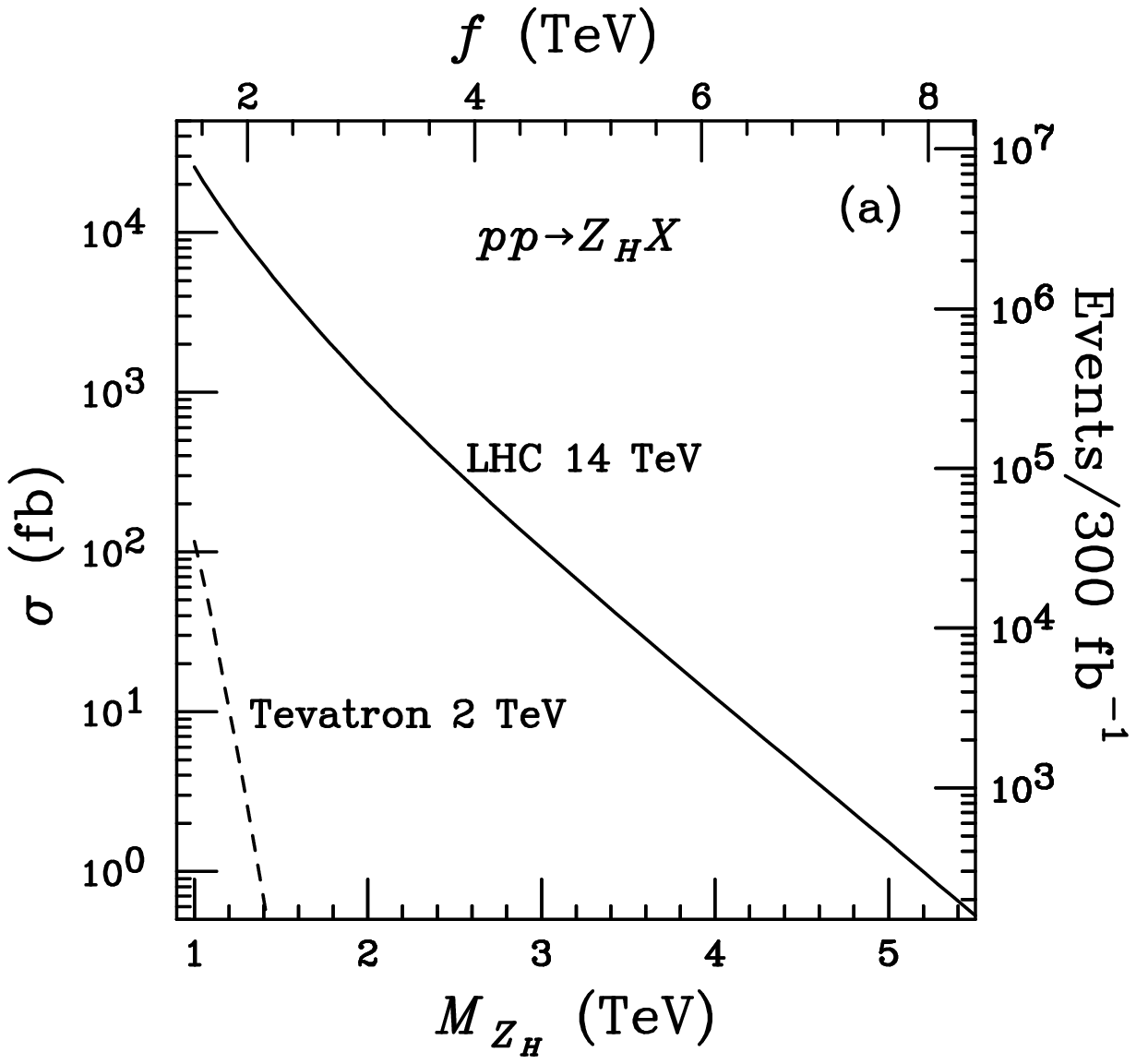}
    \includegraphics[scale=0.6]{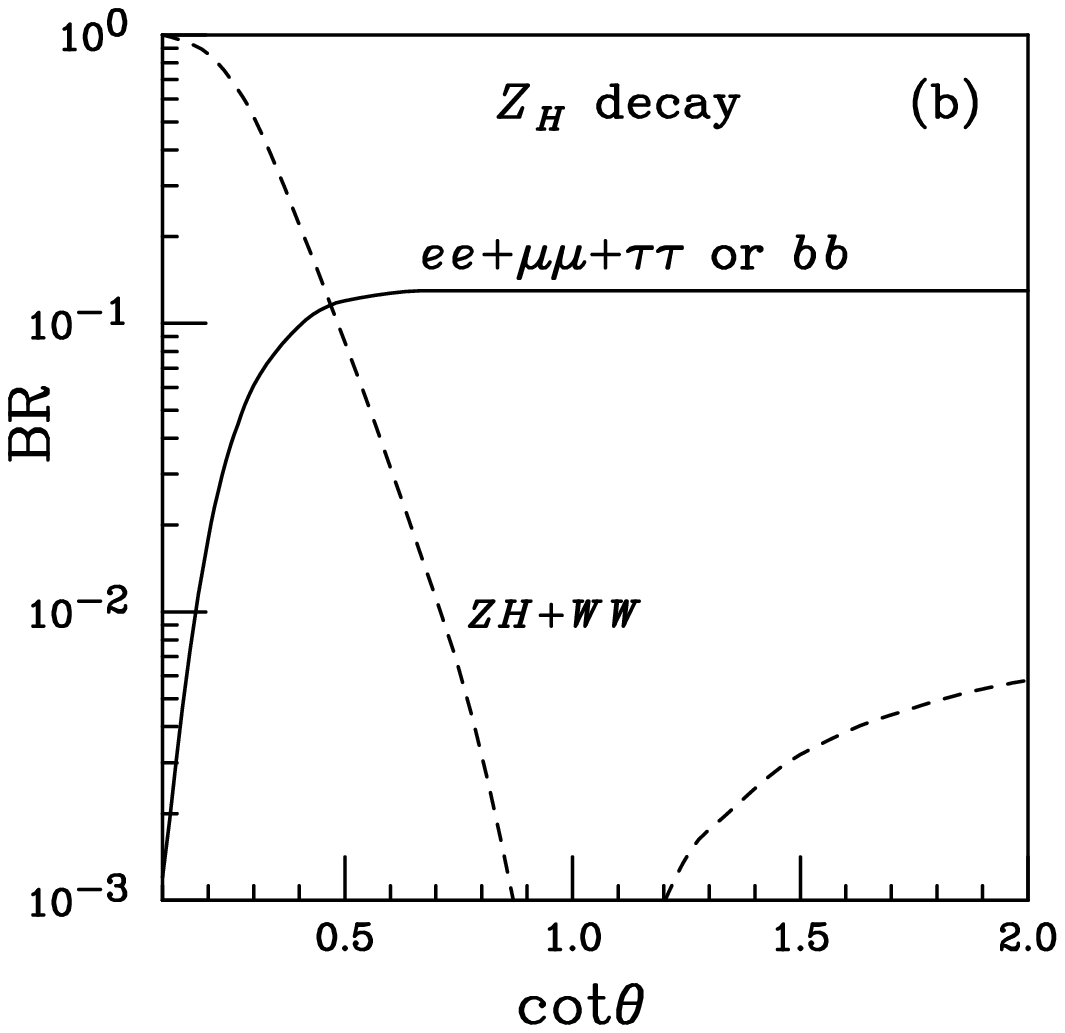}
    \caption {(a) Total cross section for $Z_H$ production 
versus its mass $M_{Z_H}$ at the Tevatron (dashed) and
the LHC (solid) for $\cot\theta=1$. The number of events expected per 
300 fb$^{-1}$ luminosity is indicated on the right-hand axis.
The scale $f$ corresponding to $\cot\theta=1$ 
is given on the top axis;
(b) $Z_H$ decay branching fractions versus $\cot\theta$.}
    \label{ZH-production}
\end{figure}

A vector boson can be produced at hadron colliders via the Drell-Yan
process $q\bar q' \to V$\cite{Burdman},
for which the production cross section is proportional
to the partial width $\Gamma(V\to q\bar q')$. We plot the $Z_H$ 
production cross sections
in Fig.~\ref{ZH-production}(a) versus its mass $M_{Z_H}$  at the Fermilab 
Tevatron and the LHC energies, where $\cot\theta=1$ has been taken
(the cross section scales as $\cot^2\theta$). We first note that 
at the Tevatron
energy, there is only a hope if $M_{Z_H}\lsim 1$ TeV and $\cot\theta$ large,
due to the severe phase space suppression. On the other hand, the LHC could 
copiously produce the heavy vector states as indicated on the right-hand 
scale of Fig.~\ref{ZH-production}(a).
For instance,  about $30,000$ $Z_H$ of a mass 3 TeV may be produced
annually at the LHC. Thus the standard search for a mass peak in the di-lepton mass
distribution of $\ell^+\ell^-$  or the transverse mass distribution of $\ell\nu$ in the 
multi-TeV range could reveal an
unambiguous signal for the vector resonant states.

It is interesting to note that there are two other competing channels 
for the heavy
gauge boson to decay, namely to its SM light gauge partner ($V_L$) 
plus the Higgs boson, and to a pair of SM light gauge bosons (i.e., 
$Z_H \to W^+_L W^-_L$ and $W_H \to W_L Z_L$). 
These bosonic decays can be best understood as decays to the components
of the Higgs doublet $h$, three of which become the longitudinal
modes of the SM light gauge bosons.
The partial width for the $V_L H$ channel is, ignoring the 
final state masses,
\begin{equation}
\Gamma(V\to V_LH)={g^2\cot^2 2\theta \over 192\pi} M_V,
\label{vtovh}
\end{equation}
and the partial width to a pair of SM light gauge bosons is the same.
We present the decay branching fractions for $Z_H$ versus $\cot\theta$ in
Fig.~\ref{ZH-production}(b). The solid curve shows the branching fraction
to the 3 generations of charged leptons, which is equal to that to one flavor of a
quark pair. The dashed curve is for the sum of the modes $Z_LH$ and 
$W^+_L W^-_L$.
We see that when $\cot\theta\gsim 1/2$, the fermionic modes dominate.
Due to the universal $SU(2)$ coupling,  the branching fraction follows the
equal-partition. The channel to the three pairs of charged leptons for instance 
approaches $1/8$ from $Z_H$ which is equal to
that to $b\bar b$, and to $t\bar t$ as well up to a phase space factor,
and $1/4$ from $W_H$. 
This is a very distinctive feature to verify once a new gauge boson is
found. On the other hand, for  $\cot\theta\lsim 0.25$, the bosonic
channels become more significant. However, one should
notice that the production cross section would be suppressed by a factor
$\cot^2\theta$ at the same time for the enhanced bosonic channels. The
branching fraction is insensitive to the heavy gauge boson mass.

\begin{figure}[tb]
    \includegraphics[scale=0.6]{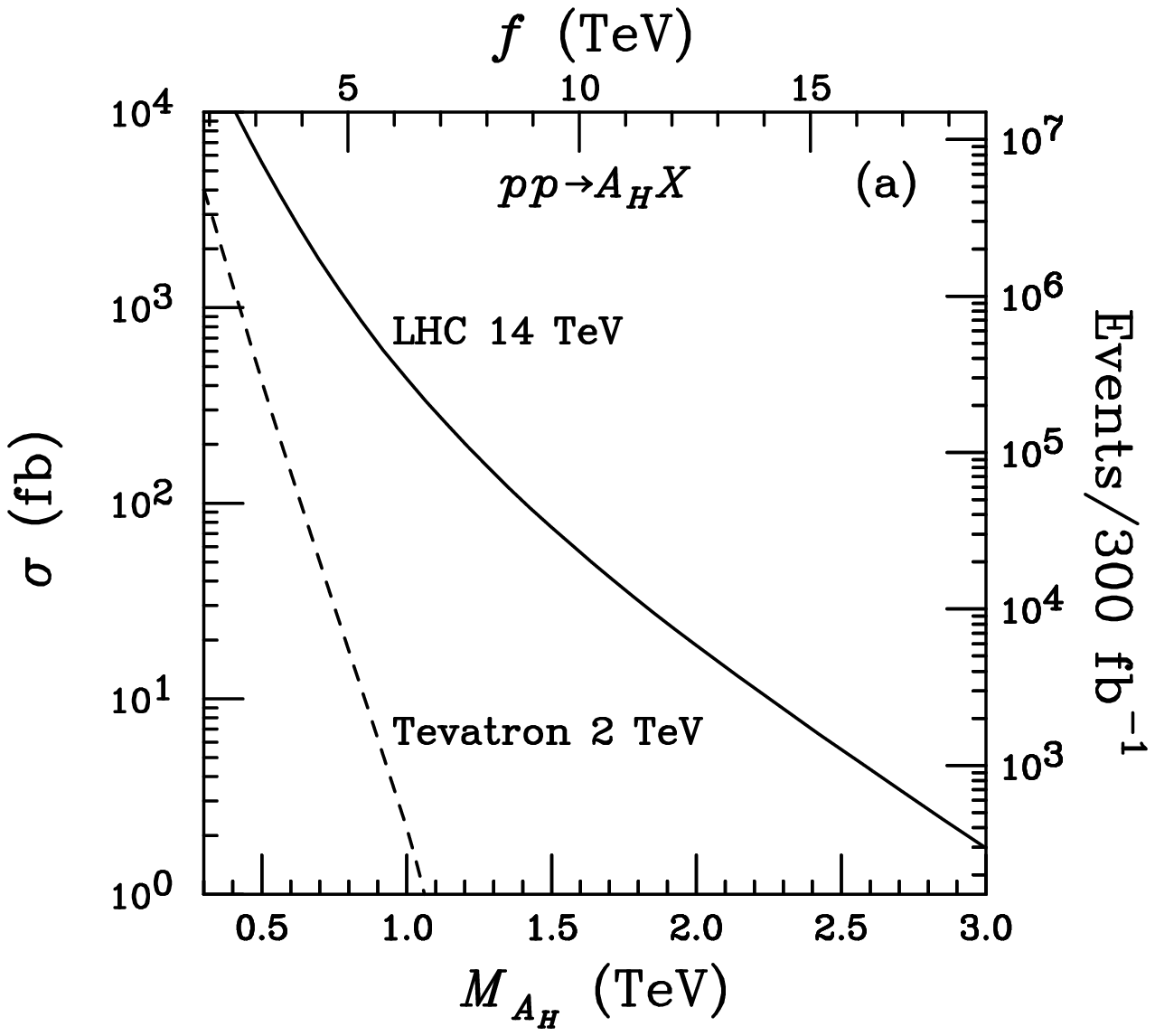}
    \includegraphics[scale=0.6]{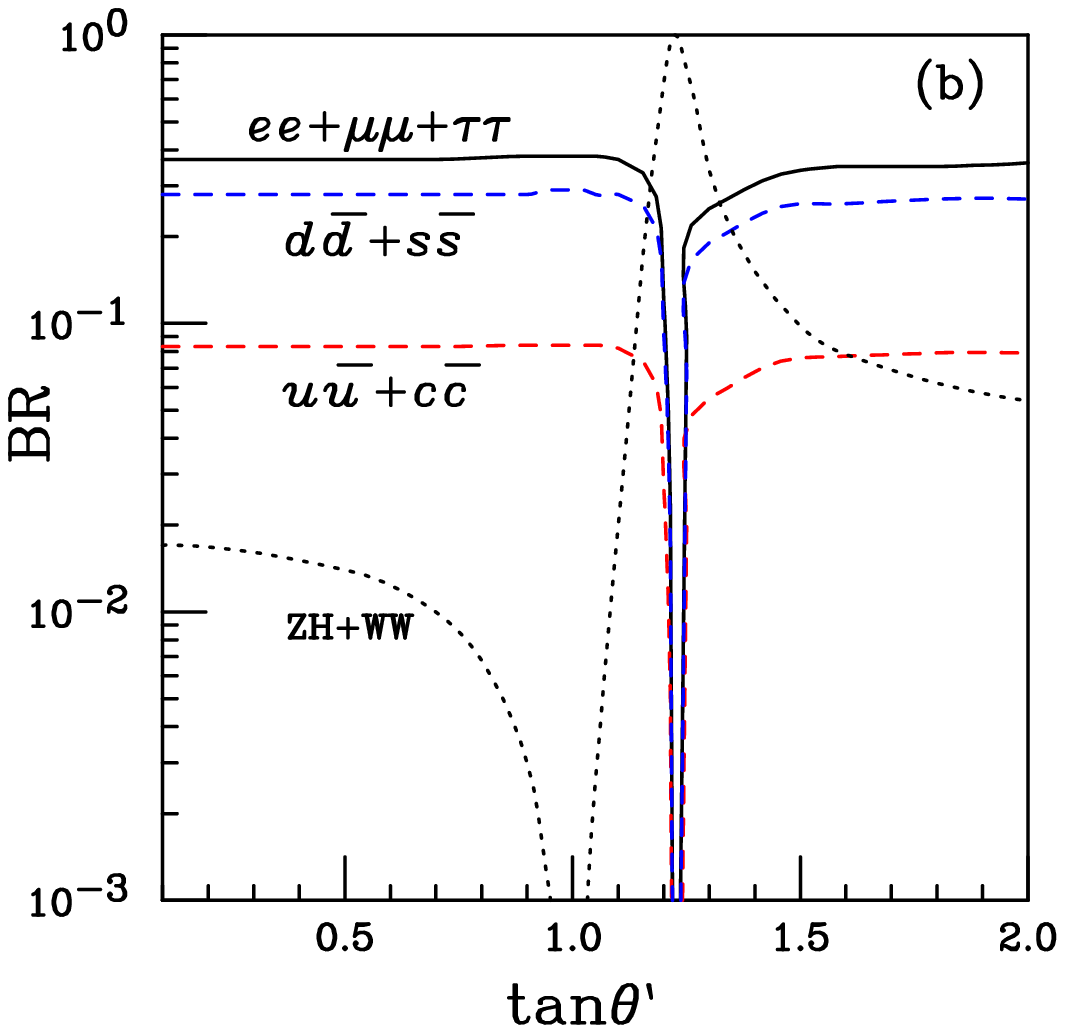}
    \caption {(a) Total cross section for $A_H$ production 
versus its mass $M_{A_H}$ at the Tevatron (dashed) and
the LHC (solid) for $\tan\theta' =1$. The number of events expected per 
300 fb$^{-1}$ luminosity is indicated on the right-hand axis. 
The scale $f$ corresponding to $\tan\theta'=1$ 
is given on the top axis;
(b) $A_H$ decay branching fractions versus $\tan\theta'$.
The fermion hypercharge assignments are fixed by the anomaly-free
condition.}
    \label{AH-production}
\end{figure}

\begin{figure}[tb]
    \includegraphics[scale=0.6]{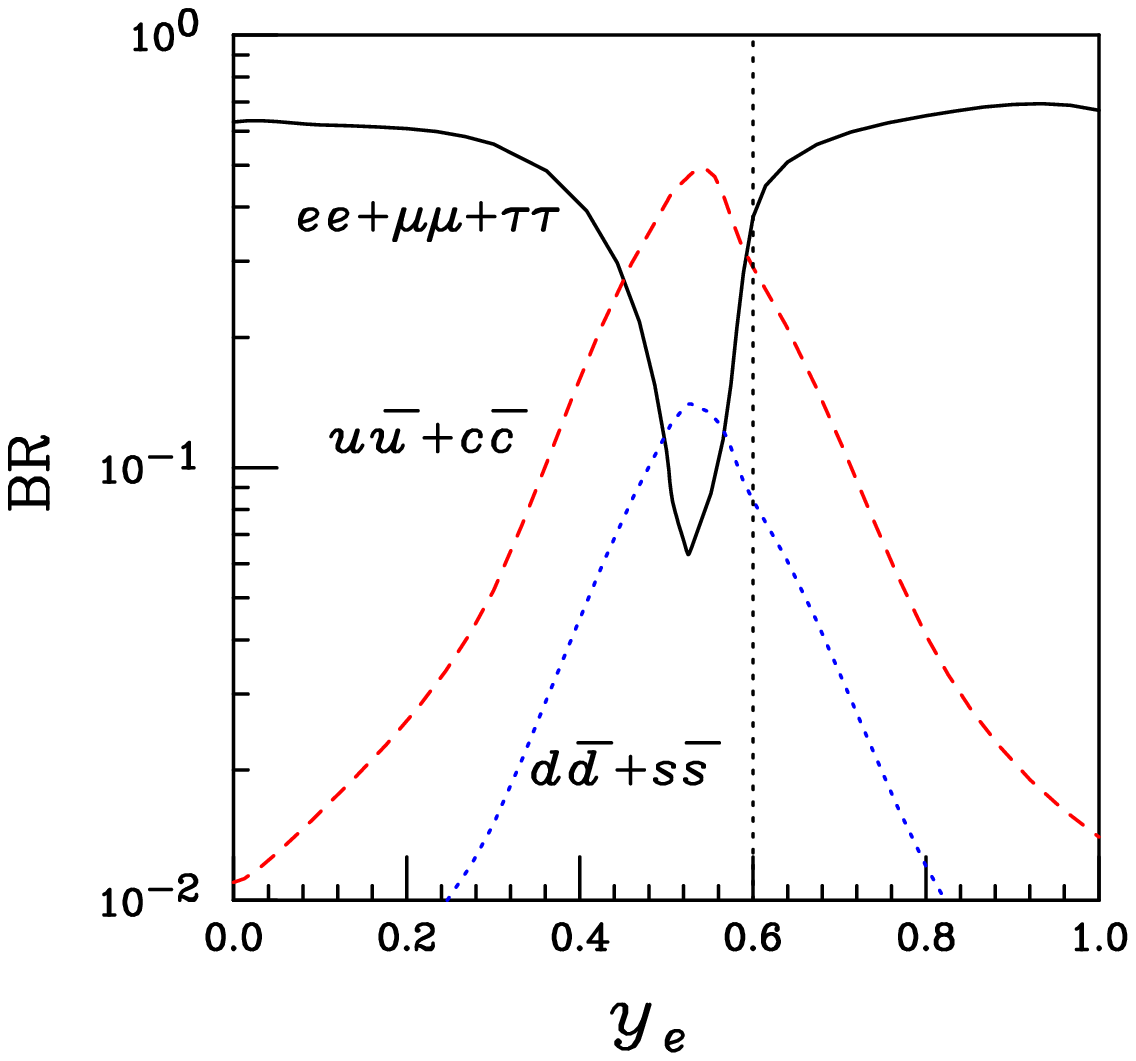}
    \includegraphics[scale=0.6]{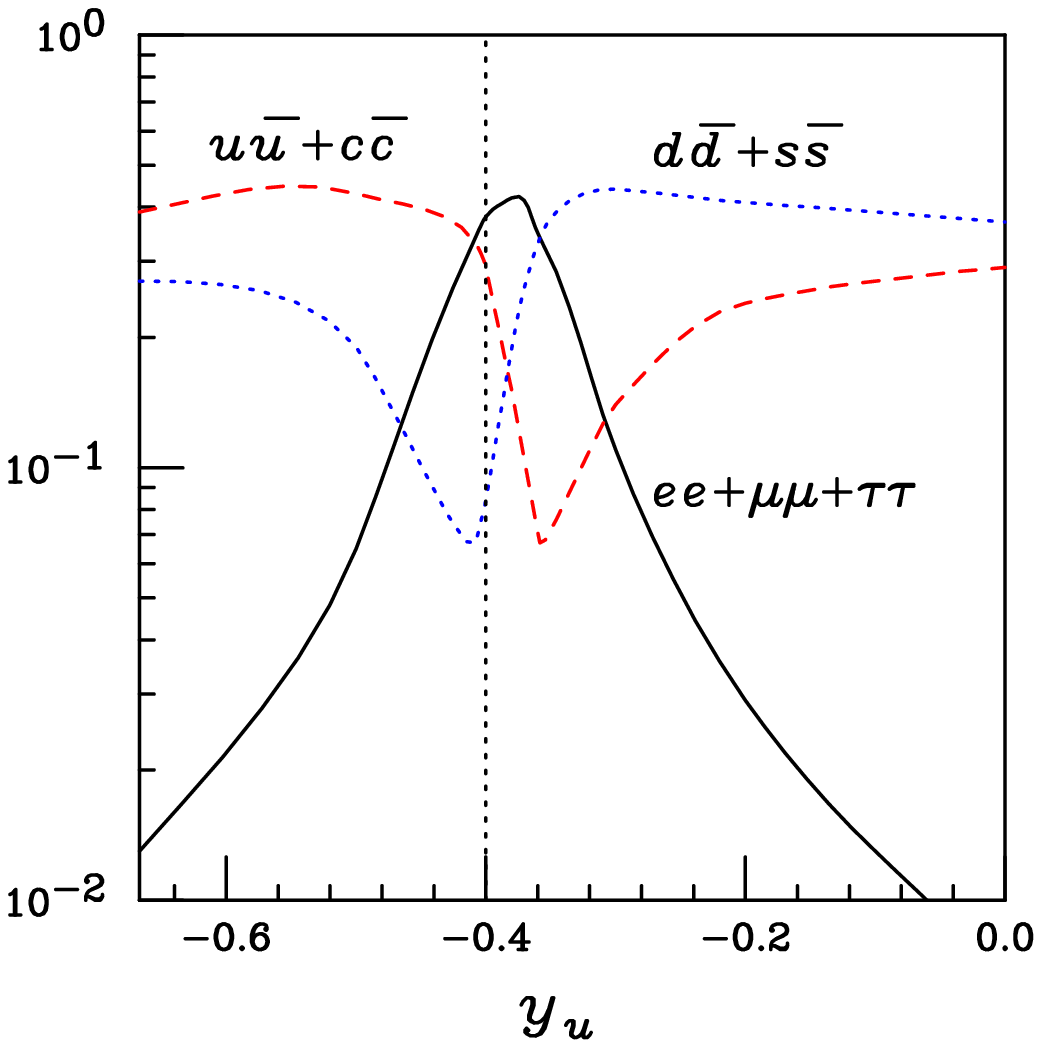}
    \caption {$A_H$ Decay branching fractions for $\tan\theta'=1$
(a) versus the charged lepton $U(1)$ hypercharge $y_e$ with 
fixed $y_u=-0.4$,  and (b) versus up-quark hypercharge $y_u$ 
with fixed $y_e=0.6$. The vertical dotted lines indicate the 
hypercharge values determined by the anomaly-free condition.}
    \label{AH-decay}
\end{figure}

In the littlest Higgs model, the $U(1)$ gauge boson $A_H$ is typically
light and could be the first signal of such a model \cite{joA}. To explore
its signature at colliders, we note first that
its decay mode to $Z_LH$ (or $W^+_L W^-_L$) 
is given by the same formula as in Eq.~(\ref{vtovh}),
but identifying the coupling and mixing as $g\to g',\ \theta \to \theta'$.
The model-dependence comes in when we consider the fermion charges
under the $U(1)$ gauge groups. As we discussed in detail in Sec.~\ref{u1s},
we take the simplest assignment with the anomaly-free
condition for illustration, where $y_e=3/5,\ y_u=-2/5$.  
Fig.~\ref{AH-production}(a) shows the total production cross section at the 
Tevatron and the LHC  energies versus its mass $M_{A_H}$ with $\tan\theta'=1$.
Fig.~\ref{AH-production}(b) gives the decay branching fractions for $A_H$ 
versus $\tan\theta'$ with the same hypercharge assignments
and for $M_{A_H}=1$ TeV. 
Due to the non-observation of resonant lepton pair events in the high mass
region, one may conclude that $A_H$ is excluded for a mass lower than 500 GeV,
which translates to a bound 
\begin{equation}
f\gsim 3\  {\rm TeV}. 
\end{equation}
However, we notice the interesting feature discussed 
earlier in Sec.~\ref{precise} that $A_H$ may decouple from the SM
fermions depending on the $U(1)$ charge assignments at a particular value 
\begin{equation}
\tan^2\theta'=Y_1/Y_2.
\end{equation} 
In this case, the only channels that $A_H$ couples to
are $Z_LH$ and $W^+_LW^-_L$.  Indeed, because of the arbitrariness
of the fermion $U(1)$ charge assignments, the prediction of the $A_H$ signal 
suffers from large theoretical uncertainty. We further explore this aspect by
considering the $A_H$ decay branching fractions when varying the $U(1)$
charge values.  
Fig.~\ref{AH-decay} gives the decay branching fractions for $A_H$ (a) versus
the $U(1)$ charge $y_e$ with fixed $y_u=-0.4$, and (b) versus $y_u$
with fixed $y_e=0.6$. We do see substantial changes
in the branching fractions for different choices of the hypercharge.
They can vary by as large as a factor of 50.
The vertical dotted lines indicate the 
hypercharge values determined by the anomaly-free condition, 
which we used in the previous figure.
In summary, 
although the relatively light $A_H$ gauge boson may give an early
signal at hadron colliders,  with the arbitrariness of the $U(1)$ charge
assignments of the SM fermions,  it cannot  serve as
a robust signature for little Higgs models. 
It could be possible even not to gauge the $U(1)$,
thus to get rid of this massive gauge boson as commented in Sec.~\ref{precise}. 
On the other hand,
if such a $U(1)$ gauge boson is observed at future collider experiments,
it could provide important insight for the gauge structure of the little 
Higgs model.

\subsection{New top quark $T$ at the LHC}
\label{Tpheno}

\begin{figure}[tb]
    \includegraphics[scale=0.8]{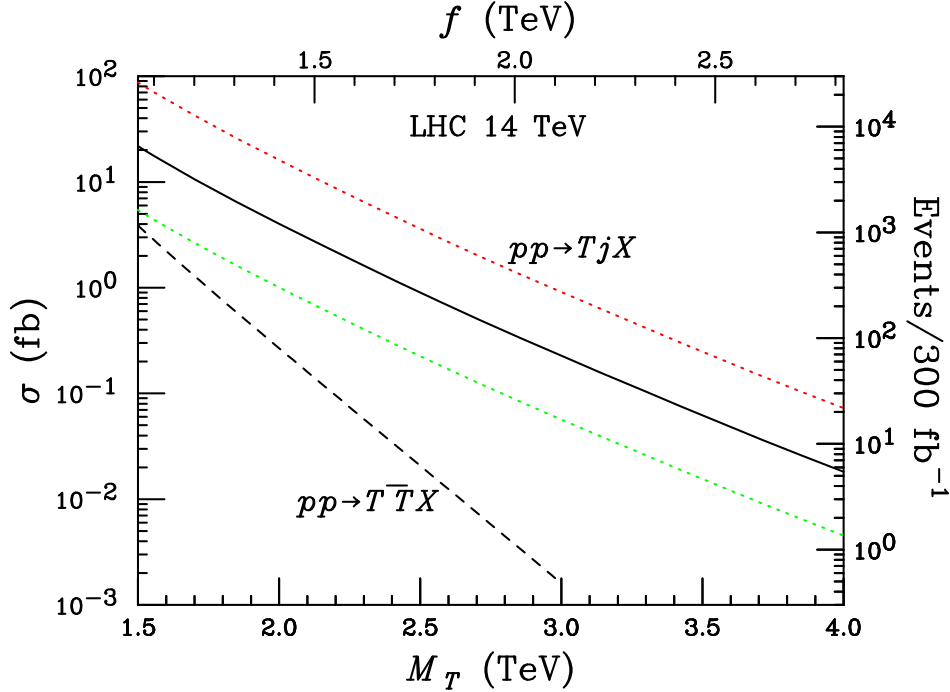}
    \caption {Total cross sections for $T\bar T$ production 
(dashed) and $T+$jet 
production (solid and dotted) via $t$-channel $W$-exchange versus mass $M_T$ 
at the LHC. The solid line is for the couplings $\lambda_1=\lambda_2$;
the dotted are for $\lambda_1/\lambda_2=2$ (upper) and 1/2 (lower).
The number of events expected per 
300 fb$^{-1}$ luminosity is indicated on the right-hand axis.
The scale $f$ corresponding to $\lambda_1=\lambda_2$
is given on the top axis. 
}
    \label{toph}
\end{figure}

The new colored vector-like heavy fermion $T$ is also a crucial prediction
in  little Higgs models. Due to its heavy mass, it may only be produced
at high energy hadron colliders. Naively, the leading contribution seems to
be from the QCD pair production  
\begin{equation}
q\bar q,\ gg \to T\bar T.
\end{equation}
However, the phase space suppression of the multi TeV mass
becomes rather severe.
In contrast, the single $T$ production via $W$-exchange
in $t$-channel (or $Wb$ fusion)
\begin{equation}
q b  \to q' T
\end{equation}
falls off much more slowly with the $T$ mass
and takes over for $M_T$ larger 
than a few hundred GeV \cite{scott}. This is also partially due to 
the enhanced
coupling of the longitudinally polarized gauge bosons at higher energies. In 
Fig.~\ref{toph} the cross sections of pair production of $T\bar T$ 
(dashed line)
and the single $T$ plus a jet production (solid and dotted) 
are presented versus its mass $M_T$ at the LHC energy. 
We see that $T+$jet production dominates throughout the mass range of current
interest. The solid line is for the choice  $\lambda_1=\lambda_2$,
while the dotted are for $\lambda_1/\lambda_2=2$ and 1/2.
We see that for a $T$ with a
3 TeV mass, the cross section can be about 0.23 fb. With an integrated
annual luminosity of 300 fb$^{-1}$, this corresponds to about 70 events
per year, as indicated on the right-hand axis.
The other processes of single $T$ production
$q \bar q'  \to \bar b T$ via  $s$-channel $W$-exchange
and the associated production $g b  \to W_L T$ are both much smaller.

Because of the unsuppressed coupling of the heavy top $T$ to the Higgs
boson, and the enhanced couplings to the longitudinally polarized gauge
bosons (Goldstone bosons)\footnote{We thank M. Perelstein \cite{maxim}
for drawing our attention to this point.}, the partial decay widths of $T$ are
\begin{equation}
\Gamma(T \to tH) = \Gamma(T \to tZ) =
{1\over 2} \Gamma(T \to bW)={\kappa^2\over 32\pi}M_T,
\end{equation}
with the coupling 
$\kappa={\lambda_1^2/\sqrt{\lambda_1^2+\lambda_2^2}}$.
Other decay channels are effectively suppressed by $v^2/f^2$.
The total width of $T$ is then given by 
\begin{equation}
\Gamma_T= {\kappa^2\over 8\pi}M_T= {1\over 8\pi}
{\lambda_1^2\over \lambda_2^2}\left({m_t\over v}\right)^2M_T.
\end{equation}
Unlike the SM top quark, whose total width scales as $m_t^3/v^2$, the
width of $T$ is linear in $M_T$. Regarding the experimental signatures
at colliders, all decay channels can be quite identifiable.
Although the final state $T\to bW$ takes $50\%$ branching fraction,
partly yielding a nice signal of $b$ jet$+\ell^\pm$ plus missing energy,
the other channels $T\to tH,\ tZ$ may lead to distinctive signatures 
as well. The $Z$ boson in the final state gives an unambiguous 
event  identification via its leptonic decay,
and the system $t(\to bW)Z$ reconstructs $M_T$. The Higgs mode can be
studied via $t(\to bW)H(\to b\bar b)$,
resulting in three $b$ jets, a charged lepton plus missing energy.
Two of the $b$ jets reconstruct $m_H$ and the
whole transverse mass system reconstructs the large $M_T$.
There is always a spectator light quark jet ($q'$), accompanying
$T$, that can be made use of as a forward tagging jet.
However, there may be substantial SM backgrounds too, such as
$t\bar tZ$ to the $T\to tZ$ signal, and  $W+4$ QCD jets, 
$t\bar t\to 2 b + 2$ jets + leptons to the $T\to tH$ signal. More detailed
simulations would be needed to make a quantitative conclusion for the
$T$ observation.

If a $T$ signal is observed at the LHC, one can deduce the mass scale
$f$ based on the relation Eq.~(\ref{MTbound}), leading to 
$f^{max}=v M_T/(2m_t) \approx  {M_T/ \sqrt2},$
which is indicated in Fig.~\ref{toph} on the top axis. More precise
determination can be made when the coupling $\lambda_1$ is
measured through the production cross section.

\subsection{The Higgs sector}

The central feature of the model is to have a relatively light neutral Higgs
boson $H$. The Higgs mass is typically of the order of $v$.
If a Higgs boson is found with a mass greater than 140 GeV, it would
imply some new physics different from weak-scale SUSY.
However, the deviation of its properties from the minimal SM is rather
small in the littlest Higgs model, 
generically of the order of $v^2/f^2$, i.e., at a percent level. 
It would thus be 
difficult to distinguish this model from the SM even when $H$ has been
observed. It has been argued that at a high luminosity $e^+e^-$ linear
collider, the determination of $WWH,\ ZZH$ can be at the $1\%$ level \cite{bobetal}.
Inspecting the gauge bosons-Higgs couplings in the littlest Higgs model, we
could anticipate a bound
\begin{equation}
f > {v\over \sqrt{3\times 1\%}}  \approx 1.4 \tev,
\end{equation}
which may not add much new knowledge to our understanding of the model.

\begin{figure}[tb]
    \includegraphics[scale=0.8]{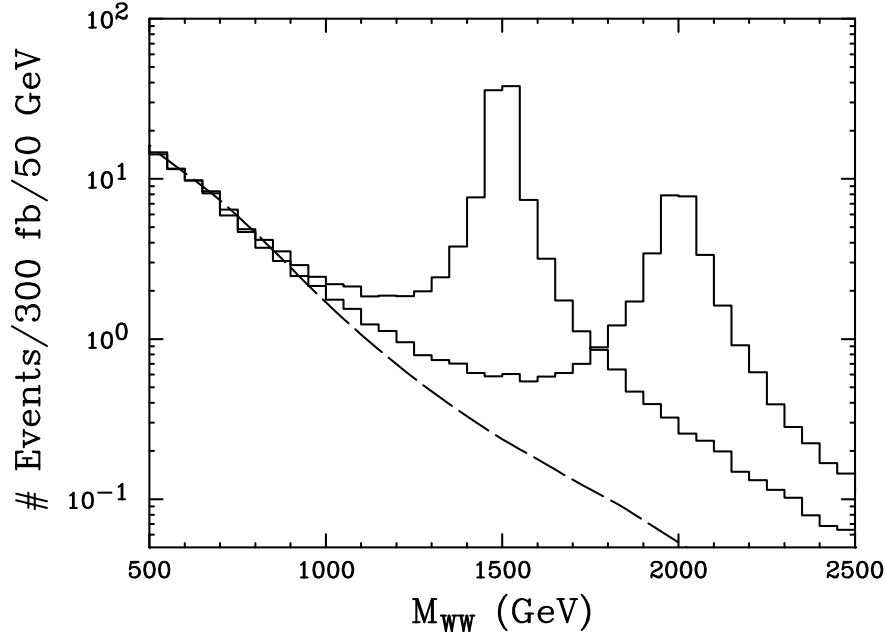}
    \caption {Invariant mass distribution of $W^+W^+$ at the LHC, in the units
of the number of events per bin (50 GeV) and per 300 fb$^{-1}$. The dashed curve
is for $W_L^+W_L^+\to W_L^+W_L^+$ in the SM 
and the histograms include the doubly charged Higgs contribution
with two representative mass values $M_{\Phi}=1.5,\ 2$ TeV, for
$v'/v=0.05$. }
    \label{MWW}
\end{figure}

The would-be Goldstone boson multiplets after the global symmetry
breaking are a necessary feature to result in the light Higgs boson,
and they generically lead to additional scalar multiplets beyond the SM
Higgs doublet. In particular, the doubly charged Higgs state $\Phi^{++}$ 
from the Higgs triplet may serve as a good signal for this class of models
if the coupling is not too small and if it is not too heavy to be accessible 
at future colliders \cite{doubly}.
We illustrate this point by considering the longitudinal $WW$ scattering
\begin{equation}
W_L^+ W_L^+ \to W_L ^+ W_L^+,
\end{equation}
which would receive a resonant contribution from $\Phi^{++}\to W^+W^+$.
Figure \ref{MWW} presents the invariant mass distribution for $M(W^+W^+)$
at the LHC energy. The histograms give the resonant structure for $M_{\Phi}=$
1.5 and 2 TeV respectively. The dashed curve is the continuum SM $W_L^+W_L^+$
background with $M_H=120$ GeV. We have used the effective $W$-boson
approximation to compute the production rates.
In the calculation, we have imposed some cuts on the $W$
transverse momentum  and the rapidity as 
\begin{equation}
p_T^{}>200\ \gev,\quad y<3. 
\end{equation}
The signal cross section is proportional to $v'^2/v^2$.
With the coupling chosen to be $v'/v=0.05$ as for Fig.~\ref{MWW}, 
there can be about 120 (30) events
near the mass peak of 1.5 (2) TeV for 300 fb$^{-1}$ luminosity. 
Although the like-sign di-leptons may
be a spectacular signal for a doubly charged resonance, 
there are SM backgrounds
to be separated. Standard techniques have been developed to identify the
$W_L^+W_L^+$ signal over the backgrounds \cite{wpwp}. We will not pursue
further quantitative evaluation for the signal observability here.

\section{Conclusions}
\label{summer}

The little Higgs models represent 
a new approach to stabilize the hierarchy between a relatively low
cutoff scale $\sim 10$ TeV and the electroweak scale.
By linearizing the ``littlest Higgs model" \cite{littlest}, 
we laid out the full structure of the theory to the order of $1/f^2$, 
and discussed its couplings
and the mass parameters for the new contents beyond the Standard 
Model (summarized in Sec.~\ref{para}).
We explored the symmetry properties in particular related to the
custodial $SU(2)$ breaking in the model (Sec.~\ref{su2}). We also discussed
the arbitrariness of the model associated with the $U(1)$ charge
assignments for the SM fermions, as well as for the Higgs doublet (Sec.~\ref{precise}).

We have studied the phenomenological consequences of the little Higgs models.
The current precision electroweak measurements can put stringent 
bounds on the model parameters, typically for the scale $f\gsim 3.5-4$ TeV,
modulo some arbitrariness of the $U(1)$ charge assignments of the SM particles.
By a clever choice of the gauge coupling parameters and fermion hypercharge 
assignments, the 
extra contributions to the electroweak precision observables may be
significantly suppressed, although even given the freedom of
assigning the fermion charges, the particular 
choice is still a fine-tuning that
needs to be justified by a suitable extension of the model.
Future precision measurements may further improve the constraints, while 
reasonable variations of the model associated with the $U(1)$ sector should
be kept in mind.

We have also studied the collider phenomenology of the little Higgs
model, concentrating on generic signatures that are robust under 
variation of the details of the model. We found that the LHC
has great potential to discover the new $SU(2)$ gauge bosons
up to the multi-TeV  mass scale. This should serve as the ``smoking gun''
signature for the little Higgs model, especially if their unique
decay branching fractions are measured to a good precision. 
The possible new $U(1)$ gauge boson may be lighter and be
observed earlier at hadron colliders, although its properties
are less robust to reflect the little Higgs idea.
The colored vector-like quark $T$ is also a unique prediction
for  little Higgs models, and it may be produced singly through
$Wb\to T$ at high energy hadron colliders. It is however typically
heavier. The doubly-charged Higgs boson may be the most
impressive member of the Higgs sector along with the SM-like 
Higgs. It can be produced singly  via the $W^+W^+\to \Phi^{++}$ 
channel and may provide interesting signatures at the LHC. 
Precision measurements on the triple gauge boson couplings
at hadron and especially at future $e^+e^-$ linear colliders
may also shed light on the symmetry breaking scale up to
$f\sim 3.5-12$ TeV.
Due to the relatively high energy scale of the little Higgs models,
multi-TeV lepton colliders would be desirable to explore the
new particles and study their properties in detail.

\begin{acknowledgments}
We would like to thank Csaba Cs\'aki, Graham Kribs, and Jay Wacker for
valuable discussions, and Piyabut Burikham and Naveen Gaur for pointing
out some typos in the Feynman rules. 
This work was supported in part by the U.S.~Department of Energy
under grant DE-FG02-95ER40896
and in part by the Wisconsin Alumni Research Foundation.
\end{acknowledgments}

\appendix
\section{The Linearized Lagrangian}\label{app_a}

We lay out the linearized Lagrangian for the littlest 
Higgs model in this appendix. 
The effective non-linear Lagrangian invariant under the local gauge group
$G_1\otimes G_2=[SU(2)_1\otimes U(1)_1]\otimes  [SU(2)_2\otimes U(1)_2]$ 
can be written as
	\begin{equation}
	{\mathcal{L}}_{eff}= {\mathcal{L}}_{G} + {\mathcal{L}}_{F} 
				+ {\mathcal{L}}_{\Sigma}
	                              + {\mathcal{L}}_Y - V_{CW}(\Sigma),
	\end{equation}
where $\mathcal{L}_{G}$ consists of the pure gauge terms;
$\mathcal{L}_{F}$ the fermion kinetic terms;  $\mathcal{L}_{\Sigma}$
the $\sigma$-Model terms of the littlest Higgs model; $\mathcal{L}_Y$ the
Yukawa couplings of fermions and pseudo-Goldstone bosons; 
and $ V_{CW}(\Sigma)$ the Coleman-Weinberg potential, generated
radiatively from $\mathcal{L}_{\Sigma},\ \mathcal{L}_Y$.  
We now discuss each individual term in detail. In order to obtain the
effective Lagrangian in terms of the physical fields, we need to
expand the nonlinear $\sigma$-model in a consistent fashion,
which corresponds to expansion in $1/f$. 

\subsection{$\mathcal{L}_{\Sigma}$: Scalar kinetic terms and the heavy gauge bosons}

At the scale $\Lambda_S\sim 4\pi f$, 
the vev associated with the spontaneous symmetry breaking proportional
to the scale $f$ is  parameterized by the $5\times 5$ symmetrical matrix \cite{littlest}
\begin{equation}
\Sigma_0 = \left( \begin{array}{ccc} 
 & & {\mathbf{1}}_{2 \times 2} \\
 &1 & \\
{\mathbf{1}}_{2 \times 2} & & 
\end{array}\right).
\end{equation}
Turning on this vev breaks the assumed global $SU(5)$ symmetry into
its subgroup $SO(5)$.
The appearance of the condensate also breaks the assumed local
gauge symmetry $[SU(2)\otimes U(1)]^2$ into its diagonal subgroup
$[SU(2)\otimes U(1)]_{SM}$. 
The scalar fields are parameterized by
\begin{equation}
\Sigma = e^{i\Pi/f}\ \Sigma_0\  e^{i\Pi^T/f},
\label{sigmaA}
\end{equation}
that transforms under the gauge group as
\begin{equation}
\Sigma \to \Sigma' = U\  \Sigma \ U^T,
\end{equation}
where $U=L_1 Y_1 L_2 Y_2$ is an element of the gauge groups.
Here $f$ is the Goldstone boson decay constant, and the Goldstone boson matrix 
$\Pi$ is expressed by
\begin{equation}
\Pi =  \left( \begin{array}{ccc}
		{} & h^\dagger/\sqrt 2 & \phi^\dagger \\
		h/\sqrt2 &  {} & h^*/\sqrt2 \\
		\phi  & h^T/\sqrt2 & {}
		\end{array} \right),
\label{Pi}
\end{equation}
where the scalar field content consists of a doublet $h$ and a triplet 
$\phi$ under the
unbroken $SU(2)_L\otimes U(1)_Y$ SM gauge group
	\begin{equation}
	h = (h^+, h^0), \qquad
	\phi = \left( \begin{array}{cc}
		\phi^{++} & \frac{\phi^+}{\sqrt{2}} \\
		\frac{\phi^+}{\sqrt{2}} & \phi^0
		\end{array} \right).
	\end{equation}
For phenomenological studies, it is important to linearize the effective
Lagrangian and write it in terms of the couplings of gauge bosons and
$h$, $\phi$. This can be achieved by expanding $\Sigma$ around its
vacuum expectation value in powers of ${1}/{f}$
\begin{equation}
	\Sigma = \Sigma_0 + \frac{2 i}{f} \left( \begin{array}{ccccc}
	\phi^{\dagger} & \frac{h^{\dagger}}{\sqrt{2}} & {\mathbf{0}}_{2\times
	2} \\
	\frac{h^{*}}{\sqrt{2}} & 0 & \frac{h}{\sqrt{2}} \\
	{\mathbf{0}}_{2\times 2} & \frac{h^{T}}{\sqrt{2}} & \phi
	\end{array} \right) 
	- \frac{1}{f^2} \left( \begin{array}{ccc}
	h^{\dagger} h^* & \sqrt{2} \phi^{\dagger} h^T
		& h^{\dagger} h + 2 \phi^{\dagger} \phi \\
	\sqrt{2} h \phi^{\dagger} & 2 h h^{\dagger}
		& \sqrt{2} h^* \phi \\
	h^T h^* + 2 \phi \phi^{\dagger} & \sqrt{2} \phi h^{\dagger}
			& h^T h
	\end{array} \right)
+ {\cal O}(\frac{1}{f^3}).
\label{eq:Sigma}
\end{equation}

The leading order dimension-two term in the non-linear $\sigma$-model 
can be written for the scalar sector as
	\begin{equation}
	\mathcal{L}_{\Sigma} = \frac{1}{2} \frac{f^2}{4} 
	{\rm Tr} | \mathcal{D}_{\mu} \Sigma |^2.
\label{SigmaA}
	\end{equation}
The numerical coefficients have been chosen so that the scalar kinetic terms
are canonically normalized. It is manifestly gauge invariant under 
$G_1\otimes G_2=[SU(2)_1\otimes U(1)_1]\otimes  [SU(2)_2\otimes U(1)_2]$
if the covariant derivative is defined as
         \begin{equation}
           \mathcal{D}_\mu \Sigma=  \partial_\mu\Sigma - i \sum_{j=1}^2\left( 
g_j( W_j\Sigma +  \Sigma W_j^T) + g'_j (B_j\Sigma + \Sigma B_j^T) \right),
	\end{equation}
where the $SU(2)$ gauge fields are $W_j=\sum_{a=1}^3 W_{\mu j}^a Q_j^a$ with
\begin{equation}
Q_1^a = \left( \begin{array}{ccc} 
\frac{\sigma^a}{2} & & \\
 & & \\
 & & {\mathbf{0}}_{3\times 3}
\end{array}\right),\qquad 
Q_2^a = \left( \begin{array}{ccc} 
{\mathbf{0}}_{3\times 3} & & \\
 & & \\
 & & -\frac{\sigma^{a*}}{2}
\end{array}\right).
\end{equation} 
Similarly, the $U(1)$ gauge fields are $B_j=B_{\mu j} Y_j$ with
\begin{equation}
Y_1 = \frac{1}{10}\left( \begin{array}{ccccc} 
-3 & &  & & \\
 &-3 & & & \\
 & & 2 & & \\
 & &  & 2 & \\
 & & & & 2 
\end{array}\right), \qquad 
Y_2 = \frac{1}{10}\left( \begin{array}{ccccc} 
-2 & &  & & \\
 &-2 & & & \\
 & & -2& & \\
 & &  & 3 & \\
 & & & & 3 
\end{array}\right).
\end{equation} 
The vacuum expectation value (vev) of the $\Sigma$ field breaks the 
$[SU(2)\otimes U(1)]_1\otimes [SU(2)\otimes U(1)]_2$ gauge symmetry down 
to the diagonal subgroup, with the broken generators (associated with $W'$)  
\begin{equation}
Q'^a=\frac{1}{\sqrt{g_1^4+g_2^4}} (g_1^2 Q_1^a - g_2^2 Q_2^a ).
\end{equation}
and the unbroken  $SU(2)_L\otimes U(1)_Y$ gauge generators  
\begin{equation}
Q^a =\frac{1}{\sqrt{2}} (Q_1^a + Q_2^a),\qquad Y=Y_1+Y_2.
\end{equation}

The spontaneous gauge symmetry breaking thereby gives
rise to mass terms of order $f$ for  the gauge bosons
\begin{eqnarray}
	\mathcal{L}_{\Sigma} (mass)&=& 
	\frac{1}{2} \frac{f^2}{4} \left[
	g_1^2 W_{1 \mu}^a W_1^{a\mu}
	+ g_2^2 W_{2\mu}^a W_2^{a\mu}
	- 2 g_1g_2 W_{1\mu}^a W_2^{a\mu}
	 \right] \nonumber \\
	&+& \frac{1}{2} \frac{f^2}{4} \frac{1}{5}
	\left[ g_1^{\prime 2} B_{1\mu} B_1^{\mu}
	+ g_2^{\prime 2} B_{2 \mu} B_2^{\mu}
	- 2 g_1^{\prime}g_2^{\prime} B_{1\mu}B_2^{\mu} \right].
\end{eqnarray}
We define
\begin{eqnarray}
	W = s W_1 + c W_2, &\qquad& W^{\prime} = -c W_1 + s W_2 \nonumber \\
	B = s^{\prime} B_1 + c^{\prime} B_2, &\qquad&
	B^{\prime} = -c^{\prime} B_1 + s^{\prime} B_2,
\end{eqnarray}
where the mixing angles are given by:
\begin{eqnarray}
	s = \frac{g_2}{\sqrt{g_1^2+g_2^2}}, &\qquad&
	c = \frac{g_1}{\sqrt{g_1^2+g_2^2}} \nonumber \\
	s^{\prime} = \frac{g_2^{\prime}}{\sqrt{g_1^{\prime 2}+g_2^{\prime 2}}},
	&\qquad&
	c^{\prime} = \frac{g_1^{\prime}}{\sqrt{g_1^{\prime 2}+g_2^{\prime 2}}}.
\end{eqnarray}
The heavy gauge boson masses are then
\begin{eqnarray}
	m_{W^{\prime}} = \frac{f}{2} \sqrt{g_1^2 + g_2^2}
	= \frac{g}{2sc} f , \qquad
	m_{B^{\prime}} =  \frac{f}{2\sqrt{5}} 
	\sqrt{g_1^{\prime 2} + g_2^{\prime 2}}
	= \frac{g^{\prime}}{2\sqrt{5}s^{\prime}c^{\prime}}f .
\end{eqnarray}
The massless states $W$ and $B$ are identified as the SM
gauge bosons, with couplings
\begin{equation}
	g = g_1 s = g_2 c, \qquad 
	g^{\prime} = g_1^{\prime} s^{\prime} = g_2^{\prime} c^{\prime}.
\end{equation}

\subsection{$V_{CW}$: Effective Higgs potential and the electroweak 
symmetry breaking}
\label{ewsb-appendix}

In the littlest Higgs model, the global symmetries prevent the appearance
of a Higgs potential at tree level.  Instead, the Higgs potential is 
generated at one-loop and higher orders due to interactions
with the gauge bosons and fermions.  The quadratically divergent contributions
to this Coleman-Weinberg potential are cut off by the scale $\Lambda_S$.
In practice, these are proportional to $\Lambda_S^2/16\pi^2 \sim f^2$.
The unknown ultraviolet physics at the cutoff scale $\Lambda_S$ is parameterized
by $\mathcal{O}(1)$ coefficients $a$ and $a^{\prime}$.

The most important terms of the Coleman-Weinberg potential can be 
parameterized as:
\begin{equation}
	V = \lambda_{\phi^2} f^2 {\rm Tr}(\phi^{\dagger} \phi)
	+ i \lambda_{h \phi h} f \left( h \phi^{\dagger} h^T
		- h^* \phi h^{\dagger} \right)
	- \mu^2 h h^{\dagger}
	+ \lambda_{h^4} ( h h^{\dagger} )^2,
\label{lambdas}
\end{equation}
where we neglect quartic terms involving $\phi^4$ and $h^2 \phi^2$ 
since they give only sub-leading contributions to the vacuum expectation
values and the scalar field masses.

The quadratically divergent contribution to the Coleman-Weinberg 
potential from vector boson loops is \cite{littlest} 
\begin{equation}
	\mathcal{L}_a = \frac{1}{2} af^4\left\{ g_j^2 \sum_a {\rm Tr} \left[
	\left( Q_j^a \Sigma \right) \left( Q_j^a \Sigma \right)^* \right]
	+ g_j^{\prime 2} {\rm Tr} \left[
	\left( Y_j \Sigma \right) \left( Y_j \Sigma \right)^* \right] \right\}.
\end{equation}
Linearizing the $\Sigma$ field, we obtain
\begin{eqnarray}
	\mathcal{L}_a &=& \frac{a}{2} (g_1^2 + g_1^{\prime 2}) \left[
	f^2 {\rm Tr} \left( \phi^{\dagger}\phi \right)
	- \frac{if}{2} \left( h \phi^{\dagger} h^T 
		- h^*\phi h^{\dagger} \right)
	+ \frac{1}{4} \left( h h^{\dagger} \right)^2 + \cdots
	 \right]
	\nonumber \\
	&+& \frac{a}{2} (g_2^2 + g_2^{\prime 2}) \left[
	f^2 {\rm Tr} \left( \phi^{\dagger}\phi \right)
	+ \frac{if}{2} \left( h \phi^{\dagger} h^T 
		- h^*\phi h^{\dagger} \right)
	+ \frac{1}{4} \left( h h^{\dagger} \right)^2 + \cdots \right].
	\label{eq:Lcexp}
\end{eqnarray}
The $[SU(2) \otimes U(1)]_1$ interactions preserve the
$SU(3)_2$ global symmetry in the lower $3\times 3$ block of $\Sigma$, 
while the 
$[SU(2) \otimes U(1)]_2$ interactions preserve the
$SU(3)_1$ global symmetry in the upper $3 \times 3$ block of $\Sigma$.

The quadratically divergent contribution to the Coleman-Weinberg 
potential from fermion loops is \cite{littlest} 
\begin{eqnarray}
	\mathcal{L}_{a^{\prime}} &=& -a^{\prime} \frac{1}{4} \lambda_1^2 f^4
	\epsilon^{wx} \epsilon_{yz} \epsilon^{ijk} \epsilon_{kmn}
	\Sigma_{iw} \Sigma_{jx} \Sigma^{*my} \Sigma^{*nz},
	\label{eq:Lc'}
\end{eqnarray}
where $i,j,k,m,n$ run over $1,2,3$ and $w,x,y,z$ run over $4,5$.
To fourth order in $h$ and second order in $\phi$, this term leads to
\begin{eqnarray}
	\mathcal{L}_{a^{\prime}} &=& 8 a^{\prime} \lambda_1^2 
	\left[
	f^2 {\rm Tr} \left( \phi^{\dagger} \phi \right)
	+ \frac{if}{2} \left( h \phi^{\dagger} h^T 
	- h^* \phi h^{\dagger} \right)
	+ \frac{1}{4} \left( h h^{\dagger} \right)^2
	+ \cdots
	\right].
\end{eqnarray}
The fermion interactions that give rise to this term
preserve the $SU(3)_1$ global symmetry in the upper $3\times 3$ block
of $\Sigma$, 
so this contribution to the potential 
must have the same form as the term proportional
to $(g_2^2 + g_2^{\prime 2})$ in Eq.~(\ref{eq:Lcexp}).

The coefficients $\lambda_{\phi^2}$, $\lambda_{h \phi h}$
and $\lambda_{h^4}$ in Eq.~(\ref{lambdas}) are therefore given by:
\begin{eqnarray}
	\lambda_{\phi^2} &=& \frac{a}{2} \left[ \frac{g^2}{s^2c^2} 
	+ \frac{g^{\prime 2}}{s^{\prime 2}c^{\prime 2}} \right]
	+ 8 a^{\prime} \lambda_1^2,
	\nonumber \\
	\lambda_{h \phi h} &=& -\frac{a}{4} 
	\left[ g^2 \frac{(c^2-s^2)}{s^2c^2}
	+ g^{\prime 2} \frac{(c^{\prime 2}-s^{\prime 2})}
	{s^{\prime 2}c^{\prime 2}} \right]
	+ 4 a^{\prime} \lambda_1^2,
	\nonumber \\
	\lambda_{h^4} &=& \frac{a}{8} \left[ \frac{g^2}{s^2c^2} 
	+ \frac{g^{\prime 2}}{s^{\prime 2}c^{\prime 2}} \right]
	+ 2 a^{\prime} \lambda_1^2 = \frac{1}{4} \lambda_{\phi^2}.
	\label{lambdas2}
\end{eqnarray}
Here we have neglected the log-divergent one-loop and quadratically divergent
two-loop contributions to the effective couplings in Eq.~(\ref{lambdas2}).  
These are suppressed
by a loop factor $1/16\pi^2$ compared to the leading terms given here.

The coefficient $\mu^2$ of the $hh^{\dagger}$ term 
is a {\it free parameter} since this term gets equally significant
contributions from the one-loop log-divergent and two-loop 
quadratically-divergent parts of the Coleman-Weinberg potential.  
At one-loop order, $\mu^2$ gets a contribution from the log-divergent
terms of order $f^2 \log(\Lambda_S^2/f^2)/(16\pi^2)$, giving a natural 
hierarchy between the TeV scale $f$ and the electroweak scale.
At two-loop order, $\mu^2$ gets a contribution from the 
quadratically-divergent
term of order $\Lambda_S^2/(16\pi^2)^2 \sim f^2/16\pi^2$, with an arbitrary
coefficient of order unity determined by the UV completion.  
We thus write the coefficient as a new free parameter 
$\mu^2 \sim f^2/16\pi^2$.

For $\mu^2 > 0$,
this scalar potential triggers electroweak symmetry breaking,
resulting in the vacuum expectation values for the $h$ and $\phi$ fields:
$\langle h^0 \rangle = v/\sqrt{2}$
and $\langle i \phi^0 \rangle = v^{\prime}$, with
\begin{equation}
	v^2 = \frac{\mu^2}
	{\lambda_{h^4} - \lambda_{h \phi h}^2 / \lambda_{\phi^2}}, \qquad
	v^{\prime} = \frac{\lambda_{h \phi h}}{2 \lambda_{\phi^2}} 
	\frac{v^2}{f}.
\end{equation}

The gauge eigenstates of the Higgs fields $h$ and $\phi$ can be written in terms of the 
mass eigenstates as follows:
\begin{eqnarray}
	h^0 &=& \left( c_0 H - s_0 \Phi^0 + v \right)/\sqrt{2}
	+ i \left( c_P G^0 - s_P \Phi^P \right) /\sqrt{2}, \nonumber \\
	\phi^0 &=& \left( s_P G^0 + c_P \Phi^P \right) /\sqrt{2}
	- i \left( s_0 H + c_0 \Phi^0 + \sqrt{2} v^{\prime} \right) /\sqrt{2},
	\nonumber \\
	h^+ &=& c_+ G^+ - s_+ \Phi^+, \qquad
	\phi^+ = \left( s_+ G^+ + c_+ \Phi^+ \right)/i  \nonumber \\
	\phi^{++} &=& \Phi^{++}/i.
\end{eqnarray}
We use the following notation for the physical mass eigenstates:
$H$ and $\Phi^0$ are neutral scalars, $\Phi^P$ is a neutral pseudoscalar,
$\Phi^+$ and $\Phi^{++}$ are the charged and doubly charged scalars, and
$G^+$ and $G^0$ are the Goldstone bosons that are eaten by the light
$W$ and $Z$ bosons, giving them mass.  Note that in defining the
mass eigenstates we have factored out an $i$ from $\phi$.

The mixing angles in the pseudoscalar and singly-charged sectors are
easily extracted in terms of the vacuum expectation values:
\begin{eqnarray}
	s_P &=& \frac{2 \sqrt{2} v^{\prime}}
	{\sqrt{v^2 + 8 v^{\prime 2}}}
	\simeq 2 \sqrt{2} \frac{v^{\prime}}{v}, \qquad
	c_P = \frac{v}{\sqrt{v^2 + 8 v^{\prime 2}}}
	\simeq 1 - 4 \frac{v^{\prime 2}}{v^2} \nonumber \\
	s_+ &=& \frac{2 v^{\prime}}{\sqrt{v^2 + 4 v^{\prime 2}}}
	\simeq 2 \frac{v^{\prime}}{v}, \qquad
	c_+ = \frac{v}{\sqrt{v^2 + 4 v^{\prime 2}}}
	\simeq 1 - 2 \frac{v^{\prime 2}}{v^2}.
\end{eqnarray}
Diagonalizing the mass terms for the neutral CP-even scalars gives
the scalar mixing angle $s_0,c_0$ to leading order in $v/f$:
\begin{equation}
	s_0 \simeq 2 \sqrt{2} \frac{v^{\prime}}{v}, \qquad
	c_0 \simeq 1 - 4 \frac{v^{\prime 2}}{v^2}.
\end{equation}
Note that to leading order $s_0 = s_P=\sqrt2 s_+ \sim \mathcal{O}(v/f)$.

To leading order, all of the triplet $\Phi$ states are degenerate in mass. 
The masses of $\Phi$ and $H$ are
\begin{equation}
	M^2_{\Phi} \simeq \lambda_{\phi^2} f^2, \qquad
	m^2_{H} \simeq 2 \left( \lambda_{h^4} 
	- \lambda_{h \phi h}^2 / \lambda_{\phi^2} \right) v^2 = 2 \mu^2.
\end{equation}

\subsection{Gauge boson masses and mixing from $\mathcal{L}_{\Sigma}$}

After EWSB, the gauge sector gets additional mass and mixing terms
due to the $h$ and $\phi$ vevs.  The full set of mass terms after EWSB is:
\begin{eqnarray}
        \mathcal{L}_\Sigma(masses) &=&
        \frac{1}{2} W_{\mu}^{\prime a} W^{\prime a \mu}
        \left[ m_{W^{\prime}}^2 - \frac{1}{4}g^2v^2 \right]
        + W_{\mu}^+ W^{- \mu}
        \left[ \frac{1}{4} g^2 v^2 \left( 1 - \frac{v^2}{6f^2} 
		+ 4 \frac{v^{\prime 2}}{v^2} \right) \right]
	\nonumber \\ &&
        + \frac{1}{2} W_{\mu}^3 W^{3 \mu}
        \left[ \frac{1}{4} g^2 v^2 \left( 1 - \frac{v^2}{6f^2} 
		+ 8 \frac{v^{\prime 2}}{v^2} \right) \right]
	\nonumber \\ &&
        - W_{\mu}^a W^{\prime a \mu}
        \left[ \frac{1}{4} g^2 v^2 \frac{(c^2-s^2)}{2sc} \right]
        + \frac{1}{2} B_{\mu}^{\prime} B^{\prime \mu}
        \left[ m_{B^{\prime}}^2 - \frac{1}{4} g^{\prime 2} v^2 \right]
	\nonumber \\ &&
        + \frac{1}{2} B_{\mu} B^{\mu}
        \left[ \frac{1}{4} g^{\prime 2} v^2 
		\left( 1 - \frac{v^2}{6f^2} 
		+ 8 \frac{v^{\prime 2}}{v^2} \right)\right]
        - B_{\mu} B^{\prime \mu}
        \left[ \frac{1}{4} g^{\prime 2} v^2 
	\frac{(c^{\prime 2}-s^{\prime 2})}{2s^{\prime}c^{\prime}} \right]
        \nonumber \\
        && + W_{\mu}^3 B^{\mu} \left[\frac{1}{4} gg^{\prime} v^2 
		\left( 1 - \frac{v^2}{6f^2} 
		+ 8 \frac{v^{\prime 2}}{v^2} \right)\right]
        + W_{\mu}^{\prime 3} B^{\prime \mu}
        \left[ -\frac{1}{8} g g^{\prime} v^2
	\left( \frac{cs^{\prime}}{sc^{\prime}} 
	+ \frac{sc^{\prime}}{cs^{\prime}} \right) \right]
        \nonumber \\ &&
        - W_{\mu}^3 B^{\prime \mu}
        \left[ \frac{1}{4} g g^{\prime} v^2 
	\frac{(c^{\prime 2}-s^{\prime 2})}{2s^{\prime}c^{\prime}} \right]
        - W_{\mu}^{\prime 3} B^{\mu}
        \left[\frac{1}{4} g g^{\prime} v^2 
	\frac{(c^2-s^2)}{2sc} \right],
\end{eqnarray}
where for the $WW$, $BB$ and $WB$ terms we have included terms up to 
order $v^4/f^2$; these will be necessary in order to find the masses
of the light gauge bosons consistently to this order.

\subsubsection{Charged gauge bosons}

Let us first consider the charged $W,W^{\prime}$ sector.
The charged $W,W^{\prime}$ mass eigenstates, to order $v^2/f^2$, are: 
\begin{equation}
        W_L = W
        + \frac{v^2}{2f^2} sc(c^2-s^2) W^{\prime},
        \qquad
        W_H = W^{\prime}
        - \frac{v^2}{2f^2} sc(c^2-s^2) W.
\end{equation}
The masses of $W_L$ (light) and $W_H$ (heavy) to the order of
$v^2/f^2$ are given by:
\begin{eqnarray}
	M_{W_L^{\pm}}^2 &=&  m_w^2 \left[
	1 - \frac{v^2}{f^2} \left( \frac{1}{6} 
	+ \frac{1}{4} (c^2-s^2)^2
	\right) + 4 \frac{v^{\prime 2}}{v^2} \right], \\
		M_{W_H^{\pm}}^2 &=& \frac{f^2g^2}{4s^2c^2}
	- \frac{1}{4} g^2v^2 
	+ \mathcal{O}(v^4/f^2)= m_w^2\left( \frac{f^2}{s^2c^2v^2}-1\right), 
\end{eqnarray}
where $m_w\equiv gv/2$.

\subsubsection{Neutral gauge bosons}

The four neutral gauge boson mass eigenstates to the order $v^2/f^2$ 
are:\footnote{We have absorbed a minus sign into the definition of $B$
in order to write $A_L$ and $Z_L$ in the standard form.}
\begin{eqnarray}
	A_L &=& s_{\w} W^3 + c_{\w} B
	\nonumber \\
	Z_L &=& c_{\w} W^3 - s_{\w} B 
	+ x_Z^{W^{\prime}} \frac{v^2}{f^2} W^{\prime 3}
		+ x_Z^{B^{\prime}} \frac{v^2}{f^2} B^{\prime}
	\nonumber \\
	A_H &=& B^{\prime} + x_H \frac{v^2}{f^2} W^{\prime 3}
	- x_Z^{B^{\prime}} \frac{v^2}{f^2} ( c_{\w} W^3 - s_{\w} B )
	\nonumber \\
	Z_H &=& W^{\prime 3} - x_H \frac{v^2}{f^2} B^{\prime}
	- x_Z^{W^{\prime}} \frac{v^2}{f^2} ( c_{\w} W^3 - s_{\w} B ), 
\label{zlzh}
\end{eqnarray}
where
\begin{eqnarray}
	x_H &=& \frac{5}{2} g g^{\prime}
	\frac{scs^{\prime}c^{\prime} (c^2s^{\prime 2} + s^2c^{\prime 2})}
	{(5g^2 s^{\prime 2} c^{\prime 2} - g^{\prime 2} s^2 c^2)},
	\nonumber \\
	x_Z^{W^{\prime}} &=& -\frac{1}{2c_{\w}} sc(c^2-s^2), \qquad
	x_Z^{B^{\prime}} = -\frac{5}{2s_{\w}} s^{\prime}c^{\prime}
	(c^{\prime 2}-s^{\prime 2}).
\label{eq:xmixing}
\end{eqnarray}
The weak mixing angle is defined as usual:
\begin{eqnarray}
	s_{\w} = \frac{g^{\prime}}{\sqrt{g^2 + g^{\prime 2}}}, &\qquad&
	c_{\w} = \frac{g}{\sqrt{g^2 + g^{\prime 2}}}.
\end{eqnarray}
The neutral gauge boson masses are:
\begin{eqnarray}
	M_{A_L}^2 &=& 0 \nonumber \\
	M_{Z_L}^2 &=& m_z^2
	\left[ 1 - \frac{v^2}{f^2} \left( \frac{1}{6} 
	+ \frac{1}{4} (c^2-s^2)^2
	+ \frac{5}{4} (c^{\prime 2}-s^{\prime 2})^2 \right) 
	+ 8 \frac{v^{\prime 2}}{v^2} \right]
	\nonumber \\
	M_{A_H}^2 &=& 
	\frac{f^2 g^{\prime 2}}{20 s^{\prime 2} c^{\prime 2}}
	- \frac{1}{4} g^{\prime 2} v^2 + g^2 v^2 \frac{x_H}{4s^2c^2}
          = m_z^2 s_{\w}^2 \left(
	\frac{ f^2 }{5 s^{\prime 2} c^{\prime 2}v^2}
	- 1 + \frac{x_H c_{\w}^2}{4s^2c^2  s_{\w}^2} \right)
	\nonumber \\
	M_{Z_H}^2 &=& \frac{f^2g^2}{4s^2c^2}
	- \frac{1}{4} g^2 v^2 
	- g^{\prime 2} v^2 \frac{x_H}{4s^{\prime 2}c^{\prime 2}}
	= m_w^2 \left( \frac{f^2}{s^2c^2 v^2}
	- 1 -  \frac{x_H s_{\w}^2}{s^{\prime 2}c^{\prime 2}c_{\w}^2}\right) ,
\end{eqnarray}
where $m_z\equiv {gv}/(2c_{\w})$.
Again, note that the $Z_L$ mass gets a correction at order $v^2/f^2$.

\subsection{Scalar-fermion couplings:
The Yukawa interactions $\mathcal{L}_{Y}$}
\label{yukawa-appen}

The scalar couplings to the top quark  can be taken as \cite{littlest}
\begin{equation}
{\mathcal{L}}_Y = \frac{1}{2} \lambda_1 f \epsilon_{ijk} \epsilon_{xy} \chi_i
\Sigma_{jx} \Sigma_{ky} u^{\prime c}_3 
+ \lambda_2 f \tilde{t} \tilde{t}^{\prime c}
+ {\rm h.c.},  
\label{Yukawat}
\end{equation}
where $\chi_i=(b_3, t_3, \tilde{t})$.  The factor of $1/2$ normalization
in front of $\lambda_1$ makes our notation simpler. 
Expanding the $\Sigma$ field generates the scalar interactions with quarks:
\begin{eqnarray}
	\mathcal{L}_{t} &=& \lambda_2 f \tilde t \tilde t^{\prime c} 
	+ i \lambda_1 \left\{
	- b_3 \left[ \sqrt{2} h^+
		+ \frac{i}{f} (\sqrt{2} h^- \phi^{++} + h^{0*} \phi^+)
		\right] u_3^{\prime c} \right. \nonumber \\
	&& \qquad - t_3 \left[ \sqrt{2} h^0
		+ \frac{i}{f} (h^- \phi^+ + \sqrt{2} h^{0*}\phi^0)
		\right] u_3^{\prime c}   
\label{top-cancel} \\
	&& \qquad \left. + \tilde t \left[ -if
		+ \frac{i}{f} (h^+h^- + h^0h^{0*}
		+ 2 \phi^{++}\phi^{--} + 2 \phi^+\phi^- + 2 \phi^0\phi^{0*})
		\right] u_3^{\prime c}	\right\}
	+ {\rm h.c.} \nonumber
\end{eqnarray}
This Lagrangian contains a mass term of order $f$ 
that couples $\tilde t$ to a linear combination of 
$\tilde t^{\prime c}$ and $u_3^{\prime c}$.  Defining mixtures 
of $\tilde t^c$ and $u_3^{\prime c}$ as follows,
\begin{eqnarray}
	\tilde t^c &=& \frac{1}{\sqrt{\lambda_1^2 + \lambda_2^2}}
	(\lambda_2 \tilde t^{\prime c} + \lambda_1 u_3^{\prime c}),
	\qquad
	u_3^c = \frac{1}{\sqrt{\lambda_1^2 + \lambda_2^2}}
	(- \lambda_1 \tilde t^{\prime c} + \lambda_2 u_3^{\prime c}),
\end{eqnarray}
diagonalizes the mass term for the heavy fermions:
\begin{equation}
	\mathcal{L}_f = f \sqrt{\lambda_1^2 + \lambda_2^2} \tilde t \tilde t^c
	= - m_{\tilde t} \tilde t \tilde t^c.
\end{equation}
The rest of the Lagrangian reads:
\begin{eqnarray}
	\mathcal{L}_f &=& 
	- \frac{\sqrt{2}i\lambda_1^2}{\sqrt{\lambda_1^2 + \lambda_2^2}}
	\left[ b_3 h^+ \tilde t^c + t_3 h^0 \tilde t^c \right]
	- \frac{\sqrt{2}i\lambda_1\lambda_2}
		{\sqrt{\lambda_1^2 + \lambda_2^2}}
	\left[ b_3 h^+ u_3^c + t_3 h^0 u_3^c \right] \nonumber \\
	&& + \frac{\lambda_1^2}{\sqrt{\lambda_1^2 + \lambda_2^2}}
	\frac{1}{f} \left[ 
	b_3 (\sqrt{2}h^-\phi^{++} + h^{0*}\phi^+)\tilde t^c
	+ t_3 (h^-\phi^+ + \sqrt{2}h^{0*}\phi^0)\tilde t^c
	\right. \nonumber \\
	&& \left.
	- \tilde t(h^+h^- + h^0h^{0*} 
		+ 2\phi^{++}\phi^{--} + 2\phi^+\phi^- + 2\phi^0\phi^{0*})
			\tilde t^c \right] \nonumber \\
	&& + \frac{\lambda_1\lambda_2}{\sqrt{\lambda_1^2 + \lambda_2^2}}
	\frac{1}{f} \left[ 
	b_3 (\sqrt{2}h^-\phi^{++} + h^{0*}\phi^+) u_3^c
	+ t_3 (h^-\phi^+ + \sqrt{2}h^{0*}\phi^0) u_3^c
	\right. \nonumber \\
	&&\left.
	- \tilde t(h^+h^- + h^0h^{0*} 
		+ 2\phi^{++}\phi^{--} + 2\phi^+\phi^- + 2\phi^0\phi^{0*})
			u_3^c \right]
	+ \mathcal{O}(1/f^2) + {\rm h.c.}
\end{eqnarray}

Electroweak symmetry breaking generates additional mass terms 
for the fermions:
\begin{eqnarray}
	\mathcal{L}_f &=& \left[ f \sqrt{\lambda_1^2 + \lambda_2^2}
	- \frac{\lambda_1^2}{\sqrt{\lambda_1^2 + \lambda_2^2}} 
	\frac{v^2}{2f} \right] \tilde t \tilde t^c
	- \frac{i\lambda_1^2 v}{\sqrt{\lambda_1^2 + \lambda_2^2}}
	\left[1 + \frac{v^2}{f^2} \left( -\frac{1}{3}
	+ \frac{fv^{\prime}}{v^2} \right) \right]
	t_3 \tilde t^c
	\nonumber \\ &&
	- \frac{\lambda_1 \lambda_2}{\sqrt{\lambda_1^2 + \lambda_2^2}}
	\frac{v^2}{2f} \tilde t u_3^c
	- \frac{i\lambda_1\lambda_2 v}{\sqrt{\lambda_1^2 + \lambda_2^2}}
	\left[1 + \frac{v^2}{f^2} \left( -\frac{1}{3} 
	+ \frac{fv^{\prime}}{v^2} \right) \right]
	t_3 u_3^c.
\end{eqnarray}
The factor of $i$ in the $t_3u_3^c$ and $t_3 \tilde t^c$ mass terms
can be absorbed into a re-phasing of the left-handed quark doublet field;
instead we keep it explicitly for simplicity.

After diagonalizing these mass terms, we obtain the physical top quark $t$
and a new heavy quark $T$:
\begin{eqnarray}
	t_L = c_L t_3 - s_L \tilde t, &\qquad & 
		t_R^c = c_R u_3^{\prime c} - s_R \tilde t^{\prime c}, 
	\nonumber \\
	T_L = s_L t_3 + c_L \tilde t, &\qquad &
		T_R^c = s_R u_3^{\tilde c} + c_R \tilde t^{\prime c},
	\nonumber
\end{eqnarray}
where
\begin{eqnarray}
	s_R &=& \frac{\lambda_1}{\sqrt{\lambda_1^2+\lambda_2^2}}
	\left[1 - \frac{v^2}{f^2}\frac{\lambda_2^2}{\lambda_1^2+\lambda_2^2}
	\left( \frac{1}{2} + \frac{\lambda_1^2}{\lambda_1^2+\lambda_2^2}
	\right) \right], 
	\nonumber \\
	c_R &=& \frac{\lambda_2}{\sqrt{\lambda_1^2+\lambda_2^2}}
	\left[1 + \frac{v^2}{f^2}\frac{\lambda_1^2}{\lambda_1^2+\lambda_2^2}
	\left( \frac{1}{2} + \frac{\lambda_1^2}{\lambda_1^2+\lambda_2^2}
	\right) \right], 
	\nonumber \\
	s_L &=& -i \frac{\lambda_1^2}{\lambda_1^2+\lambda_2^2} \frac{v}{f}
	\left[ 1 - \frac{v^2}{f^2} 
	\left( \frac{5}{6} - \frac{fv^{\prime}}{v^2} 
	- \frac{1}{2} \frac{\lambda_1^4}{(\lambda_1^2+\lambda_2^2)^2}
	\right) \right],
	\nonumber \\
	c_L &=& 1 - \frac{v^2}{2f^2} 
	\frac{\lambda_1^4}{(\lambda_1^2+\lambda_2^2)^2}.
	\label{eq:Ttmixing}
\end{eqnarray}
The corresponding masses are:
\begin{eqnarray}
	m_t &=& \frac{i\lambda_1\lambda_2}{\sqrt{\lambda_1^2 + \lambda_2^2}}
	v \left\{ 1 + 
	\frac{v^2}{f^2} \left[ -\frac{1}{3} + \frac{fv^{\prime}}{v^2}
	+ \frac{1}{2}\frac{\lambda_1^2}{\lambda_1^2+\lambda_2^2}
	\left(1 - \frac{\lambda_1^2}{\lambda_1^2+\lambda_2^2}
	\right) \right] \right\},
	\nonumber \\
	M_T &=& -f\sqrt{\lambda_1^2 + \lambda_2^2}
	\left[ 1 + \mathcal{O}(v^2/f^2) \right].
\end{eqnarray}

The  scalar interactions with the up-type quarks of the first two generations
take the same form as in Eq.~(\ref{Yukawat}), except that there is no need for
the extra vector-like  quarks $\tilde t, \tilde t^{\prime c}$.
The interactions with the down-type quarks $d^c$ and leptons of the three
generations are generated by a similar Lagrangian, again without  the extra 
quarks, and can be written as
\begin{equation}
{\mathcal{L}}_Y = \frac{1}{2} \lambda_d f \epsilon_{ijk} \epsilon_{xy} \chi_i
\Sigma_{jx}^* \Sigma_{ky}^* d^c + {\rm h.c.}	
\end{equation}
with the isospin index $i=1,2$ only, and similarly for the leptons.

\subsection{Fermion kinetic terms $\mathcal{L}_{F} $}

The fermion gauge interactions take the generic form
	\begin{eqnarray}
	\mathcal{L}_{F} &=& 
	\sum_f \overline\psi_f i \gamma^\mu D_\mu \psi_f,\quad 
	f={\rm flavor\ and\ chirality}\\
	D_\mu &=& \partial_\mu - i \sum_{j=1}^2(g_j W_{j\mu} + g'_j B_{j\mu}),
         \label{Fermion}
	\end{eqnarray}
where $W_j=W_j^a Q^a$ and $B_j=B_jY_j$.

The Lagrangian must be gauge invariant under the gauge groups
$[SU(2)\otimes U(1)]_1\otimes[SU(2)\otimes U(1)]_2$.
In particular, the gauge invariance of the scalar couplings to fermions
discussed in the previous section requires that the 
Standard Model quark and lepton doublets transform as doublets under $SU(2)_1$
and as singlets under $SU(2)_2$.

Because all the Standard Model fermions except the top quark
have small Yukawa couplings, their quadratically
divergent contributions to the Higgs mass do not constitute a hierarchy
problem if the cutoff $\Lambda_S$ is around a few tens of TeV.  Thus, in
the littlest Higgs model one does not have to introduce extra 
vector-like quarks to cancel the divergences due to 
the first two generations of quarks or the 
$b$ quark, or due to any of the leptons.  
Thus, except for the top quark, there will be no mixing between the 
$SU(2)$ doublet fermions and vector fermions.
We first write down the gauge couplings to all fermions except the top
quark; we will later write the top quark and vector-like quark 
gauge couplings, including the mixing.

The $SU(2)_1$ gauge couplings to SM fermions are given by:
\begin{equation}
	\mathcal{L} = g_1 W_{1\mu}^a Q_1^a \bar Q_L \gamma^{\mu} Q_L
	= g_1 \left(W_{1\mu}^+ J^{+\mu} + W_{1\mu}^- J^{-\mu} 
	+ W_{1\mu}^3 J_{W_1^3}^{\mu} \right),
	\label{eq:W1ff}
\end{equation}
where the charged and neutral currents are:
\begin{equation}
	J^{+\mu} = \frac{1}{\sqrt{2}} \bar u_L \gamma^{\mu} d_L,  \qquad
	J^{-\mu} = \frac{1}{\sqrt{2}} \bar d_L \gamma^{\mu} u_L, \qquad
	J_{W_1^3}^{\mu} = \bar Q_L \gamma^{\mu} (-T^3) Q_L = -J_{W^3}^{\mu},
\end{equation}
where $W_1^{\pm} = (W_1^1 \pm i W_1^2)/\sqrt{2}$,
and similarly for the lepton doublet.

\subsubsection{Charged currents }

The couplings of the $W_L^{\pm}$ and $W_H^{\pm}$ gauge bosons
are found by writing $g_1W_1^{\pm}$ in terms of the mass eigenstates:
\begin{equation}
	g_1 W_1 = g W_L \left[ 1 - \frac{v^2}{2f^2} c^2 (c^2-s^2) \right] 
	- g \frac{c}{s} W_H \left[ 1 + \frac{v^2}{2f^2} s^2 (c^2-s^2) \right],
\end{equation}
and inserting this expression into Eq.~(\ref{eq:W1ff}) above.

For the gauge couplings involving the top quark, we must include the 
mixing between $t_3$ and $\tilde t$.
The charged current gets modified as follows:
\begin{equation}
	J^{+\mu} = \frac{1}{\sqrt{2}} \left[
	c_L \bar t_L \gamma^{\mu} b_L 
	+ s_L \bar T_L \gamma^{\mu} b_L \right].
\end{equation}

Because of the mixing of the $SU(2)$ doublet state $t_3$ into the heavier
mass eigenstate $T_L$ as a result of EWSB, 
the CKM matrix involving only the usual three
generations of quarks is no longer unitary; it deviates from unitarity
at order $(v^2/f^2)$.
The modification is as follows:
\begin{eqnarray}
	V_{tb} &=& c_L V_{tb}^{\rm SM}
	= V_{tb}^{\rm SM} \left( 1 - \frac{v^2}{2f^2}
	\frac{\lambda_1^4}{(\lambda_1^2 + \lambda_2^2)^2}
	\right)
	= V_{tb}^{\rm SM} \left(1 - \frac{1}{2}
	\frac{\lambda_1^2}{\lambda_2^2} \frac{m_t^2}{M_T^2} \right), 
	\nonumber \\
	V_{Tb} &=& s_L V_{tb}^{\rm SM}
	= -i V_{tb}^{\rm SM} \frac{\lambda_1^2}{\lambda_1^2 + \lambda_2^2}
	\frac{v}{f} 
	= V_{tb}^{\rm SM} \frac{\lambda_1}{\lambda_2} \frac{m_t}{M_T}.
\end{eqnarray}

\subsubsection{Neutral currents }

The neutral gauge boson couplings to fermions are somewhat more 
complicated, since they depend on both
the isospin and the hypercharge of the fermions.
The quantum numbers of the fermion fields under $U(1)_1 \otimes U(1)_2$
are determined by requiring that the scalar couplings to fermions
are gauge invariant, using
the $U(1)$ quantum number assignments of the $\Sigma$ fields specified by 
$Y_1$ and $Y_2$.  The resulting hypercharges are given in 
Table~\ref{tab:hypercharge} in terms of the free parameters
$y_u$ and $y_e$.
If one further requires that both of the $U(1)$ gauge groups are anomaly
free, then $y_u = -2/5$ and $y_e = 3/5$.

Note that the hypercharge assignments of $u^{\prime c}$ 
and $\tilde t^{\prime c}$ are different, so that
the $t^c_R$ and $T^c_R$ mass eigenstates are mixtures of states
of different hypercharge.
For the first two generations of quarks there is no mixing with an
extra vector-like quark, so the hypercharges of the right-handed
charm and up quarks are equal to those of $u^{\prime c}$.
In particular, the hypercharge of the right-handed top quark is now
different from the hypercharges of the right-handed charm and up quarks
under the two $U(1)$ groups.

The neutral gauge boson couplings to fermions take the form:
\begin{equation}
	\mathcal{L} = -g_1 W_{1\mu}^3 J_{W^3}^{\mu}
	+ g_1^{\prime} B_{1\mu} J_{B_1}^{\mu} 
	+ g_2^{\prime} B_{2\mu} J_{B_2}^{\mu},
\end{equation}
where $J_{B_{1,2}}^{\mu} = \bar f \gamma^{\mu} Y_{1,2} f$, with $Y_{1,2}$ 
given in Table~\ref{tab:hypercharge}.  The couplings of the neutral gauge
boson mass eigenstates $A_L$, $Z_L$, $A_H$ and $Z_H$ are given by:
\begin{eqnarray}
	\mathcal{L} &=& - A_{L\mu} e J^{\mu}_{\rm EM}
	+ A_{H\mu} g^{\prime} 
	\frac{(-c^{\prime 2} J_{B_1}^{\mu} + s^{\prime 2} J_{B_2}^{\mu})}
	{s^{\prime}c^{\prime}}
	+ Z_{H\mu} g 
	\frac{c}{s} J_{W^3}^{\mu}
		\nonumber \\
	&-& Z_{L\mu} \frac{g}{c_\w}
	\left\{ (J_{W^3}^{\mu} - s^2_\w J^{\mu}_{\rm EM})
	- \frac{v^2}{f^2} \left[c_\w x_Z^{W^{\prime}} \frac{c}{s} J^{\mu}_{W^3}
	+ s_\w x_Z^{B^{\prime}} 
	\frac{(-c^{\prime 2} J_{B_1}^{\mu} + s^{\prime 2} J_{B_2}^{\mu})}
	{s^{\prime}c^{\prime}}
	\right] \right\},
\label{zcoupling}
\end{eqnarray}
where the mixing coefficients 
$x_Z^{W^{\prime}}$, $x_Z^{B^{\prime}}$, and $x_H$ are given in Eq.~(\ref{eq:xmixing}).
The electromagnetic current is 
$J^{\mu}_{\rm EM} = J_{W^3}^{\mu} + J_{B_1}^{\mu} + J_{B_2}^{\mu}$; 
note that
the photon coupling to charge $Q = T^3 + Y_1 + Y_2$ is not modified from 
its SM value.  The $Z_L$ boson coupling gets modified from its SM form,
$(g/c_\w)(T^3 - s^2_\w Q)$, by terms of order $v^2/f^2$.
Finally, the $A_H$ and $Z_H$ couplings to fermions are essentially those
of $B^{\prime}$ and $W^{\prime 3}$, respectively, up to terms of
order $v^2/f^2$ that we have neglected here.

The mixing between fermions with different $SU(2)\otimes U(1)$ quantum numbers 
(i.e., $t_3$ and $\tilde t$) will lead to flavor changing neutral currents
mediated by the $Z_L$ boson.
The flavor-preserving gauge couplings will also be anomalous at
order $v^2/f^2$ because of the mixing.

\subsection{Gauge kinetic terms $\mathcal{L}_{G} $}
The gauge kinetic terms take the standard form:
	\begin{equation}
	\mathcal{L}_{G} = 
	-{1\over 4} \sum_{j=1}^2(W_{ja}^{\mu\nu} W_{j\mu\nu}^a + 
	B_{ja}^{\mu\nu} B_{j\mu\nu}^a)  .
         \label{Gauge}
	\end{equation}
These terms yield 3- and 4-particle interactions among the $SU(2)$ 
gauge bosons.
Of course, the $U(1)$ gauge bosons have no self-couplings or couplings to the
$SU(2)$ 
gauge bosons. The explicit couplings are listed in Appendix \ref{app_b}.


\newpage

\section{Feynman rules: interaction vertices}\label{app_b}

For the convenience of further phenomenological exploration, we 
list the Feynman rules of the interaction vertices in unitary gauge
among the new scalar sector, the new gauge bosons, the new vector-like fermion 
and the SM particles. All particles are the mass eigenstates.
In the Feynman rules, all particles are assumed to be outgoing, and
we adopt the convention Feynman rule $= i \mathcal{L}$.

\subsection{Couplings between gauge bosons and scalars}

\subsubsection{Three-point vertices in Tables  \ref{gauge-scalar3-1} and
 \ref{gauge-scalar3-2} }

\begin{table}[h]
\begin{tabular}{|c|c||c|c|}
\hline
particles & vertices & particles & vertices \\
\hline
$W^+_{L\mu} W^-_{L\nu} H$ & 
	$\frac{i}{2} g^2 v g_{\mu\nu} \left( 1 
	- \frac{v^2}{3f^2}+ \right.$  
	& $W^+_{H\mu} W^-_{H\nu} H$ & $-\frac{i}{2} g^2 v g_{\mu\nu} $ \\
  & $\left. \frac{1}{2} (c^2-s^2)^2 \frac{v^2}{f^2}
 - \frac{1}{2} s_0^2 - 2 \sqrt{2} s_0 \frac{v^{\prime}}{v}\right) $ &  &  \\
\hline
$Z_{L\mu} Z_{L\nu} H$ &
	$\frac{i}{2} \frac{g^2}{c^2_\w} v g_{\mu\nu}
	\left( 1 - \frac{v^2}{3f^2} 
	- \frac{1}{2} s_0^2 + 4 \sqrt{2} s_0 \frac{v^{\prime}}{v}\right. $
	& $Z_{H\mu} Z_{H\nu} H$ & $-\frac{i}{2} g^2 v g_{\mu\nu}$ \\
  & $\left.  -\frac{1}{2} \left( (c^2-s^2)^2 
	+ 5 (c^{\prime 2}-s^{\prime 2})^2 \right) \frac{v^2}{f^2}\right)$
	& 
$A_{H\mu} A_{H\nu} H$ & 
	$-\frac{i}{2} g^{\prime 2} v g_{\mu\nu}$ \\
\hline
$W_{L\mu}^+ W_{H\nu}^- H$ &
	$-\frac{i}{2} g^2 \frac{(c^2-s^2)}{2sc} v g_{\mu\nu}$ &
$Z_{L\mu} Z_{H\nu} H$ &
	$-\frac{i}{2} \frac{g^2}{c_\w} \frac{(c^2-s^2)}{2sc}
		v g_{\mu\nu}$ \\
\hline
$Z_{L\mu} A_{H\nu} H$ &
	$-\frac{i}{2} \frac{gg^{\prime}}{c_\w} 
	\frac{(c^{\prime 2}-s^{\prime 2})}{2s^{\prime}c^{\prime}}
		v g_{\mu\nu}$ &
$Z_{H\mu} A_{H\nu} H$ &
	$-\frac{i}{4} g g^{\prime}
	\frac{(c^2s^{\prime 2} + s^2c^{\prime 2})}{scs^{\prime}c^{\prime}}
		v g_{\mu\nu}$  \\
\hline 
$W_{L\mu}^+ W_{L\nu}^- \Phi^0$ &
	$-\frac{i}{2} g^2 ( s_0 v - 2 \sqrt{2} v^{\prime} ) g_{\mu\nu}$ &
$W_{H\mu}^+ W_{H\nu}^- \Phi^0$ &
	$\frac{i}{2} g^2 ( s_0 v - 2 \sqrt{2} v^{\prime} ) g_{\mu\nu}$ \\
\hline
$W_{L\mu}^+ W_{H\nu}^- \Phi^0$ &
	$\frac{i}{2} g^2 \frac{(c^2-s^2)}{2sc} 
	( s_0 v - 2 \sqrt{2} v^{\prime} ) g_{\mu\nu}$ &
$$ &
	$$ \\
\hline
$Z_{L\mu} Z_{L\nu} \Phi^0$ &
	$-\frac{i}{2} \frac{g^2}{c^2_\w} ( v s_0 - 4 \sqrt{2} v^{\prime} )
	g_{\mu\nu}$ &
$Z_{H\mu} Z_{H\nu} \Phi^0$ &
	$\frac{i}{2} g^2 \left( v s_0 
	+ \frac{(c^2-s^2)^2}{s^2c^2} \sqrt{2} v^{\prime} \right) g_{\mu\nu}$ \\
\hline
$Z_{L\mu} Z_{H\nu} \Phi^0$ &
	$\frac{i}{2} \frac{g^2}{c_\w} \frac{(c^2-s^2)}{2sc}
	( v s_0 - 4 \sqrt{2} v^{\prime} ) g_{\mu\nu}$ &
$Z_{L\mu} A_{H\nu} \Phi^0$ &
	$\frac{i}{2} \frac{gg^{\prime}}{c_\w} 
	\frac{(c^{\prime 2}-s^{\prime 2})}{2s^{\prime}c^{\prime}}
	( v s_0 - 4 \sqrt{2} v^{\prime} ) g_{\mu\nu}$ \\
\hline
$A_{H\mu} Z_{H\nu} \Phi^0$ &
	$\frac{i}{4} g g^{\prime} \frac{1}{scs^{\prime}c^{\prime}} \left(
	(c^2s^{\prime 2} + s^2c^{\prime 2}) v s_0 \right.$ &
$A_{H\mu} A_{H\nu} \Phi^0$ &
	$\frac{i}{2} g^{\prime 2} \left( v s_0 
	+ \frac{(c^{\prime 2}-s^{\prime 2})^2}{s^{\prime 2}c^{\prime 2}}
	\sqrt{2} v^{\prime} \right) g_{\mu\nu}$ \\
$$ &
	$\left. + 2 \sqrt{2} (c^2-s^2)(c^{\prime 2}-s^{\prime 2}) v^{\prime} 
	\right) g_{\mu\nu}$ &
$$ &
	$$ \\
\hline
$W_{L\mu}^+ A_{L\nu} \Phi^-$ &
	$0$ &
$W_{H\mu}^+ A_{L\nu} \Phi^-$ &
	$0$ \\
\hline
$W_{L\mu}^+ Z_{L\nu} \Phi^-$ &
	$-i \frac{g^2}{c_\w} v^{\prime} g_{\mu\nu}$ &
$W_{H\mu}^+ Z_{L\nu} \Phi^-$ &
	$ i \frac{g^2}{c_\w} \frac{(c^2-s^2)}{2sc} v^{\prime} g_{\mu\nu}$ \\
\hline
$W_{L\mu}^+ A_{H\nu} \Phi^-$ &
	$-\frac{i}{2} g g^{\prime}
	\frac{(c^{\prime 2}-s^{\prime 2})}{2s^{\prime}c^{\prime}}
	( v s_+ - 4 v^{\prime} ) g_{\mu\nu}$ &
$W_{H\mu}^+ A_{H\nu} \Phi^-$ &
	$-\frac{i}{2} g g^{\prime}
	\frac{(c^2c^{\prime 2}+s^2s^{\prime 2})}{scs^{\prime}c^{\prime}}
	v^{\prime} g_{\mu\nu}$ \\
\hline
$W_{L\mu}^+ Z_{H\nu} \Phi^-$ &
	$ i g^2 \frac{(c^2-s^2)}{2sc} v^{\prime} g_{\mu\nu}$ &
$W_{H\mu}^+ Z_{H\nu} \Phi^-$ &
	$-i g^2 \frac{(c^4+s^4)}{2s^2c^2} v^{\prime} g_{\mu\nu}$ \\
\hline
$W_{L\mu}^+ W_{L\nu}^+ \Phi^{--}$ &
	$2 i g^2 v^{\prime} g_{\mu\nu}$ &
$W_{H\mu}^+ W_{H\nu}^+ \Phi^{--}$ &
	$2 i g^2 \frac{(c^4+s^4)}{2s^2c^2} v^{\prime} g_{\mu\nu}$ \\
\hline
$W_{L\mu}^+ W_{H\nu}^+ \Phi^{--}$ &
	$-2 i g^2 \frac{(c^2-s^2)}{2sc} v^{\prime} g_{\mu\nu}$ &
$$ &
	$$ \\
\hline
\end{tabular}
\caption{Three-point couplings of two gauge bosons to one scalar. 
All particles are the mass eigenstates.
\label{gauge-scalar3-1}
}
\end{table}

\begin{table}[h]
\begin{tabular}{|c|c||c|c|}
\hline
particles & vertices & particles & vertices \\
\hline
$W_{L\mu}^+ H \Phi^-$ &
	$-\frac{ig}{2} \left( \sqrt{2} s_0 - s_+ \right) (p_1-p_2)_{\mu}$ &
$W_{H\mu}^+ H \Phi^-$ &
	$\frac{ig}{2} \frac{(c^2-s^2)}{2sc}
	\left( \sqrt{2} s_0 - s_+ \right) (p_1-p_2)_{\mu}$ \\
\hline
$W_{L\mu}^+ \Phi^0 \Phi^-$ &
	$-\frac{ig}{\sqrt{2}} (p_1-p_2)_{\mu}$ &
$W_{H\mu}^+ \Phi^0 \Phi^-$ &
	$\frac{ig}{\sqrt{2}} \frac{(c^2-s^2)}{2sc} (p_1-p_2)_{\mu}$ \\
\hline
$W_{L\mu}^+ \Phi^P \Phi^-$ &
	$\frac{g}{\sqrt{2}} (p_1-p_2)_{\mu}$ &
$W_{H\mu}^+ \Phi^P \Phi^-$ &
	$-\frac{g}{\sqrt{2}} \frac{(c^2-s^2)}{2sc} (p_1-p_2)_{\mu}$ \\
\hline
$W_{L\mu}^+ \Phi^+ \Phi^{--}$ &
	$- i g (p_1-p_2)_{\mu}$ &
$W_{H\mu}^+ \Phi^+ \Phi^{--}$ &
	$i g \frac{(c^2-s^2)}{2sc} (p_1-p_2)_{\mu}$ \\
\hline
$A_{L\mu} H \Phi^P$ & $0$ &
$A_{H\mu} H \Phi^P$ &
	$-\frac{1}{2} 
	g^{\prime} \frac{(c^{\prime 2}-s^{\prime 2})}{2s^{\prime}c^{\prime}}
	(s_P - 2 s_0) (p_1-p_2)_{\mu}$ \\
\hline
$A_{L\mu} \Phi^0 \Phi^P$ & $0$ &
$A_{H\mu} \Phi^0 \Phi^P$ &
	$g^{\prime} \frac{(c^{\prime 2}-s^{\prime 2})}{2s^{\prime}c^{\prime}}
	(p_1-p_2)_{\mu}$ \\
\hline
$A_{L\mu} \Phi^+ \Phi^-$ &
	$-i e (p_1 - p_2)_{\mu}$ &
$A_{H\mu} \Phi^+ \Phi^-$ &
	$i 
	g^{\prime} \frac{(c^{\prime 2}-s^{\prime 2})}{2s^{\prime}c^{\prime}}
	(p_1-p_2)_{\mu}$ \\
\hline
$A_{L\mu} \Phi^{++} \Phi^{--}$ &
	$-2 i e (p_1-p_2)_{\mu}$ &
$A_{H\mu} \Phi^{++} \Phi^{--}$ &
	$i 
	g^{\prime} \frac{(c^{\prime 2}-s^{\prime 2})}{2s^{\prime}c^{\prime}}
	(p_1-p_2)_{\mu}$ \\
\hline
$Z_{L\mu} H \Phi^P$ &
	$\frac{1}{2} \frac{g}{c_\w} \left( s_P - 2 s_0 \right)
	(p_1-p_2)_{\mu}$ &
$Z_{H\mu} H \Phi^P$ &
	$-\frac{1}{2} g \frac{(c^2-s^2)}{2sc}
	(s_P - 2 s_0) (p_1-p_2)_{\mu}$ \\
\hline
$Z_{L\mu} \Phi^0 \Phi^P$ &
	$- \frac{g}{c_\w} (p_1-p_2)_{\mu}$ &
$Z_{H\mu} \Phi^0 \Phi^P$ &
	$g \frac{(c^2-s^2)}{2sc} (p_1-p_2)_{\mu}$ \\
\hline
$Z_{L\mu} \Phi^+ \Phi^-$ &
	$i \frac{g}{c_\w} s^2_\w (p_1-p_2)_{\mu}$ &
$Z_{H\mu} \Phi^+ \Phi^-$ &
	$\mathcal{O}(v^2/f^2)$ \\
\hline
$Z_{L\mu} \Phi^{++} \Phi^{--}$ &
	$-i \frac{g}{c_\w} \left( 1 - 2 s^2_\w \right) (p_1-p_2)_{\mu}$ &
$Z_{H\mu} \Phi^{++} \Phi^{--}$ &
	$i g \frac{(c^2-s^2)}{2sc} (p_1-p_2)_{\mu}$ \\
\hline
\end{tabular}
\caption{Three-point couplings of one gauge boson to two scalars. 
The momenta are assigned according to $V_{\mu} S_1(p_1) S_2(p_2)$.
All particles are the mass eigenstates and all momenta are out-going.
\label{gauge-scalar3-2}
}
\end{table}


\vspace*{4in}

\newpage

\subsubsection{Four-point vertices in Tables \ref{GH4-1} and \ref{GH4-2}. }

\begin{table}[h!]
\begin{tabular}{|c|c||c|c|}
\hline
particles & vertices & particles & vertices \\
\hline
$W_{L\mu}^+ W_{L\nu}^- H H$ &
	$\frac{i}{2} g^2 g_{\mu\nu}
	+ \mathcal{O}(v^2/f^2)$ &
$W_{H\mu}^+ W_{H\nu}^- H H$ &
	$-\frac{i}{2} g^2 g_{\mu\nu}$ \\
\hline
$Z_{L\mu} Z_{L\nu} H H$ &
	$\frac{i}{2} \frac{g^2}{c^2_\w} g_{\mu\nu}
	+ \mathcal{O}(v^2/f^2)$ &
$Z_{H\mu} Z_{H\nu} H H$ &
	$- \frac{i}{2} g^2 g_{\mu\nu}$ \\
$$ &
	$$ &
$A_{H\mu} A_{H\nu} H H$ &
	$- \frac{i}{2} g^{\prime 2} g_{\mu\nu}$ \\
\hline
$W_{L\mu}^+ W_{H\nu}^- H H$ &
	$-\frac{i}{2} g^2 \frac{(c^2-s^2)}{2sc} g_{\mu\nu}$ &
$Z_{L\mu} Z_{H\nu} H H$ &
	$-\frac{i}{2} \frac{g^2}{c_\w} \frac{(c^2-s^2)}{2sc}
		g_{\mu\nu}$ \\
\hline
$Z_{L\mu} A_{H\nu} H H$ &
	$-\frac{i}{2} \frac{gg^{\prime}}{c_\w} 
	\frac{(c^{\prime 2}-s^{\prime 2})}{2s^{\prime}c^{\prime}}
		g_{\mu\nu}$ &
$Z_{H\mu} A_{H\nu} H H$ &
	$-\frac{i}{4} g g^{\prime}
	\frac{(c^2s^{\prime 2} + s^2c^{\prime 2})}{scs^{\prime}c^{\prime}}
		g_{\mu\nu}$ \\
\hline
$W_{L\mu}^+ W_{L\nu}^- H \Phi^0$ &
	$\frac{i}{2} g^2 s_0 g_{\mu\nu}$ &
$W_{H\mu}^+ W_{H\nu}^- H \Phi^0$ &
	$-\frac{i}{2} g^2 s_0 g_{\mu\nu}$ \\
\hline
$Z_{L\mu} Z_{L\nu} H \Phi^0$ &
	$\frac{3i}{2} \frac{g^2}{c^2_\w} s_0 g_{\mu\nu}$ &
$Z_{H\mu} Z_{H\nu} H \Phi^0$ &
	$\frac{i}{2} g^2 \left[ 1 + \frac{(c^2-s^2)^2}{s^2c^2} \right]
	s_0 g_{\mu\nu}$ \\
$$ &
	$$ &
$A_{H\mu} A_{H\nu} H \Phi^0$ &
	$\frac{i}{2} g^{\prime 2}
	\left[ 1 + 
	\frac{(c^{\prime 2}-s^{\prime 2})^2}{s^{\prime 2}c^{\prime 2}}
	\right] s_0 g_{\mu\nu}$ \\
\hline
$W_{L\mu}^+ W_{H\nu}^- H \Phi^0$ &
	$-\frac{i}{2} g^2 \frac{(c^2-s^2)}{2sc} s_0 g_{\mu\nu}$ &
$Z_{L\mu} Z_{H\nu} H \Phi^0$ &
	$-\frac{3i}{2} \frac{g^2}{c_\w} \frac{(c^2-s^2)}{2sc}
		s_0 g_{\mu\nu}$ \\
\hline
$Z_{L\mu} A_{H\nu} H \Phi^0$ &
	$-\frac{3i}{2} \frac{gg^{\prime}}{c_\w} 
	\frac{(c^{\prime 2}-s^{\prime 2})}{2s^{\prime}c^{\prime}}
		s_0 g_{\mu\nu}$ &
$Z_{H\mu} A_{H\nu} H \Phi^0$ &
	$\frac{i}{4} g g^{\prime} \frac{1}{scs^{\prime}c^{\prime}} \left[
	(c^2s^{\prime 2} + s^2c^{\prime 2}) \right.$ \\
$$ &
	$$ &
$$ &
	$\left. + 2 (c^2-s^2)(c^{\prime 2}-s^{\prime 2}) \right] 
		s_0 g_{\mu\nu}$ \\
\hline
$W_{L\mu}^+ A_{L\nu} H \Phi^-$ &
	$-\frac{i}{2} e g (s_+ - \sqrt{2} s_0) g_{\mu\nu}$ &
$W_{H\mu}^+ A_{L\nu} H \Phi^-$ &
	$\frac{i}{2} e g \frac{(c^2-s^2)}{2sc} 
	(s_+ - \sqrt{2} s_0) g_{\mu\nu}$ \\
\hline
$W_{L\mu}^+ Z_{L\nu} H \Phi^-$ &
	$\frac{i}{2} \frac{g^2}{c_\w}
	\left[ s_+ s^2_\w 
	\right. $ &
$W_{H\mu}^+ Z_{L\nu} H \Phi^-$ &
	$-\frac{i}{2} \frac{g^2}{c_\w} \frac{(c^2-s^2)}{2sc}
	\left[ s_+ s^2_\w
	\right. $ \\
$$ &
	$\left. - \sqrt{2} s_0 ( 1 + s^2_\w ) \right]
	g_{\mu\nu}$ &
$$ &
	$\left. - \sqrt{2} s_0 ( 1 + s^2_\w ) \right]
	g_{\mu\nu}$ \\
\hline
$W_{L\mu}^+ A_{H\nu} H \Phi^-$ &
	$-\frac{i}{2} g g^{\prime} 
	\frac{(c^{\prime 2}-s^{\prime 2})}{2s^{\prime}c^{\prime}}
	( s_+ - 2 \sqrt{2} s_0 )
	g_{\mu\nu}$ &
$W_{H\mu}^+ A_{H\nu} H \Phi^-$ &
	$-\frac{i}{4} g g^{\prime}
	\frac{1}{scs^{\prime}c^{\prime}}
	\left[ (c^2s^{\prime 2} + s^2c^{\prime 2})s_+ \right.$ \\
$$ &
	$$ &
$$ &
	$\left. + \sqrt{2}(c^2-s^2)(c^{\prime 2}-s^{\prime 2})s_0 \right]
		g_{\mu\nu}$ \\
\hline
$W_{L\mu}^+ Z_{H\nu} H \Phi^-$ &
	$\frac{i}{2} g^2 \frac{(c^2-s^2)}{2sc} s_0
	g_{\mu\nu}$ &
$W_{H\mu}^+ Z_{H\nu} H \Phi^-$ &
	$-\frac{i}{2} g^2 \frac{(c^4+s^4)}{2s^2c^2} s_0
	g_{\mu\nu}$ \\
\hline
$W_{L\mu}^+ W_{L\nu}^+ H \Phi^{--}$ &
	$\sqrt{2} i g^2 s_0 g_{\mu\nu}$ &
$W_{H\mu}^+ W_{H\nu}^+ H \Phi^{--}$ &
	$\sqrt{2} i g^2 \frac{(c^4+s^4)}{2s^2c^2} s_0 g_{\mu\nu}$ \\
\hline
$W_{L\mu}^+ W_{H\nu}^+ H \Phi^{--}$ &
	$-\sqrt{2} i g^2 \frac{(c^2-s^2)}{2sc} s_0 g_{\mu\nu}$ &
$$ &
	$$ \\
\hline
$W_{L\mu}^+ W_{L\nu}^- \Phi^0 \Phi^0$ &
	$i g^2 g_{\mu\nu}$ &
$W_{H\mu}^+ W_{H\nu}^- \Phi^0 \Phi^0$ &
	$-i g^2 g_{\mu\nu}$ \\
\hline
$Z_{L\mu} Z_{L\nu} \Phi^0 \Phi^0$ &
	$2 i \frac{g^2}{c^2_\w} g_{\mu\nu}$ &
$Z_{H\mu} Z_{H\nu} \Phi^0 \Phi^0$ &
	$2 i g^2 \frac{(c^2-s^2)^2}{4s^2c^2} g_{\mu\nu}$ \\
$$ &
	$$ &
$A_{H\mu} A_{H\nu} \Phi^0 \Phi^0$ &
	$2 i g^{\prime 2}
	\frac{(c^{\prime 2}-s^{\prime 2})^2}{4s^{\prime 2}c^{\prime 2}}
		g_{\mu\nu}$ \\
\hline
$W_{L\mu}^+ W_{H\nu}^- \Phi^0 \Phi^0$ &
	$-i g^2 \frac{(c^2-s^2)}{2sc} g_{\mu\nu}$ &
$Z_{L\mu} Z_{H\nu} \Phi^0 \Phi^0$ &
	$-2 i \frac{g^2}{c_\w} \frac{(c^2-s^2)}{2sc}
		g_{\mu\nu}$ \\
\hline
$Z_{L\mu} A_{H\nu} \Phi^0 \Phi^0$ &
	$-2 i \frac{gg^{\prime}}{c_\w} 
	\frac{(c^{\prime 2}-s^{\prime 2})}{2s^{\prime}c^{\prime}}
		g_{\mu\nu}$ &
$Z_{H\mu} A_{H\nu} \Phi^0 \Phi^0$ &
	$2 i g g^{\prime}
	\frac{(c^2-s^2)(c^{\prime 2}-s^{\prime 2})}{4scs^{\prime}c^{\prime}}
		g_{\mu\nu}$ \\
\hline
$W_{L\mu}^+ A_{L\nu} \Phi^0 \Phi^-$ &
	$-\frac{i}{\sqrt{2}} e g g_{\mu\nu}$ &
$W_{H\mu}^+ A_{L\nu} \Phi^0 \Phi^-$ &
	$\frac{i}{\sqrt{2}} e g \frac{(c^2-s^2)}{2sc} g_{\mu\nu}$ \\
\hline
$W_{L\mu}^+ Z_{L\nu} \Phi^0 \Phi^-$ &
	$-\frac{i}{\sqrt{2}} \frac{g^2}{c_\w}
	(1 + s^2_\w) g_{\mu\nu}$ &
$W_{H\mu}^+ Z_{L\nu} \Phi^0 \Phi^-$ &
	$\frac{i}{\sqrt{2}} \frac{g^2}{c_\w} \frac{(c^2-s^2)}{2sc}
	(1 + s^2_\w) g_{\mu\nu}$ \\
\hline
$W_{L\mu}^+ A_{H\nu} \Phi^0 \Phi^-$ &
	$\sqrt{2} i g g^{\prime}
	\frac{(c^{\prime 2}-s^{\prime 2})}{2s^{\prime}c^{\prime}}
	g_{\mu\nu}$ &
$W_{H\mu}^+ A_{H\nu} \Phi^0 \Phi^-$ &
	$-\frac{i}{2\sqrt{2}} g g^{\prime}
	\frac{(c^2-s^2)(c^{\prime 2}-s^{\prime 2})}{scs^{\prime}c^{\prime}}
	g_{\mu\nu}$ \\
\hline
$W_{L\mu}^+ Z_{H\nu} \Phi^0 \Phi^-$ &
	$\frac{i}{\sqrt{2}} g^2 \frac{(c^2-s^2)}{2sc} g_{\mu\nu}$ &
$W_{H\mu}^+ Z_{H\nu} \Phi^0 \Phi^-$ &
	$-\frac{i}{\sqrt{2}} g^2 \frac{(c^4+s^4)}{2s^2c^2} g_{\mu\nu}$ \\
\hline
$W_{L\mu}^+ W_{L\nu}^+ \Phi^0 \Phi^{--}$ &
	$\sqrt{2} i g^2 g_{\mu\nu}$ &
$W_{H\mu}^+ W_{H\nu}^+ \Phi^0 \Phi^{--}$ &
	$\sqrt{2} i g^2 \frac{(c^4+s^4)}{2s^2c^2} g_{\mu\nu}$ \\
\hline
$W_{L\mu}^+ W_{H\nu}^+ \Phi^0 \Phi^{--}$ &
	$-\sqrt{2} i g^2 \frac{(c^2-s^2)}{2sc} g_{\mu\nu}$ &
$$ &
	$$ \\
\hline
\end{tabular}
\caption{Four-point gauge boson-scalar couplings. 
\label{GH4-1}
}
\end{table}


%
\begin{table}[h]
\begin{tabular}{|c|c||c|c|}
\hline
particles & vertices & particles & vertices \\
\hline
$W_{L\mu}^+ W_{L\nu}^- \Phi^P \Phi^P$ &
	$i g^2 g_{\mu\nu}$ &
$W_{H\mu}^+ W_{H\nu}^- \Phi^P \Phi^P$ &
	$-i g^2 g_{\mu\nu}$ \\
\hline
$Z_{L\mu} Z_{L\nu} \Phi^P \Phi^P$ &
	$2 i \frac{g^2}{c^2_\w} g_{\mu\nu}$ &
$Z_{H\mu} Z_{H\nu} \Phi^P \Phi^P$ &
	$2 i g^2 \frac{(c^2-s^2)^2}{4s^2c^2} g_{\mu\nu}$ \\
$$ &
	$$ &
$A_{H\mu} A_{H\nu} \Phi^P \Phi^P$ &
	$2 i g^{\prime 2}
	\frac{(c^{\prime 2}-s^{\prime 2})^2}{4s^{\prime 2}c^{\prime 2}}
		g_{\mu\nu}$ \\
\hline
$W_{L\mu}^+ W_{H\nu}^- \Phi^P \Phi^P$ &
	$-i g^2 \frac{(c^2-s^2)}{2sc} g_{\mu\nu}$ &
$Z_{L\mu} Z_{H\nu} \Phi^P \Phi^P$ &
	$-2 i \frac{g^2}{c_\w} \frac{(c^2-s^2)}{2sc}
		g_{\mu\nu}$ \\
\hline
$Z_{L\mu} A_{H\nu} \Phi^P \Phi^P$ &
	$-2 i \frac{gg^{\prime}}{c_\w} 
	\frac{(c^{\prime 2}-s^{\prime 2})}{2s^{\prime}c^{\prime}}
		g_{\mu\nu}$ &
$Z_{H\mu} A_{H\nu} \Phi^P \Phi^P$ &
	$2 i g g^{\prime}
	\frac{(c^2-s^2)(c^{\prime 2}-s^{\prime 2})}{4scs^{\prime}c^{\prime}}
		g_{\mu\nu}$ \\
\hline
$W_{L\mu}^+ A_{L\nu} \Phi^P \Phi^-$ &
	$\frac{1}{\sqrt{2}} e g g_{\mu\nu}$ &
$W_{H\mu}^+ A_{L\nu} \Phi^P \Phi^-$ &
	$\frac{1}{\sqrt{2}} e g \frac{(c^2-s^2)}{2sc} g_{\mu\nu}$ \\
\hline
$W_{L\mu}^+ Z_{L\nu} \Phi^P \Phi^-$ &
	$\frac{1}{\sqrt{2}} \frac{g^2}{c_\w}
	(1 + s^2_\w) g_{\mu\nu}$ &
$W_{H\mu}^+ Z_{L\nu} \Phi^P \Phi^-$ &
	$-\frac{1}{\sqrt{2}} \frac{g^2}{c_\w} \frac{(c^2-s^2)}{2sc}
	(1 + s^2_\w) g_{\mu\nu}$ \\
\hline
$W_{L\mu}^+ A_{H\nu} \Phi^P \Phi^-$ &
	$-\sqrt{2} g g^{\prime} 
	\frac{(c^{\prime 2}-s^{\prime 2})}{2s^{\prime}c^{\prime}}
	g_{\mu\nu}$ &
$W_{H\mu}^+ A_{H\nu} \Phi^P \Phi^-$ &
	$\frac{1}{2\sqrt{2}} g g^{\prime}
	\frac{(c^2-s^2)(c^{\prime 2}-s^{\prime 2})}{scs^{\prime}c^{\prime}}
	g_{\mu\nu}$ \\
\hline
$W_{L\mu}^+ Z_{H\nu} \Phi^P \Phi^-$ &
	$-\frac{1}{\sqrt{2}} g^2\frac{(c^2-s^2)}{2sc} g_{\mu\nu}$ &
$W_{H\mu}^+ Z_{H\nu} \Phi^P \Phi^-$ &
	$\frac{1}{\sqrt{2}} g^2\frac{(c^4+s^4)}{2s^2c^2} g_{\mu\nu}$ \\
\hline
$W_{L\mu}^+ W_{L\nu}^+ \Phi^P \Phi^{--}$ &
	$-\sqrt{2} g^2 g_{\mu\nu}$ &
$W_{H\mu}^+ W_{H\nu}^+ \Phi^P \Phi^{--}$ &
	$-\sqrt{2} g^2 \frac{(c^4+s^4)}{2s^2c^2} g_{\mu\nu}$ \\
\hline
$W_{L\mu}^+ W_{H\nu}^+ \Phi^P \Phi^{--}$ &
	$\sqrt{2} g^2 \frac{(c^2-s^2)}{2sc} g_{\mu\nu}$ &
$$ &
	$$ \\
\hline
$W_{L\mu}^+ W_{L\nu}^- \Phi^{+} \Phi^{-}$ &
	$2 i g^2 g_{\mu\nu}$ &
$W_{H\mu}^+ W_{H\nu}^- \Phi^{+} \Phi^{-}$ &
	$2 i g^2 \frac{(c^2-s^2)^2}{4s^2c^2} g_{\mu\nu}$ \\
\hline
$Z_{L\mu} Z_{L\nu} \Phi^+ \Phi^-$ &
	$2 i \frac{g^2}{c^2_\w} s^4_\w g_{\mu\nu}$ &
$Z_{H\mu} Z_{H\nu} \Phi^+ \Phi^-$ &
	$- 2 i g^2 \frac{1}{4s^2c^2} g_{\mu\nu}$ \\
$A_{L\mu} A_{L\nu} \Phi^+ \Phi^-$ &
	$2 i e^2 g_{\mu\nu}$ &
$A_{H\mu} A_{H\nu} \Phi^+ \Phi^-$ &
	$2 i g^{\prime 2}
	\frac{(c^{\prime 2}-s^{\prime 2})^2}{4s^{\prime 2}c^{\prime 2}}
		g_{\mu\nu}$ \\
\hline
$A_{L\mu} Z_{L\nu} \Phi^+ \Phi^-$ &
	$- 2 i e \frac{g}{c_\w} s^2_\w g_{\mu\nu}$ &
$A_{H\mu} Z_{H\nu} \Phi^+ \Phi^-$ &
	$\mathcal{O}(v^2/f^2)$ \\
\hline
$W_{L\mu}^+ W_{H\nu}^- \Phi^{+} \Phi^{-}$ &
	$-2 i g^2 \frac{(c^2-s^2)}{2sc} g_{\mu\nu}$ &
$A_{L\mu} A_{H\nu} \Phi^+ \Phi^-$ &
	$-2 i e g^{\prime} 
	\frac{(c^{\prime 2}-s^{\prime 2})}{2s^{\prime}c^{\prime}} 
	g_{\mu\nu}$ \\
$$ &
	$$ &
$Z_{L\mu} Z_{H\nu} \Phi^+ \Phi^-$ &
	$\mathcal{O}(v^2/f^2)$ \\
\hline
$A_{L\mu} Z_{H\nu} \Phi^+ \Phi^-$ &
	$\mathcal{O}(v^2/f^2)$ &
$Z_{L\mu} A_{H\nu} \Phi^+ \Phi^-$ &
	$2 i \frac{gg^{\prime}}{c_\w} 
	\frac{(c^{\prime 2}-s^{\prime 2})}{2s^{\prime}c^{\prime}}
		s^2_\w g_{\mu\nu}$ \\
\hline
$W_{L\mu}^+ A_{L\nu} \Phi^+ \Phi^{--}$ &
	$3 i e g g_{\mu\nu}$ &
$W_{H\mu}^+ A_{L\nu} \Phi^+ \Phi^{--}$ &
	$-3 i e g \frac{(c^2-s^2)}{2sc} g_{\mu\nu}$ \\
\hline
$W_{L\mu}^+ Z_{L\nu} \Phi^+ \Phi^{--}$ &
	$i \frac{g^2}{c_\w}
	(1 - 3 s^2_\w) g_{\mu\nu}$ &
$W_{H\mu}^+ Z_{L\nu} \Phi^+ \Phi^{--}$ &
	$-i \frac{g^2}{c_\w} \frac{(c^2-s^2)}{2sc}
	(1 - 3 s^2_\w) g_{\mu\nu}$ \\
\hline
$W_{L\mu}^+ A_{H\nu} \Phi^+ \Phi^{--}$ &
	$2 i g g^{\prime}
	\frac{(c^{\prime 2}-s^{\prime 2})}{2s^{\prime}c^{\prime}}
	g_{\mu\nu}$ &
$W_{H\mu}^+ A_{H\nu} \Phi^+ \Phi^{--}$ &
	$- \frac{i}{2} g g^{\prime}
	\frac{(c^2-s^2)(c^{\prime 2}-s^{\prime 2})}{scs^{\prime}c^{\prime}}
	g_{\mu\nu}$ \\
\hline
$W_{L\mu}^+ Z_{H\nu} \Phi^+ \Phi^{--}$ &
	$-\frac{1}{\sqrt{2}} g^2\frac{(c^2-s^2)}{2sc} g_{\mu\nu}$ &
$W_{H\mu}^+ Z_{H\nu} \Phi^+ \Phi^{--}$ &
	$\frac{1}{\sqrt{2}} g^2\frac{(c^4+s^4)}{2s^2c^2} g_{\mu\nu}$ \\
\hline
$W_{L\mu}^+ W_{L\nu}^- \Phi^{++} \Phi^{--}$ &
	$i g^2 g_{\mu\nu}$ &
$W_{H\mu}^+ W_{H\nu}^- \Phi^{++} \Phi^{--}$ &
	$-i g^2 g_{\mu\nu}$ \\
\hline
$Z_{L\mu} Z_{L\nu} \Phi^{++} \Phi^{--}$ &
	$2 i \frac{g^2}{c^2_\w} (1 - 2 s^2_\w)^2 g_{\mu\nu}$ &
$Z_{H\mu} Z_{H\nu} \Phi^{++} \Phi^{--}$ &
	$2 i g^2 \frac{(c^2-s^2)^2}{4s^2c^2} g_{\mu\nu}$ \\
$A_{L\mu} A_{L\nu} \Phi^{++} \Phi^{--}$ &
	$8 i e^2 g_{\mu\nu}$ &
$A_{H\mu} A_{H\nu} \Phi^{++} \Phi^{--}$ &
	$2 i g^{\prime 2}
	\frac{(c^{\prime 2}-s^{\prime 2})^2}{4s^{\prime 2}c^{\prime 2}}
		g_{\mu\nu}$ \\
\hline
$A_{L\mu} Z_{L\nu} \Phi^{++} \Phi^{--}$ &
	$4 i e \frac{g}{c_\w} (1 - 2 s^2_\w) g_{\mu\nu}$ &
$A_{H\mu} Z_{H\nu} \Phi^{++} \Phi^{--}$ &
	$- 2 i g g^{\prime}
	\frac{(c^2-s^2)(c^{\prime 2}-s^{\prime 2})}{4scs^{\prime}c^{\prime}}
		g_{\mu\nu}$ \\
\hline
$W_{L\mu}^+ W_{H\nu}^- \Phi^{++} \Phi^{--}$ &
	$-i g^2 \frac{(c^2-s^2)}{2sc} g_{\mu\nu}$ &
$A_{L\mu} A_{H\nu} \Phi^{++} \Phi^{--}$ &
	$-4 i e g^{\prime}
	\frac{(c^{\prime 2}-s^{\prime 2})}{2s^{\prime}c^{\prime}} 
	g_{\mu\nu}$ \\
$$ &
	$$ &
$Z_{L\mu} Z_{H\nu} \Phi^{++} \Phi^{--}$ &
	$2 i \frac{g^2}{c_\w} \frac{(c^2-s^2)}{2sc}
		( 1 - 2 s^2_\w ) g_{\mu\nu}$ \\
\hline
$A_{L\mu} Z_{H\nu} \Phi^{++} \Phi^{--}$ &
	$4 i e g \frac{(c^2-s^2)}{2sc} g_{\mu\nu}$ &
$Z_{L\mu} A_{H\nu} \Phi^{++} \Phi^{--}$ &
	$-2 i \frac{gg^{\prime}}{c_\w} 
	\frac{(c^{\prime 2}-s^{\prime 2})}{2s^{\prime}c^{\prime}}
		( 1 - 2 s^2_\w ) g_{\mu\nu}$ \\
\hline
\end{tabular}
\caption{Four-point gauge boson-scalar couplings, continued. 
\label{GH4-2}
}
\end{table}


\subsubsection{Gauge boson self-interactions}

The gauge boson self-couplings are given as follows,
with all momenta out-going.  The three-point couplings take the form:
\begin{equation}
	V_1^{\mu}(k_1) V_2^{\nu}(k_2) V_3^{\rho}(k_3): \ \ 
	- i g_{V_1V_2V_3} \left[ g^{\mu\nu}(k_1-k_2)^{\rho}
	+ g^{\nu\rho}(k_2-k_3)^{\mu}
	+ g^{\rho\mu}(k_3-k_1)^{\nu} \right].
\end{equation}
The four-point couplings take the form:
\begin{eqnarray}
	W_1^{+\mu} W_2^{+\nu} W_3^{-\rho} W_4^{-\sigma}:
	&& - i g_{W_1^+W_2^+W_3^-W_4^-} ( 2 g^{\mu\nu} g^{\rho\sigma}
	- g^{\mu\rho}g^{\nu\sigma} - g^{\nu\rho}g^{\mu\sigma} ) \nonumber \\
	V_1^{\mu} V_2^{\nu} W_1^{+ \rho} W_2^{- \sigma}:
	&&  i g_{V_1V_2W_1^+W_2^-} ( 2 g^{\mu\nu} g^{\rho\sigma}
	- g^{\mu\rho} g^{\nu\sigma} - g^{\nu\rho}g^{\mu\sigma} ).
\end{eqnarray}
The coefficients $g_{V_1V_2V_3}$,
$g_{V_1V_2W_1^+W_2^-}$ and $g_{W_1^+W_2^+W_3^-W_4^-}$
are given in Table~\ref{gauges}.

%
\vskip -0.2cm
\begin{table}[h!]
\begin{tabular}{|c|c||c|c|}
\hline
particles & $g_{WWV}$ & particles & $g_{WWV}$ \\
\hline
$W_L^+ W_L^- A_L$ & $-e$ &
$W_L^+ W_L^- Z_L$ & $-ec_\w/s_\w$ \\
\hline
$W_L^+ W_L^- A_H$ & $\frac{e}{s_\w} \frac{v^2}{f^2} c_\w x_Z^{B^{\prime}}$ &
$W_L^+ W_L^- Z_H$ & $\frac{e}{s_\w} \frac{v^2}{f^2} \left( c_\w x_Z^{W^{\prime}}
	+ s c (c^2-s^2)\right)$ \\
\hline
$W_L^+ W_H^- A_L$ & $0$ & 
$W_L^+ W_H^- Z_L$ & $-\frac{e}{s_\w} \frac{v^2}{f^2} x_Z^{W^{\prime}}$ \\
\hline
$W_L^+ W_H^- A_H$ & $- \frac{e}{s_\w} \frac{v^2}{f^2} x_H$ &
$W_L^+ W_H^- Z_H$ & $-e/s_\w$ \\
\hline
$W_H^+ W_H^- A_L$ & $-e$ &
$W_H^+ W_H^- Z_L$ & $-e c_\w/s_\w$ \\
\hline
$W_H^+ W_H^- A_H$ & $\frac{e}{s_\w} \frac{v^2}{f^2} \left(
	x_H \frac{(c^2-s^2)}{sc}
	+ c_\w x_Z^{B^{\prime}} \right)$ &
$W_H^+ W_H^- Z_H$ & $\frac{e}{s_\w} \frac{(c^2-s^2)}{sc}$ \\
\hline \hline
particles & $g_{W_1^+W_2^+W_3^-W_4^-}$ 
	& particles & $g_{W_1^+W_2^+W_3^-W_4^-}$ \\
\hline
$W^+_L W^+_L W^-_L W^-_L$ &
	$- g^2$ &
$W^+_L W^+_L W^-_L W^-_H$ &
	$- g^2 s c (c^2-s^2) v^2 / 4 f^2$ \\
\hline
$W^+_L W^+_L W^-_H W^-_H$ &
	$-g^2$ &
$W^+_L W^+_H W^-_L W^-_H$ &
	$-g^2/4$ \\
\hline
$W^+_H W^+_H W^-_L W^-_H$ &
	$g^2 (c^2-s^2)/ 2 s c$ &
$W^+_H W^+_H W^-_H W^-_H$ &
	$- g^2 (c^6+s^6)/s^2c^2$ \\
\hline \hline
particles & $g_{V_1V_2W_1^+W_2^-}$ 
	& particles & $g_{V_1V_2W_1^+W_2^-}$ \\
\hline
$A_L A_L W^+_L W^-_L$ &
	$- g^2 s^2_{\w}$ &
$A_L A_L W^+_H W^-_H$ &
	$- g^2 s^2_{\w}$ \\
\hline
$Z_L Z_L W^+_L W^-_L$ &
	$-g^2 c^2_{\w}$ &
$Z_L Z_L W^+_H W^-_H$ &
	$- g^2 c^2_{\w}$ \\
\hline
$A_L Z_L W^+_L W^-_L$ &
	$-g^2 s_{\w} c_{\w}$ &
$A_L Z_L W^+_H W^-_H$ &
	$-g^2 s_{\w} c_{\w}$ \\
\hline
$A_L A_H W^+_L W^-_L$ &
	$g^2 s_{\w} c_{\w} x_Z^{B^{\prime}} v^2/f^2$ &
$A_L A_H W^+_H W^-_H$ &
	$g^2 s_{\w} c_{\w} x_Z^{B^{\prime}} v^2/f^2$ \\
$$ &
	$$ &
$$ & 
	$+ g^2 s_{\w} x_H \frac{v^2}{f^2} (c^2-s^2)/sc$ \\
\hline
$A_L Z_H W^+_L W^-_L$ &
	$g^2 s_{\w} c_{\w} x_Z^{W^{\prime}} v^2/f^2$ &	
$A_L Z_H W^+_H W^-_H$ &
	$g^2 s_{\w} (c^2-s^2)/sc$ \\
$$ &
	$- g^2 s_{\w} s c (c^2-s^2) v^2 / 2f^2$ &
$$ &
	$$ \\
\hline
$Z_L Z_H W^+_L W^-_L$ &
	$g^2 (c^2_{\w} - s^2_{\w}) x_Z^{W^{\prime}} v^2/f^2$ &
$Z_L Z_H W^+_H W^-_H$ &
	$g^2 c_{\w} (c^2-s^2)/sc$ \\
\hline
$Z_L A_H W^+_L W^-_L$ &
	$g^2 c^2_{\w} x_Z^{B^{\prime}} v^2/f^2$ &
$Z_L A_H W^+_H W^-_H$ &
	$g^2 c^2_{\w} x_Z^{B^{\prime}} v^2/f^2$ \\
$$ &
	$$ &
$$ &
	$+ g^2 c_{\w} x_H \frac{v^2}{f^2} (c^2-s^2)/sc$ \\
\hline
$A_H A_H W^+_L W^-_L$ &
	$\mathcal{O}(v^4/f^4)$ &
$A_H A_H W^+_H W^-_H$ &
	$\mathcal{O}(v^4/f^4)$ \\
\hline
$Z_H Z_H W^+_L W^-_L$ &
	$-g^2$ &
$Z_H Z_H W^+_H W^-_H$ &
	$-g^2 (c^6+s^6)/s^2c^2$ \\
\hline
$Z_H A_H W^+_L W^-_L$ &
	$-g^2 x_H v^2/f^2$ &
$Z_H A_H W^+_H W^-_H$ &
	$-g^2 x_H \frac{v^2}{f^2} (c^6+s^6)/s^2c^2$ \\
$$ &
	$$ &
$$ &
	$- g^2 c_{\w} x_Z^{B^{\prime}} \frac{v^2}{f^2} (c^2-s^2)/sc$ \\
\hline
$A_L A_L W^+_L W^-_H$ &
	$0$ &
$A_H A_H W^+_L W^-_H$ &
	$\mathcal{O}(v^4/f^4)$ \\
\hline
$Z_L Z_L W^+_L W^-_H$ &
	$-2 g^2 c_{\w} x_Z^{W^{\prime}} v^2/f^2$ &
$Z_H Z_H W^+_L W^-_H$ &
	$g^2 (c^2-s^2)/sc$ \\
\hline
$Z_L A_L W^+_L W^-_H$ &
	$-g^2 s_{\w} x_Z^{W^{\prime}} v^2 / 2f^2$ &
$Z_H A_H W^+_L W^-_H$ &
	$g^2 x_H \frac{v^2}{f^2} (c^2-s^2)/sc$ \\
$$ &
	$$ &
$$ &
	$+ g^2 c_{\w} x_Z^{B^{\prime}} v^2/f^2$ \\
\hline
$A_L A_H W^+_L W^-_H$ &
	$-g^2 s_{\w} x_H v^2/f^2$ &
$A_L Z_H W^+_L W^-_H$ &
	$-g^2 s_{\w}$ \\
\hline
$Z_L A_H W^+_L W^-_H$ &
	$-g^2 c_{\w} x_H v^2/f^2$ &
$Z_L Z_H W^+_L W^-_H$ &
	$-g^2 c_{\w}$ \\
\hline
\end{tabular}
\caption{Gauge boson self-couplings. }
\label{gauges}
\end{table}
%

\subsection{Couplings between gauge bosons and fermions}

The couplings between gauge bosons and fermions are given in 
Tables~\ref{gauge-fermion} and \ref{gauge-fermion2}.  The charged
gauge boson couplings to fermions in Table~\ref{gauge-fermion} 
are all left-handed, and the projection operator 
$P_L=(1-\gamma^5)/2$ is implied.
We define $x_L \equiv \lambda_1^2/(\lambda_1^2+\lambda_2^2)$
to shorten the notation.
\begin{table}[h]
\begin{tabular}{|c|c||c|c|}
\hline
particles & vertices & particles & vertices \\
\hline
$W_L^{+\mu} \bar u_L d_L$ &
	$\frac{ig}{\sqrt{2}}  
	\left[ 1 
	- \frac{v^2}{2f^2} c^2 (c^2-s^2) \right]
	\gamma^{\mu} V_{ud}^{\rm SM}$ &
$W_H^{+\mu} \bar u_L d_L$ &
	$-\frac{ig}{\sqrt{2}} \frac{c}{s}
	\gamma^{\mu} V_{ud}^{\rm SM}$ \\
\hline
$W_L^{+\mu} \bar t_L b_L$ &
	$\frac{ig}{\sqrt{2}}
	\left[ 1 - \frac{v^2}{f^2} \left( \frac{1}{2} x_L^2
	+ \frac{1}{2} c^2 (c^2-s^2) \right) \right]
	\gamma^{\mu} V_{tb}^{\rm SM}$ &
$W_H^{+\mu} \bar t_L b_L$ &
	$-\frac{ig}{\sqrt{2}} \frac{c}{s}
	\gamma^{\mu} V_{tb}^{\rm SM}$ \\
\hline
$W_L^{+\mu} \bar T_L b_L$ &
	$\frac{g}{\sqrt{2}}
	\frac{v}{f} x_L
	\gamma^{\mu} V_{tb}^{\rm SM}$ &
$W_H^{+\mu} \bar T_L b_L$ &
	$-\frac{g}{\sqrt{2}}
	\frac{v}{f} x_L \frac{c}{s}
	\gamma^{\mu} V_{tb}^{\rm SM}$ \\
\hline
\end{tabular}
\caption{Charged gauge boson-fermion couplings. They are purely left-handed,
and the projection operator $P_L=(1-\gamma^5)/2$ is implied.}
\label{gauge-fermion}
\end{table}

For the neutral gauge bosons in Table~\ref{gauge-fermion2},
we write the couplings to fermions in the
form $i \gamma^{\mu} (g_V + g_A \gamma^5)$.
The fermion charge assignments under the two $U(1)$ groups are 
given in Table \ref{tab:hypercharge}, 
requiring only 
gauge invariance of the scalar-fermion couplings in Eq.~(\ref{yuk})
for the top quark and similar equations for the other fermions.
The additional requirement that the two $U(1)$ groups be anomaly-free
fixes $y_u = -2/5$ and $y_e = 3/5$.

\newpage

%
\begin{table}
\begin{tabular}{|c|c|c|}
\hline
particles & $g_V$ & $g_A$ \\
\hline
$A_L \bar f f$ & $-e Q_f$ & $0$ \\
\hline
$Z_L \bar u u$ & 
	$-\frac{g}{2c_{\w}} \left\{ (\frac{1}{2} - \frac{4}{3} s^2_{\w})
	- \frac{v^2}{f^2} \left[ c_{\w} x_Z^{W^{\prime}} c/2s
	\right. \right.$ &
	$-\frac{g}{2c_{\w}} \left\{ -\frac{1}{2}
	- \frac{v^2}{f^2} \left[ -c_{\w} x_Z^{W^{\prime}} c/2s
	\right. \right.$ \\
$$ &
	$\left. \left.
	+ \frac{s_{\w} x_Z^{B^{\prime}}}{s^{\prime}c^{\prime}}
	\left( 2y_u + \frac{7}{15} - \frac{1}{6} c^{\prime 2}
	\right) \right] \right\} $
	&
	$\left. \left.
	+ \frac{s_{\w} x_Z^{B^{\prime}}}{s^{\prime}c^{\prime}}
	\left( \frac{1}{5} - \frac{1}{2} c^{\prime 2} \right) 
	\right] \right\}$ \\
\hline
$Z_L \bar d d$ &
	$-\frac{g}{2c_{\w}} \left\{ (-\frac{1}{2} + \frac{2}{3} s^2_{\w})
	- \frac{v^2}{f^2} \left[ -c_{\w} x_Z^{W^{\prime}} c/2s
	\right. \right.$ &
	$-\frac{g}{2c_{\w}} \left\{ \frac{1}{2}
	- \frac{v^2}{f^2} \left[ c_{\w} x_Z^{W^{\prime}} c/2s
	\right. \right.$ \\
$$ &
	$\left. \left.
	+ \frac{s_{\w} x_Z^{B^{\prime}}}{s^{\prime}c^{\prime}}
	\left( 2y_u + \frac{11}{15} + \frac{1}{6} c^{\prime 2}
	\right) \right] \right\}$ &
	$\left. \left.
	+ \frac{s_{\w} x_Z^{B^{\prime}}}{s^{\prime}c^{\prime}}
	\left( -\frac{1}{5} + \frac{1}{2} c^{\prime 2} \right)
	\right] \right\}$ \\
\hline
$Z_L \bar e e$ &
	$-\frac{g}{2c_{\w}} \left\{ (-\frac{1}{2} + 2 s^2_{\w})
	- \frac{v^2}{f^2} \left[ -c_{\w} x_Z^{W^{\prime}} c/2s
	\right. \right.$ &
	$-\frac{g}{2c_{\w}} \left\{ \frac{1}{2}
	- \frac{v^2}{f^2} \left[ c_{\w} x_Z^{W^{\prime}} c/2s
	\right. \right.$ \\
$$ &
	$\left. \left.
	+ \frac{s_{\w} x_Z^{B^{\prime}}}{s^{\prime}c^{\prime}}
	\left( 2y_e - \frac{9}{5} + \frac{3}{2} c^{\prime 2} 
	\right) \right] \right\}$ &
	$\left. \left.
	+ \frac{s_{\w} x_Z^{B^{\prime}}}{s^{\prime}c^{\prime}}
	\left( -\frac{1}{5} + \frac{1}{2} c^{\prime 2} \right)
	\right] \right\}$ \\
\hline
$Z_L \bar \nu \nu$ &
	$-\frac{g}{2c_{\w}} \left\{ \frac{1}{2}
	- \frac{v^2}{f^2} \left[ c_{\w} x_Z^{W^{\prime}} c/2s
	\right. \right.$ &
	$-\frac{g}{2c_{\w}} \left\{ -\frac{1}{2}
	- \frac{v^2}{f^2} \left[ -c_{\w} x_Z^{W^{\prime}} c/2s
	\right. \right.$ \\
$$ &
	$\left.\left.
	+ \frac{s_{\w} x_Z^{B^{\prime}}}{s^{\prime}c^{\prime}}
	\left( y_e - \frac{4}{5} + \frac{1}{2} c^{\prime 2}
	\right) \right] \right\}$ &
	$\left. \left.
	+ \frac{s_{\w} x_Z^{B^{\prime}}}{s^{\prime}c^{\prime}}
	\left( -y_e + \frac{4}{5} - \frac{1}{2} c^{\prime 2}
	\right) \right] \right\}$ \\
\hline
$Z_L \bar t t$ &
	$-\frac{g}{2c_{\w}} \left\{ (\frac{1}{2} - \frac{4}{3} s^2_{\w})
	- \frac{v^2}{f^2} \left[ - x_L^2/2 
	+ c_{\w} x_Z^{W^{\prime}} c/2s
	\right. \right.$ &
	$-\frac{g}{2c_{\w}} \left\{ -\frac{1}{2}
	- \frac{v^2}{f^2} \left[ x_L^2/2
	- c_{\w} x_Z^{W^{\prime}} c/2s
	\right. \right.$ \\
$$ &
	$\left. \left.
	+ \frac{s_{\w} x_Z^{B^{\prime}}}{s^{\prime}c^{\prime}}
	\left( 2y_u + \frac{9}{5} - \frac{3}{2} c^{\prime 2}
	+ \left( \frac{7}{15} - \frac{2}{3} c^{\prime 2} \right)
	\frac{\lambda_1^2}{\lambda_1^2 + \lambda_2^2}
	\right) \right] \right\}$ &
	$\left. \left.
	+ \frac{s_{\w} x_Z^{B^{\prime}}}{s^{\prime}c^{\prime}}
	\left( \frac{1}{5} - \frac{1}{2} c^{\prime 2}
	- \frac{1}{5} \frac{\lambda_1^2}{\lambda_1^2 + \lambda_2^2}
	\right) \right] \right\}$ \\
\hline
$Z_L \bar T T$ &
	$2gs^2_{\w} / 3c_{\w}$ &
	$\mathcal{O}(v^2/f^2)$ \\
$Z_L \bar T t$ &
	$ig x_L v / 4 f c_{\w}$ &
	$-ig x_L v / 4 f c_{\w}$ \\
\hline
$A_H \bar u u$ & 
	$\frac{g^{\prime}}{2s^{\prime}c^{\prime}}
	\left( 2y_u + \frac{17}{15} - \frac{5}{6} c^{\prime 2} \right)$ &
	$\frac{g^{\prime}}{2s^{\prime}c^{\prime}}
	\left( \frac{1}{5} - \frac{1}{2} c^{\prime 2} \right)$ \\
$A_H \bar d d$ &
	$\frac{g^{\prime}}{2s^{\prime}c^{\prime}}
	\left( 2y_u + \frac{11}{15} + \frac{1}{6} c^{\prime 2} \right)$ &
	$\frac{g^{\prime}}{2s^{\prime}c^{\prime}}
	\left( -\frac{1}{5} + \frac{1}{2} c^{\prime 2} \right)$ \\
$A_H \bar e e$ &
	$\frac{g^{\prime}}{2s^{\prime}c^{\prime}}
	\left( 2y_e - \frac{9}{5} + \frac{3}{2} c^{\prime 2} \right)$ &
	$\frac{g^{\prime}}{2s^{\prime}c^{\prime}}
	\left( -\frac{1}{5} + \frac{1}{2} c^{\prime 2} \right)$ \\
$A_H \bar \nu \nu$ &
	$\frac{g^{\prime}}{2s^{\prime}c^{\prime}}
	\left( y_e - \frac{4}{5} + \frac{1}{2} c^{\prime 2} \right)$ &
	$\frac{g^{\prime}}{2s^{\prime}c^{\prime}}
	\left( -y_e + \frac{4}{5} - \frac{1}{2} c^{\prime 2} \right)$ \\
\hline
$A_H \bar t t$ &
	$\frac{g^{\prime}}{2s^{\prime}c^{\prime}}
	\left( 2y_u + \frac{17}{15} - \frac{5}{6} c^{\prime 2}
	- \frac{1}{5} \frac{\lambda_1^2}{\lambda_1^2 + \lambda_2^2} \right)$ &
	$\frac{g^{\prime}}{2s^{\prime}c^{\prime}}
	\left( \frac{1}{5} - \frac{1}{2} c^{\prime 2}
	- \frac{1}{5} \frac{\lambda_1^2}{\lambda_1^2 + \lambda_2^2} \right)$ \\
$A_H \bar T T$ &
	$\frac{g^{\prime}}{2s^{\prime}c^{\prime}}
	\left( 2y_u + \frac{14}{15} - \frac{4}{3} c^{\prime 2}
	+ \frac{1}{5} \frac{\lambda_1^2}{\lambda_1^2 + \lambda_2^2} \right)$ &
	$\frac{g^{\prime}}{2s^{\prime}c^{\prime}}
	\frac{1}{5} \frac{\lambda_1^2}{\lambda_1^2 + \lambda_2^2}$ \\
$A_H \bar T t$ &
	$\frac{g^{\prime}}{2s^{\prime}c^{\prime}}
	\frac{1}{5} \frac{\lambda_1 \lambda_2}{\lambda_1^2 + \lambda_2^2}$ &
	$\frac{g^{\prime}}{2s^{\prime}c^{\prime}}
	\frac{1}{5} \frac{\lambda_1 \lambda_2}{\lambda_1^2 + \lambda_2^2}$ \\
\hline
$Z_H \bar u u$ & $gc/4s$ & $-gc/4s$ \\
$Z_H \bar d d$ & $-gc/4s$ & $gc/4s$ \\
$Z_H \bar e e$ & $-gc/4s$ & $gc/4s$ \\
$Z_H \bar \nu \nu$ & $gc/4s$ & $-gc/4s$ \\
\hline
$Z_H \bar t t$ & $gc/4s$ & $-gc/4s$ \\
$Z_H \bar T T$ & $\mathcal{O}(v^2/f^2)$ & $\mathcal{O}(v^2/f^2)$ \\
$Z_H \bar T t$ &
	$-i g x_L vc/4fs$ &
	$i g x_L vc/4fs$ \\
\hline
\end{tabular}
\caption{Neutral gauge boson-fermion couplings. The $U(1)$
charge assignments are given in Table \ref{tab:hypercharge}.
Anomaly cancellation requires $y_u = -2/5$ and $y_e = 3/5$.
}
\label{gauge-fermion2}
\end{table}

\vspace*{.5in}

\subsection{Couplings between scalars and fermions}

The scalar-fermion couplings are listed in Table \ref{scalar-fermion}.

\begin{table}[h]
\begin{tabular}{|c|c||c|c|}
\hline
particles & vertices & particles & vertices \\
\hline
$H \bar u u$ &
	$-i \frac{m_u}{v} \left( 1 - \frac{1}{2} s_0^2
	+ \frac{v}{f} \frac{s_0}{\sqrt{2}} - \frac{2 v^2}{3 f^2} \right)$ &
$H\bar d d$ &
	$-i \frac{m_d}{v} \left( 1 - \frac{1}{2} s_0^2
	+ \frac{v}{f} \frac{s_0}{\sqrt{2}} - \frac{2 v^2}{3 f^2} \right)$ \\
\hline
$H \bar t t$ &
	$- i \frac{m_t}{v} \left[ 1 - \frac{1}{2} s_0^2
	+ \frac{v}{f} \frac{s_0}{\sqrt{2}} - \frac{2 v^2}{3 f^2} \right.$ &
$H \bar T T$ &
	$- i \frac{\lambda_1^2}{\sqrt{\lambda_1^2 + \lambda_2^2}}
	\left( 1 + \frac{\lambda_1^2}{\lambda_1^2 + \lambda_2^2}\right)
	\frac{v}{f}$ \\
 &
	$\left. 
	+ \frac{v^2}{f^2} \frac{\lambda_1^2}{\lambda_1^2 + \lambda_2^2}
	\left( 1 + \frac{\lambda_1^2}{\lambda_1^2 + \lambda_2^2} \right)
	\right]$ &
 & \\
\hline
$H \bar T t$ &
	$\frac{m_t}{v} \frac{v}{f} 
	\left( 1 + \frac{\lambda_1^2}{\lambda_1^2 + \lambda_2^2} \right) P_R
	+ \frac{\lambda_1^2}{\sqrt{\lambda_1^2 + \lambda_2^2}} P_L$ &
$H \bar t T$ &
	$-\frac{m_t}{v} \frac{v}{f} 
	\left( 1 + \frac{\lambda_1^2}{\lambda_1^2 + \lambda_2^2} \right) P_L
	- \frac{\lambda_1^2}{\sqrt{\lambda_1^2 + \lambda_2^2}} P_R$ \\
\hline
$H H \bar t t$ &
	$i \frac{2m_t}{f^2} \left[1 - \frac{2fv^{\prime}}{v^2}
	- \frac{1}{2} \frac{\lambda_1^2}{\lambda_1^2+\lambda_2^2} \right]$ &
$H H \bar T T$ &
	$-\frac{i}{f} \frac{\lambda_1^2}{\sqrt{\lambda_1^2+\lambda_2^2}}$ \\
\hline
$H H \bar T t$ &
	$- \frac{2v}{f^2} \frac{\lambda_1^2}{\sqrt{\lambda_1^2+\lambda_2^2}}
	\left[1 - \frac{2fv^{\prime}}{v^2}
	- \frac{1}{2} \frac{\lambda_1^2}{\lambda_1^2+\lambda_2^2} \right]
	P_L$ &
$H H \bar t T$ &
	$\frac{2v}{f^2} \frac{\lambda_1^2}{\sqrt{\lambda_1^2+\lambda_2^2}}
	\left[1 - \frac{2fv^{\prime}}{v^2}
	- \frac{1}{2} \frac{\lambda_1^2}{\lambda_1^2+\lambda_2^2} \right]
	P_R$ \\
 & $+ \frac{m_t}{vf} P_R$ &
 & $- \frac{m_t}{vf} P_L$ \\
\hline
$\Phi^0 \bar u u$ &
	$-\frac{im_u}{\sqrt{2}v} 
	\left( \frac{v}{f} - \sqrt{2} s_0 \right)$ &
$\Phi^0 \bar d d$ &
	$-\frac{im_d}{\sqrt{2}v} 
	\left( \frac{v}{f} - \sqrt{2} s_0 \right)$ \\
\hline
$\Phi^P \bar u u$ &
	$- \frac{m_u}{\sqrt{2}v} 
	\left( \frac{v}{f} - \sqrt{2} s_P \right) \gamma^5$ &
$\Phi^P \bar d d$ &
	$\frac{m_d}{\sqrt{2}v} 
	\left( \frac{v}{f} - \sqrt{2} s_P \right) \gamma^5$ \\
\hline
$\Phi^+ \bar u d$ &
	$-\frac{i}{\sqrt{2}v}
	\left( m_u P_L + m_d P_R \right) \left( \frac{v}{f} - 2 s_+ \right)$ &
$\Phi^- \bar d u$ &
	$-\frac{i}{\sqrt{2}v}
	\left( m_u P_R + m_d P_L \right) \left( \frac{v}{f} - 2 s_+ \right)$ \\
\hline
$\Phi^0 \bar T t$ &
	$-\frac{im_t}{\sqrt{2}v} \left(\frac{v}{f} - \sqrt{2} s_0 \right)
	\frac{\lambda_1}{\lambda_2} P_L$ &
$\Phi^0 \bar t T$ &
	$-\frac{im_t}{\sqrt{2}v} \left(\frac{v}{f} - \sqrt{2} s_0 \right)
	\frac{\lambda_1}{\lambda_2} P_R$ \\
\hline
$\Phi^P \bar T t$ &
	$\frac{m_t}{\sqrt{2}v} \left(\frac{v}{f} - \sqrt{2} s_0 \right)
	\frac{\lambda_1}{\lambda_2} P_L$ &
$\Phi^P \bar t T$ &
	$\frac{m_t}{\sqrt{2}v} \left(\frac{v}{f} - \sqrt{2} s_0 \right)
	\frac{\lambda_1}{\lambda_2} P_R$ \\
\hline
$\Phi^+ \bar T b$ &
	$-\frac{im_t}{\sqrt{2}v} \left(\frac{v}{f} - 2 s_+ \right)
	\frac{\lambda_1}{\lambda_2} P_L$ &
$\Phi^- \bar b T$ &
	$-\frac{im_t}{\sqrt{2}v} \left(\frac{v}{f} - 2 s_+ \right)
	\frac{\lambda_1}{\lambda_2} P_R$ \\
\hline
\end{tabular}
\caption{Scalar-fermion couplings. }
\label{scalar-fermion}
\end{table}

\end{document}